\documentclass[useAMS,usenatbib]{mn2e}
\usepackage{graphicx,rotate,url,mathptmx,lscape,times}
\def\gtsim{\mathrel{\hbox{\rlap{\hbox{\lower4pt\hbox{$\sim$}}}\hbox{$>$}}}}
\def\lesssim{\mathrel{\hbox{\rlap{\hbox{\lower4pt\hbox{$\sim$}}}\hbox{$<$}}}}

\def\Msun{M$_{\odot}$}
\def\cm{{\rm\thinspace cm}}

\def\erg{{\rm\thinspace erg}}
\def\eV{{\rm\thinspace eV}}

\def\K{{\rm\thinspace K}}

\def\km{{\rm\thinspace km}}
\def\kpc{{\rm\thinspace kpc}}

\def\Msun{\mbox{$\rm\thinspace M_{\odot}$}}

\def\pc{{\rm\thinspace pc}}
\def\s{{\rm\thinspace s}}
\def\ps{{\rm\thinspace s^{-1}}}
\def\pcc{{\rm\thinspace cm^{-3}}}

\def\yr{{\rm\thinspace yr}}
\def\ergps{\mbox{$\erg\s^{-1}\,$}}

\def\kmps{\mbox{$\km\ps\,$}}

\def\pscm{\mbox{$\cm^{-2}\,$}}

\def\pscm{\mbox{$\cm^{-2}\,$}}

\def\hi{\mbox{{\rm H~{\sc i}}}}
\def\hei{\mbox{{\rm He~{\sc i}}}}
\def\heii{\mbox{{\rm He~{\sc ii}}}}
\def\ci{\mbox{{\rm C~{\sc i}}}}
\def\cii{\mbox{{\rm C~{\sc ii}}}}

\def\ni{\mbox{{\rm N~{\sc i}}}}
\def\nii{\mbox{{\rm N~{\sc ii}}}}
\def\oi{\mbox{{\rm O~{\sc i}}}}
\def\oii{\mbox{{\rm O~{\sc ii}}}}
\def\oiii{\mbox{{\rm O~{\sc iii}}}}
\def\neii{\mbox{{\rm Ne~{\sc ii}}}}
\def\neiii{\mbox{{\rm Ne~{\sc iii}}}}
\def\mgii{\mbox{{\rm Mg~{\sc ii}}}}
\def\sii{\mbox{{\rm S~{\sc ii}}}}
\def\siii{\mbox{{\rm S~{\sc iii}}}}
\def\ariii{\mbox{{\rm Ar~{\sc iii}}}}
\def\caii{\mbox{{\rm Ca~{\sc ii}}}}
\def\CO{\mbox{{\rm CO}}}
\def\htwo{\mbox{{\rm H}$_2$}}
\def\hplus{\mbox{{\rm H}$^+$}}
\def\hminus{\mbox{{\rm H}$^-$}}
\def\h0{\mbox{{\rm H}$^0$}}
\def\he0{\mbox{{\rm He}$^0$}}
\DeclareMathAlphabet{\vib}{OML}{cmm}{m}{it}
\title[Collisional heating of cool cluster filaments]{Collisional heating as
the origin of filament emission in galaxy clusters\thanks{Contains material \copyright\ British Crown copyright 2008/MoD}}

\author[G.J. Ferland et al.]
       {\parbox[]{6.0in}
        {G.J. Ferland$^{1,2}$\thanks{E-mail: gary@pa.uky.edu},
        A.C. Fabian$^1$,
        N.A. Hatch$^{3}$,
        R.M. Johnstone$^1$,\\
        R.L. Porter$^{1,2}$,
        P. A. M. van Hoof$^4$
        and R.J.R. Williams$^{5}$\\
        \footnotesize
        $^1$Institute of Astronomy, University of Cambridge, Madingley
        Road, Cambridge CB3 0HA\\
        $^2$Department of Physics, University of Kentucky, Lexington, KY 40506, USA\\
        $^3$Leiden Observatory, University of Leiden, P.B. 9513, Leiden 2300 RA, The Netherlands\\
        $^4$Royal Observatory of Belgium, Ringlaan 3, 1180 Brussels, Belgium\\
        $^5$AWE plc, Aldermaston, Reading RG7 4PR}}

\date{%Accepted .
      Received }

\pagerange{\pageref{firstpage}--\pageref{lastpage}}
\pubyear{}

\begin{document}

\maketitle

\label{firstpage}

\begin{abstract}
\noindent
It has long been known that photoionization, whether by starlight or other sources,
has difficulty in accounting for the observed spectra of the optical filaments
that often surround central galaxies in large clusters.
This paper builds on the first of this series in which we examined
whether heating by energetic particles or dissipative MHD wave
can account for the observations.
The first paper focused on the molecular regions
which produce strong \htwo\ and CO lines.
Here we extend the calculations to include atomic and low-ionization regions.
Two major improvements to the previous calculations have been made.
The model of the hydrogen atom, along with all elements of the H-like
iso-electronic sequence, is now fully \emph{nl}-resolved.
This allows us to predict the hydrogen emission-line spectrum
including excitation by suprathermal secondary electrons
and thermal electrons or nuclei.
We show how the predicted \hi\ spectrum
differs from the pure recombination case.
The second update is to the rates for H$^0$ -- \htwo\ inelastic collisions.
We now use the values computed by Wrathmall et al.
The rates are often much larger and allow the
ro-vibrational \htwo\ level populations to achieve a thermal
distribution at substantially lower densities than previously thought.

We calculate the chemistry, ionization, temperature, gas pressure,
and emission-line spectrum for
a wide range of gas densities and collisional heating rates.
We assume that the filaments are magnetically confined.
The gas is free to move along field lines so that
the gas pressure is equal to that of the surrounding hot gas.
A mix of clouds, some being dense and cold and
others hot and tenuous, can exist.
The observed spectrum will be the integrated
emission from clouds with different densities and
temperatures but the same pressure $P/k = nT$.
We assume that the gas filling factor is given by
a power law in density.
The power-law index, the only free parameter in this theory,
is set by matching the observed intensities of IR
\htwo\ lines relative to optical \hi\ lines.
We conclude that the filaments are heated by ionizing
particles, either conducted in from surrounding regions or
produced \emph{in situ} by processes related to MHD waves.
\end{abstract}


\section{Introduction}
\label{intro}

The central galaxies in large clusters are frequently surrounded by a
system of filaments that emit strong
molecular, atomic, and low-ionization emission lines.
Understanding the origin of this line emission
has been a long-standing challenge \citep{JohnstoneEtAl07}.
The line luminosities are too great for the filaments to be powered by
known sources of radiation such as the central AGN
or the diffuse emission from the surrounding hot gas.
The evolutionary state of the gas is totally unknown.
Two possibilities are that they form from surrounding
hot gas or by ejection from
the central galaxy in the cluster.
Star formation may occur in the filaments although the emission-line spectra
do not resemble those of Galactic H~{\sc ii} regions near early-type stars.
Questions concerning the origin, energy source, and evolutionary history
are important, in part because of the large mass that may be involved,
as high as
$\sim 4 \times 10^{10}\Msun$ according to \citet{SalomeEtAl06}.

Infrared spectra show \htwo\ lines which are
far stronger relative to \hi\ lines
than those emitted by molecular gas near O stars.
Attempts at reproducing the spectra assuming starlight photoionization are largely
unsuccessful, as reviewed by \citet{JohnstoneEtAl07}.
Hybrid models, in which starlight produces the optical emission
while other energy sources produce the molecular emission,
appear necessary.
The fact that optical and molecular emission luminosities trace one another
\citep{JaffeEtAl05}
would require that these independent energy sources have
correlated luminosities.

Since photoionization by O and B stars does not appear able to account for the spectrum,
Paper 1 \citep{FerlandEtAl08}
considered whether purely collisional heating sources can
reproduce the observed spectrum.
Magnetic fields occur in the environment and wave energy is likely to be
associated with the field.
Dissipation of this wave energy could heat the gas.
Cosmic rays are also present \citep{SandersFabian07}.
These would heat and ionize the gas and produce strong low-ionization emission.
MHD waves may also accelerate low-energy cosmic rays within a filament.
Similar particle and wave processes
occur in stellar coronae where they are collectively
referred to as non-radiative energy sources, a term we shall use in the
remainder of this paper.

Paper 1 focused on \htwo\ lines produced in
molecular regions that are well shielded from light.
\citet{JohnstoneEtAl07} detected \htwo\ lines with a wide range of excitation potential
and found a correlation between excitation and the derived population temperature.
We found that non-radiative heating with a range of heating rates can reproduce
the observed \htwo\ spectrum.
In this paper we concentrate on atomic and low-ionization emission
and develop the methodology needed to predict the spectrum of gas with a range
of densities but a single gas pressure.
We find that cosmic-ray heating produces a spectrum that is in general
agreement with a wide range of observations.
Purely thermalized
energy injection cannot reproduce the spectrum.
This does not rule out MHD wave heating but does require that they produce
or accelerate high-energy particles in addition to other forms of energy.
The resulting model, while empirical, points the way for a physical model of
the origin of the filament emission.

\section{Spectral simulations}

\subsection{The basic model}
\label{sec:TheBasicModel}

Starlight photoionization has long been known to
have difficulty in reproducing observations of cluster filaments.
Hybrid models, in which different energy sources produce the molecular
and atomic emission are more successful but have problems accounting
for why different energy sources would correlate with one another.
Here we assume that \emph{only}
non-radiative heating, by either cosmic rays or dissipative MHD waves is important.
The entire spectrum is produced by these energy sources.

For simplicity we assume that the gas is well shielded from
significant sources of radiative heating.
In reality light from the central galaxy or the surrounding
hot gas will photoionize a thin skin on the surface of a cloud but will
have little effect on the majority of the cloud's core.
The emission lines emitted by the ionized layer will be faint.
In the calculations that follow we include the metagalactic radiation background,
including the CMB, so that the continuum
is fully defined from the gamma-ray through
the radio.
As described in Paper 1 this external continuum is extinguished by a cold neutral
absorber with a column density of $10^{21}$~cm$^{-2}$ to approximate the
radiation field deep within the filaments.
This continuum is faint enough to have little effect on the
predictions in this paper.

In keeping with our assumption that the regions we model are well shielded,
resonance lines such as the Lyman lines of H~{\sc i} or the Lyman-Werner electronic
systems of \htwo\ are assumed to be optically thick.
Because of this continuum fluorescent excitation of \h0 and \htwo\ does not occur.
With these assumptions the conditions in the cloud are mainly
determined by the non-radiative heating
sources which are the novel aspect of this work.

For simplicity we assume that the chemical composition is the same as
the Local ISM.  The detailed gas-phase abundances are based on
emission-line observations of the Orion star-forming environment and
are given in Table~\ref{tab:abund}.  A Galactic dust-to-gas ratio is
assumed.  Refractory elements are depleted from the gas phase in
keeping with our assumption that dust is present.

\begin{table}
\caption{Assumed gas-phase abundances by log nucleon number density
[cm$^{-3}$] relative to hydrogen.}
\begin{center}
\begin{tabular}{ll|ll|ll}
\hline
He& $-1.022$& Li&$-10.268$& Be&$-20.000$\\
B&$-10.051$& C&$ -3.523$&N&$ -4.155$\\
O&$ -3.398$& F&$-20.000$& Ne&$ -4.222$\\
Na&$ -6.523$&Mg&$ -5.523$& Al&$ -6.699$\\
 Si&$ -5.398$& P&$ -6.796$& S&$ -5.000$\\
Cl&$ -7.000$& Ar&$ -5.523$& K&$ -7.959$\\
Ca&$ -7.699$& Sc&$-20.000$&Ti&$ -9.237$\\
 V&$-10.000$& Cr&$ -8.000$& Mn&$ -7.638$\\
 Fe&$ -5.523$&Co&$-20.000$& Ni&$ -7.000$\\
 Cu&$ -8.824$& Zn&$ -7.6990$ \\
\hline
\end{tabular}
\end{center}
\label{tab:abund}
\end{table}

Grains have several effects on the gas.
\htwo\ forms by catalytic reactions
on grain surfaces in dusty environments.
Collisions between gas and dust
tend to bring them to the same temperature.
This process either heats or cools the gas
depending on whether the grain temperature is above or below the gas temperature.
Molecules can
condense as ices coating the grains if the dust becomes cold enough.
Each of these processes is considered in detail in Cloudy,
the spectral synthesis code we use here,
but the underlying grain theory depends on knowing the grain material,
its size distribution, and the dust to gas ratio.
The grain temperature depends on the UV--IR radiation field within the core,
which in turn depends on whether {\em in situ}\/ star formation has occurred.
Rather than introduce all of these as additional free parameters
we simply adopt the grain \htwo\ catalysis rate measured in the Galactic ISM
\citep{Jura75}.
We do not consider grain-gas energy exchange and neglect
condensations of molecules onto grain ices.
Tests show that these assumptions mainly affect the details of the \h0 -- \htwo\ transition.
One goal of our work is to develop a physical model that accounts for
the spectral observations.
A long-term goal is to use such a model to determine the grain properties
from observations of line extinction and the infrared continuum.

Substantial uncertainties are introduced by the need to assume
a specific gas-phase composition and dust properties.
It would be surprising if the gas and dust composition happened to match
that of the local ISM,
although it would also be surprising if it were greatly different.
The molecular collision rates, described in Section \ref{sec:H2CollisionRates},
have their own substantial uncertainties, probably 0.3 dex or more.
These considerations suggest that there is roughly a factor-of-two uncertainty in
the results we present below.
This is intended as a `proof of concept' calculation aimed at identifying
what physical processes may power the observed emission.
If successful, we can then invert the problem and determine the composition
or evolutionary history from the spectrum.

\subsection{Non-radiative energy sources}

Paper 1 considered two cases, heating by dissipative MHD wave energy,
which we assumed to be deposited as thermal energy,
and cosmic rays, which both heat and ionize the gas.
We refer to these as the `extra heating' and `cosmic ray' cases below.
The effects of these energy sources on the microphysics are fundamentally different.

Supersonic line widths,
often thought to be due to MHD waves associated with the magnetic field,
are observed in the ISM of our galaxy \citep{DysonWilliams97}.
The field and wave kinetic energy are often in rough energy
equipartition \citep{MyersGoodman98} although the details remain uncertain.
\citet{HeilesCrutcher05} review the extensive numerical simulations
of MHD waves.
Waves can be damped by processes such as ion-neutral drift,
which convert wave energy into other forms of kinetic energy,
although the details are uncertain and the process may be unstable \citep{TytarenkoWilliamsFalle02}.
In this exploratory work we simply want to quantify the effects of such
heating on otherwise well-shielded gas.
We adopt a heating rate that is proportional to density,
\begin{equation}
H = H_0 \left[ {n({\rm H})/n_0 } \right] \ \ [{\rm erg} \pcc \ps ]  .
\end{equation}
This form, modified from that used in Paper 1, includes
the ratio of the hydrogen density $n$(H) to a scale density
$n_0$ which we take as $1{\rm\,cm^{-3}}$.
This density dependence causes the wave-heating rate
to go to zero as the gas density goes to zero.

We parameterize the extra heating rate
as the leading coefficient in the heating rate $H_0$.
We assume that the heating simply adds to the thermal energy of the gas
so that the velocity distribution remains a Maxwellian with
a well-defined kinetic temperature.
With these assumptions the only collisional processes which occur are those which
are energetically possible at the local gas kinetic temperature.
This is a major distinction between the extra-heating case
and the cosmic-ray case described next.

The second case we consider is energy deposition
by ionizing particles.  These particles could be related to the high-energy
particles which are known to exist in the hot gas surrounding
the filaments \citep{SandersFabian07},
or could be caused by MHD-related phenomena
like magnetic reconnection \citep{Lazarian05}.
Whatever their fundamental source we will refer to this as the
cosmic-ray case for simplicity.
As in Paper 1 we specify the ionizing particle density
in terms of the equivalent cosmic-ray density
relative to the Galactic background.
We adopt the background cosmic ray \htwo\ dissociation rate of
$3\times 10^{-17} \ps $ \citep{WilliamsElAl98}.
\citet{SandersFabian07} find an electron energy density that is
roughly 10$^3$ times the galactic background in inner regions of the Perseus cluster.
This value guided our choice of the range of cosmic-ray densities shown
in the calculations which follow.

There is some evidence that this Galactic background cosmic ray
\htwo\ dissociation rate may be substantially too low.
\citet{ShawEtal08} found a cosmic-ray ionization rate forty times higher along the
sight line to Zeta Per from detailed modeling while
\citet{IndrioloEtAl07} derived a value ten times higher from the
chemistry of H$_3^+$ along fourteen different sight lines.
If these newer, substantially higher, Galactic background rates
are accepted as typical then the
ratio of the cosmic ray rate to the Galactic background
given below would
be reduced by about an order of magnitude.
This is clearly an area of active research
\citep{Dalragno96}.
We adopt the Galactic cosmic-ray background
quoted in Paper 1 for consistency with that work.
We will express the particle ionization rate in terms
of the Galactic cosmic ray background rate.

Cosmic rays both heat and ionize the emitting gas.
Their interactions with low-density gas are described in
\citet{SpitzerTomasko68}, \citet{FerlandMushotzkyCR84},
\citet{AGN3}, \citet{XuMcCray91}, \citet{XuMcCray91},
\citet{DalgarnoEtAl99}, \citet{TineEtAl97}, and many more papers.
\citet{AbelEtal05} describe our implementation of this physics in
the current version of Cloudy.
Briefly, if the gas is ionized (the electron fraction $n_e / n_{\rm{H}} > 0.9$)
cosmic rays give most of their energy to free electrons which are then thermalized
by elastic collisions with other electrons.
In this highly-ionized limit cosmic rays mainly heat the gas.
In neutral gas, with low electron fraction,
some of the cosmic-ray energy goes into heating the gas
but much goes into the creation of a population of suprathermal secondary electrons.
These secondaries cause direct ionization
as well as excitation of UV resonance lines.
These excitation and ionization rates depend on the cosmic-ray density and the
electron fraction but not on the kinetic temperature of the thermal gas.

These two cases behave quite differently in the cold molecular limit
and in the nature of the transition from the molecular to atomic and ionized states.
In a cosmic-ray energized gas the suprathermal secondary
electrons will cause ionization and excite
resonance lines even when the gas is quite cold.
As the cosmic-ray density is increased the
transition from the fully molecular limit to the atomic or ionized states
is gradual since the ionization rate,
in the low electron-fraction limit, is proportional to the cosmic-ray density
and has little dependence on temperature.
Thus a very cold cosmic-ray energized gas will have a significant level of
ionization, dissociation, and excitation of the H~{\sc i}, He~{\sc i},
and \htwo\ resonance transitions.
The UV lines, being optically thick in a well-shielded medium,
undergo multiple scatterings with most lines being absorbed by dust or gas.
Some emission in infrared and optical subordinate lines,
and in ro-vibrational transitions in the ground electronic state of \htwo , occurs.

In the extra-heating case the thermal gas has a single kinetic temperature.
As the extra-heating rate increases the temperature rises
but the gas remains fully molecular
until the kinetic temperature rises to
the point where dissociation is energetically possible.
The transition from molecular to atomic phase occurs abruptly when the
gas kinetic temperature approaches the \htwo\ dissociation energy.
The gas will be in one of three distinct phases, with essentially all H
in the form of \htwo, \h0, or \hplus .
As the heating rate and kinetic temperature increase the gas will go from
one form to another in abrupt transitions which occur where the
temperature reaches the appropriate value.

These two models mainly differ in how energy is
deposited in molecular or atomic gas.
The energy of a cosmic ray will eventually appear as heat, internal
excitation, or as an ionizing particle, while
in the wave-heating case the energy deposition is purely as heat.
The cosmic-ray case does not exclude MHD effects as the fundamental
energy source since MHD waves can produce
low-energy cosmic rays within a filament \citep{Lazarian05}.

\subsection{A fully \emph{l}-resolved H-like iso-sequence}
\label{sec:AFullyLResolvedHLikeIsoSequence}

The calculations presented here use version C08.00 of the
spectral synthesis code Cloudy,
last described by \citet{FerlandCloudy98}.
There have been several improvements to the simulations since Paper 1.
These are described next.

The hydrogen recombination spectrum is special because it can be predicted
with great precision.
Two of the most important rates, radiative recombination and
transition probabilities between bound levels, are known to a
precision of several percent.
If the \hi\ lines form by recombination in an ionized gas that is optically
thick in the Lyman continuum then the emission-line spectrum should be
close to Case B of \citet{BakerMenzel38}.
Collisional excitation of excited levels from the ground state of \h0\
are a complication since their contribution to the observed \hi\
spectrum depends on density, temperature, and ionization
in ways that are fundamentally different from the recombination Case B.

The computed \hi\ emission-line spectrum presented here is the result of
a full calculation of the formation of the lines, including radiative
and three-body recombination, collisional ionization, collisional coupling
between $nl$ terms, and spontaneous and induced transitions between levels.
For much of its history Cloudy has used the compact model of the H-like iso sequence
described by \citet{FergusonFerland97} with the
rate formalism described by \citet{FerlandRees84}.
That model resolved the 2\emph{s} and 2\emph{p} levels but assumed that
\emph{l}-terms within higher-\emph{n} configurations were populated
according to their statistical weights.
This is formally correct only at high densities.
Various approximations
were introduced to allow the
model to work well in the low-density limit.

Our treatment of the entire He-like iso-electronic sequence was recently enhanced to
resolve any number of $nl$ levels, resulting in a far better representation
of the physics of the resulting emission.
Applications to He~I are discussed by \citet{PorterEtAl05} while \citet{PorterFerland07}
describe ions of the He-like sequence.
This same methodology has now been applied to the H-like iso sequence.
Levels can now be fully \emph{l}-resolved and any number can be considered.
The methods used to calculate rates for \emph{l}-changing collisions are given in \citet{PorterEtAl05}.

A very large number of levels, and resulting high precision,
were used in \citet{PorterEtAl05}.
The was possible because of the overall simplicity of that study.
The temperature and electron density were specified so a single calculation
of the level populations was sufficient to obtain the \hei\ spectrum.
The calculations presented below self-consistently
solve for the temperature, ionization, and
chemistry, which requires that the model atom be reevaluated a very large
number of times for each point in the cloud.
An intermediate model must be adopted with a small enough number of
levels to make the problem solvable with
today's computers but yet expandable to have more levels and
become more accurate when more power becomes available
or high precision is needed.
A hybrid approach was chosen.
The lowest \emph{n'} configurations are \emph{nl} resolved.
They are supplemented with another \emph{n''} configurations,
referred to as collapsed levels,
which do not resolve the $nl$ terms.
The number of resolved and collapsed levels can be adjusted with
a tradeoff between accuracy and speed.
In the following calculations levels $n \leq 15$ are $nl$ resolved with
another ten collapsed levels representing higher states.
This is sufficient to achieve a convergence accuracy of better than 2\% for
the \hi\ lines we predict.
Larger errors are introduced by uncertainties in the collisional rates.
These are difficult to quantify but probably produce errors of $\sim 20\%$
in the line emissivities and $\sim 5\%$ in relative intensities of \hi\ lines.

Resolving the \emph{nl} terms
makes it possible to predict the detailed effects of
thermal and cosmic-ray excitation of
optical and infrared H~{\sc i} lines
because the resulting emission spectrum is sensitive to the
precise \emph{nl} populations.
High-energy particles will mainly excite $np$ levels from the ground $1s$ term
because most atoms are in the ground state and electric-dipole allowed
transitions have larger collisional excitation cross sections
at high energies \citep{SpitzerTomasko68}.
After excitation, the $np$ level will decay to
a variety of lower $s$ or $d$ terms
because of selection rules for optically-allowed transitions.
The \hi\ Lyman lines
are assumed to be optically thick so
that $np\to1s$ transitions scatter often enough to
be degraded into Balmer lines plus L$\alpha$.
The resulting L$\alpha$ photons will largely be absorbed by grains
rather than emerge from the cloud.

State-specific cosmic-ray excitation rates to $np$ from ground
are derived from the Born approximation.
This is appropriate due to the high energy of a typical secondary electron
\citep{SpitzerTomasko68}.
The theory described in \citet{AbelEtal05} and
\citet{ShawEtal08} is used to derive a secondary
excitation rate of L$\alpha$, $q_{1s\to2p}^{sec }$.
The secondary excitation rate of any permitted line, $q_{E1}^{sec }$ is given
in terms of the L$\alpha$ rate by
\begin{equation}
q_{E1}^{sec }  = q_{1s\to2p}^{sec } \left( {\frac{{gf_{E1} }}{{gf_{1s\to2p} }}} \right)\left( {\frac{{\varepsilon _{1s\to2p} }}{{\varepsilon _{E1} }}} \right)
\cm ^3 \ps
\end{equation}
where $\varepsilon$ is the energy of the transition and $gf$ is its oscillator strength \citep{ShemanskyEtal85}.

Quantal calculations of collisional excitation rates for excitation of
$nl$ terms by thermal electrons are used \citep{Anderson00}.
There is no favored final $l$ term, another distinction between
the cosmic ray and extra heating case.
% data is in file h_coll_str.dat in data directory
Details of the resulting optical and infrared H~{\sc i} emission
will be discussed below.

\subsection{Revised H -- \htwo\ collision rates}

\label{sec:H2CollisionRates}
The rates for collisions between H and \htwo \ have been updated
from those used in \citet{ShawEtal05}.
We had used the fits given by \citet{LeBourletEtAlCollFits99}
in Paper 1.
\citet{AllersEtal05} noted that observations of \htwo\ emission from the
Orion Bar suggested that these rates were two small by nearly 2 dex.
\citet{WrathmallEtAl07} confirmed this and presented a new set of collisional
rate coefficients using an improved interaction potential and scattering theory.
These rate coefficients for excited vibrational levels are systematically larger,
often by 2 to 3 dex.
These rates are employed in this work.

The critical density of a level is the density were radiative and collisional
deexcitation are equally probable \citep{AGN3}.
The total radiative rate out of level $i$, $A_i$,
is the sum of the transition probabilities out of the level,
$A_i  = \sum\limits_{l < i} {A_{i,l} }$.
If the total deexcitation rate [s$^{-1}]$ due to collisions with species $S$
with density $n(S)$ is
$n(S)q_i  = \sum\limits_{l < i} {n(S) q_{i,l} }$,
where $q_{i,l}$ is the collisional deexcitation rate coefficient
(units cm$^3 \ps$),
then the critical density of a molecule in level $i$ colliding with species $S$ is given by
\begin{equation}
n\left( {S,i} \right) = A_i /q_i  = {{\sum\limits_{l < i} {A_{i,l} } } \mathord{\left/
 {\vphantom {{\sum\limits_{l < i} {A_{i,l} } } {\sum\limits_{l < i} {q_{i,l} } }}} \right.
 \kern-\nulldelimiterspace} {\sum\limits_{l < i} {q_{i,l} } }}
\end{equation}
The large increase in the \h0 -- \htwo\ collision rate
coefficients for vibrationally-excited levels
in the new data lowers the critical densities
of most levels by a large amount.
Table~\ref{tab:CriticalDensities} gives critical densities for the
upper levels of the \htwo\ lines analyzed in this paper.
The rates are evaluated at 1000 K, a typical \htwo\ temperature.
Each collider considered in our calculations is listed.
This table should be compared with Table 1 of \citet{LeBourletEtAlCollFits99}
or \citet{SternbergDalgarno89}.
Our critical densities are substantially smaller.
This means that a Boltzmann level population
distribution will be established at considerably lower densities than
was previously thought necessary.

\begin{table*}
\centering
\caption{\htwo\ lines, levels, and critical densities at 1000 K}
\begin{tabular}{ c c c c c c c c c c }
%{|p{48pt}|p{48pt}|p{48pt}|p{48pt}|p{48pt}|p{48pt}|p{48pt}|p{48pt}|p{48pt}|p{48pt}|p{48pt}|}
%\begin{tabular}{||l|r|}\hline
\hline
WL& ID& $v_{up}$& $J_{up}$& $T_{exc}$(K)& H$^0$& He& H$_2^o$& H$_2^p$& H$^+$ \\
\hline
28.21& 0-0 S(0)& 0& 2& 509.8& 1.11E+01& 1.61E+00& 3.76E+00& 5.16E+00& 2.36E-02 \\
%\hline
17.03& 0-0 S(1)& 0& 3& 1015& 1.54E+02& 2.70E+01& 6.18E+01& 8.01E+01& 4.14E-01 \\
%\hline
12.28& 0-0 S(2)& 0& 4& 1681& 7.49E+02& 2.51E+02& 5.17E+02& 6.29E+02& 1.52E+00 \\
%\hline
9.66& 0-0 S(3)& 0& 5& 2504& 2.40E+03& 1.43E+03& 2.91E+03& 3.50E+03& 5.96E+00 \\
%\hline
8.02& 0-0 S(4)& 0& 6& 3474& 6.37E+03& 6.47E+03& 1.28E+04& 1.59E+04& 1.27E+01 \\
\hline
6.907& 0-0 S(5)& 0& 7& 4586& 1.51E+04& 2.50E+04& 4.45E+04& 5.67E+04& 3.02E+01 \\
%\hline
%2.406& %1-0 Q(1)& %1& %1& %6149& %1.15E+05& %2.52E+08& %6.79E+08& %1.03E+09& %3.09E+02 \\
%\hline
2.223& 1-0 S(0)& 1& 2& 6471& 4.78E+04& 3.61E+04& 9.14E+04& 1.07E+05& 2.76E+02 \\
%\hline
2.121& 1-0 S(1)& 1& 3& 6951& 3.60E+04& 3.53E+04& 8.52E+04& 1.07E+05& 2.57E+02 \\
%\hline
2.423& 1-0 Q(3)& 1& 3& 6951& 3.60E+04& 3.53E+04& 8.52E+04& 1.07E+05& 2.57E+02 \\
\hline
2.033& 1-0 S(2)& 1& 4& 7584& 3.16E+04& 5.53E+04& 1.15E+05& 1.39E+05& 2.15E+02 \\
%\hline
1.957& 1-0 S(3)& 1& 5& 8365& 3.05E+04& 6.77E+04& 1.78E+05& 2.10E+05& 2.17E+02 \\
%\hline
1.891& 1-0 S(4)& 1& 6& 9286& 3.14E+04& 1.25E+05& 2.69E+05& 3.21E+05& 1.96E+02 \\
%\hline
2.248& 2-1 S(1)& 2& 3& 12550& 1.84E+04& 5.61E+04& 1.25E+05& 1.67E+05& 3.19E+02 \\
%\hline
1.748& 1-0 S(7)& 1& 9& 12817& 3.81E+04& 4.44E+05& 8.37E+05& 1.04E+06& 1.35E+02 \\

\hline
\end{tabular}

\label{tab:CriticalDensities}
\end{table*}

There is likely to be a substantial uncertainty in all molecular collision rates.
This is partially because molecular collisions are complicated many-body problems.
The current \h0 - \htwo\ rates,
while sometimes substantially larger than the previously published set,
do not yet include reactive channels \citep{WrathmallEtAl07}.
The rates will become larger still, at temperatures where these reactions can occur,
when this process is included.

These critical densities can give an indication of the
density of the molecular gas \citep{JaffeEtAl01}.
The distribution of level populations within \htwo\ ro-vibrational levels
can be determined from relative emission-line intensities.
\citet{JaffeEtAl01} find that low $v,J$ populations can be fitted as a single
Boltzmann distribution corresponding to excitation temperatures $\sim 2000$ K.
In this paper we fit the observed \htwo\ line intensities and do not
present \htwo\ level-excitation diagrams.
The results are fully equivalent but have the simplification that we
are working with the directly-observed quantity.

A thermal population distribution
will result if the gas density is greater
than the critical densities of the levels involved.
\citet{JaffeEtAl01} argue that the density in the filaments
must be $\geq 10^6 \pcc$ if the gas
is predominantly molecular.
They based this on a modified scaling of the \citet{MandyMartin93} semi-classical
\htwo\ -- \htwo\ collision rates.
The rate coefficients presented by \citet{LeBourletEtAlCollFits99}
would suggest a significantly higher density ($\ge 10^9 \pcc$).
Paper 1 showed that there are other ways to establish a
quasi-thermal distribution at fairly low densities.
Table~\ref{tab:CriticalDensities} shows that the well-observed 1-0 S(1)
line at 2.121 \micron\ has a critical density ranging from $3\times 10^2 \pcc$,
for collisions with H$^+$,
to $1.9\times 10^5 \pcc$, for collisions with \htwo .
In practice the critical density will depend
on whether ions are mixed with the molecular gas.
We show below that
the extra heating and cosmic ray cases have very different mixtures of
molecules, atoms, and ions, and this may provide a discriminant between
the energy sources.
Note that only collisions with H$^0$ and H$^+$ are capable of inducing
a reactive ortho-para transition in \htwo \
\citep{DalgarnoEtAl73}.

\subsection{L$\alpha$ pumping of \htwo }

L$\alpha$ fluorescence can be a significant \htwo \
excitation source when ions and molecules are mixed
as discussed by \citet{BlackvanDishoeck87} and \citet{LupuEtAl06}.
In effect \htwo\ offers a second channel, competing with dust,
to remove L$\alpha$ photons created by collisional excitation and recombination.
The L$\alpha$ energy is converted into internal \htwo\
excitation energy and eventually
dissociation or \htwo\ line emission.

We determine the full set of \h0 level populations by solving the equations of
statistical equilibrium.
The population of the \h0\ $2p$ and $1s$ terms determine the
excitation temperature
\begin{equation}
T_{ex}(2p,1s) = \left[ \ln {\frac{{n(1s)/g(1s)}}{{n(2p)/g(2p)}}} \right]\frac{\chi }{k}
\end{equation}
where $\chi$ and $k$ are the excitation potential of $2p$ and the
Boltzmann constant respectively.
This establishes the source function within the L$\alpha$ transition,
which we treat following \citet{ElitzurFerland86}.
It follows that within the optically thick part of the line
$S_\nu = B_\nu (T_{ex})$ where $S_\nu$ is the line source function and
$B_\nu (T_{ex})$ is the Planck function evaluated at the level excitation temperature.

Numerical solutions of L$\alpha$ transfer in the limits of
very large optical depths show that the line source function is
constant over a line width that depends on the total optical depth
\citep{Adams72,Harrington73,FerlandNetzer79}.
The results of \citet{Adams72} are adopted.
The resulting expression for the full line width in the case of
large optical depths, $\tau  \gg 1$, is
\begin{equation}
\Delta \nu  = \Delta \nu _{Dop} \;2.3\,\left( {a\tau } \right)^{1/3}
\end{equation}
where $a$ is the line damping constant and $\Delta \nu _{Dop}$ is the Doppler width.
With these definitions the photon occupation number within the L$\alpha$ line is
\begin{equation}
\eta _\nu   \equiv S_\nu  \left/\left( {2h\nu ^3 /c^2 } \right)\right. \equiv \left[ {\exp \left( {h\nu /kT_{ex} } \right) - 1} \right]^{ - 1} .
\end{equation}
The rate of induced radiative excitation
of \htwo\ by L$\alpha$ photons is given by
\begin{equation}
r_{l,u}  = n_l B_{l,u} J_{l,u}  = n_l A_{u,l} \frac{{J_{l,u} }}{{2h\nu ^3 /c^2 }}\frac{{g_u }}{{g_l }} = n_l A_{u,l} \eta _\nu  \frac{{g_u }}{{g_l }}
\pcc \ps.
\end{equation}
This is included in the \htwo\
balance equations as a general level excitation process \citep{ShawEtal05}.

%\subsection{Dissociative recombination of H$_2^+$}
%
%Takagi 2002PhST...96...52T - I have only read the abstract since Physica Scripta is not %freely accessible from my house. I'll try via work later.

\section{The density -- heating plane}

\subsection{General considerations}

The magnetic field and cosmic-ray density within the filaments are not known.
The gas density is low enough for the optical [S~II] lines to be near their
low-density limit \citep{HatchEtAl06}, $n < 10^3 \pcc$,
but the density of the molecular
gas is only constrained from the form of the \htwo\ level-population distribution
\citep{JaffeEtAl01}.
Given these uncertainties we explore a broad range of non-radiative
heating rates and gas densities, compute the thermal and chemical state of the gas,
and predict the resulting emission-line spectrum to compare with observations.
Results are presented below as a series of contour diagrams showing
various predictions.

The non-radiative energy sources are the dominant heating agents for most
conditions shown here.
For the very lowest values of the heating rates the
background metagalactic radiation field, described above, becomes significant.
The resulting emission will be unobservably faint in this limit, however,
as shown in diagrams below and in Paper1.

\begin{figure}
\begin{center}
\includegraphics[clip=on,width=\columnwidth,height=0.8\textheight,keepaspectratio]{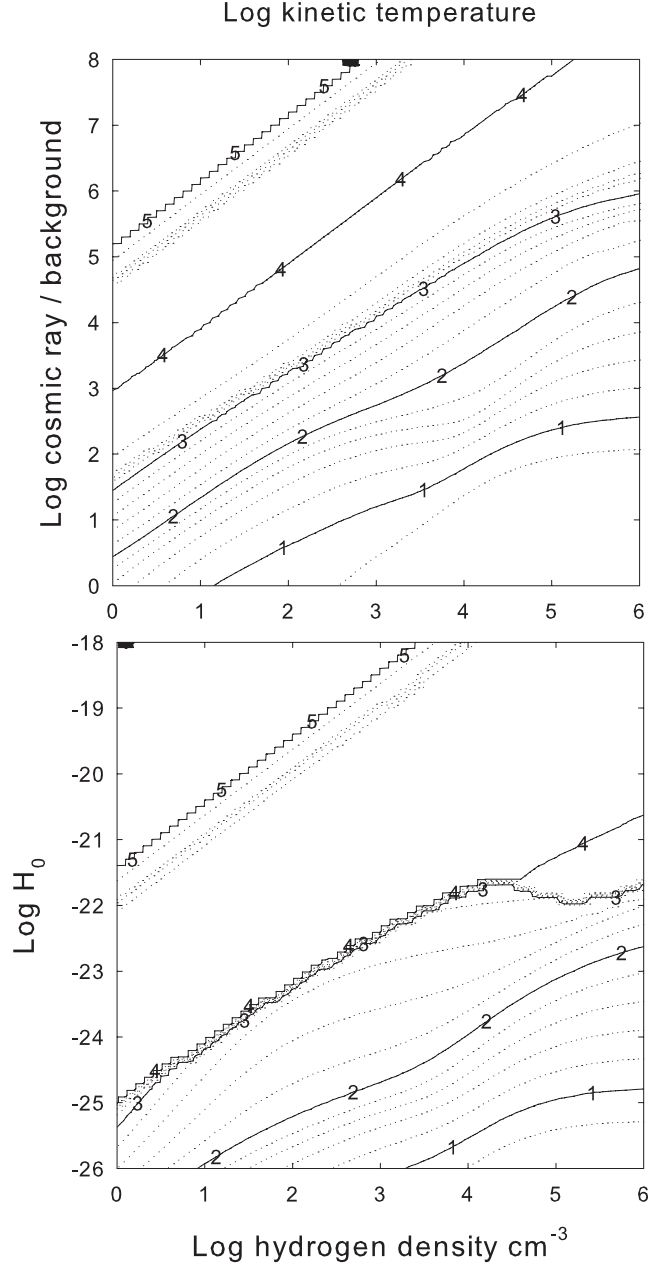}
\end{center}
\caption{The log of the gas kinetic temperature is shown as a function of the hydrogen
density, the independent variable,
and (top panel) the cosmic-ray density relative to the Galactic background
and (bottom panel) the leading term in the extra-heating rate.
The temperature is near the CMB in the lower right corner and
rises along a diagonal from lower right to upper left as the
heating increases and the gas density decreases.
There are discontinuous jumps in
the temperature as the gas changes thermal phase.
}
\label{fig:HeatCrTempA}
\end{figure}

\begin{figure}
\begin{center}
\includegraphics[clip=on,width=\columnwidth,height=0.8\textheight,keepaspectratio]{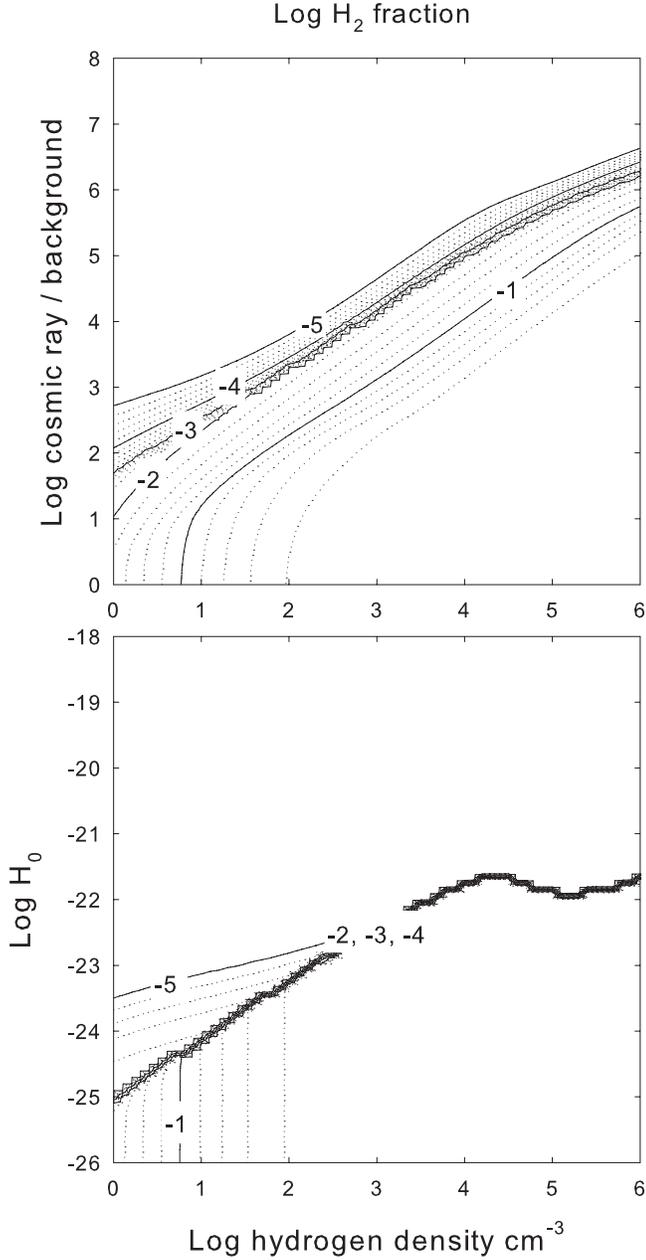}
\end{center}
\caption{The log of the \htwo\ fraction is shown as a function of the hydrogen
density and (top panel) the cosmic-ray
and (bottom panel) the extra-heating rates.
Gas is molecular in the lower right corner
where the density is high and the non-radiative heating rates are low.
The gas is nearly fully molecular in the lower right corner and
becomes increasingly
ionized along a diagonal running to the upper left.
There are discontinuous jumps in the
chemistry as the gas changes phases.}
\label{fig:HydroMoleFrac}
\end{figure}

\begin{figure}
\begin{center}
\includegraphics[clip=on,width=\columnwidth,height=0.8\textheight,keepaspectratio]{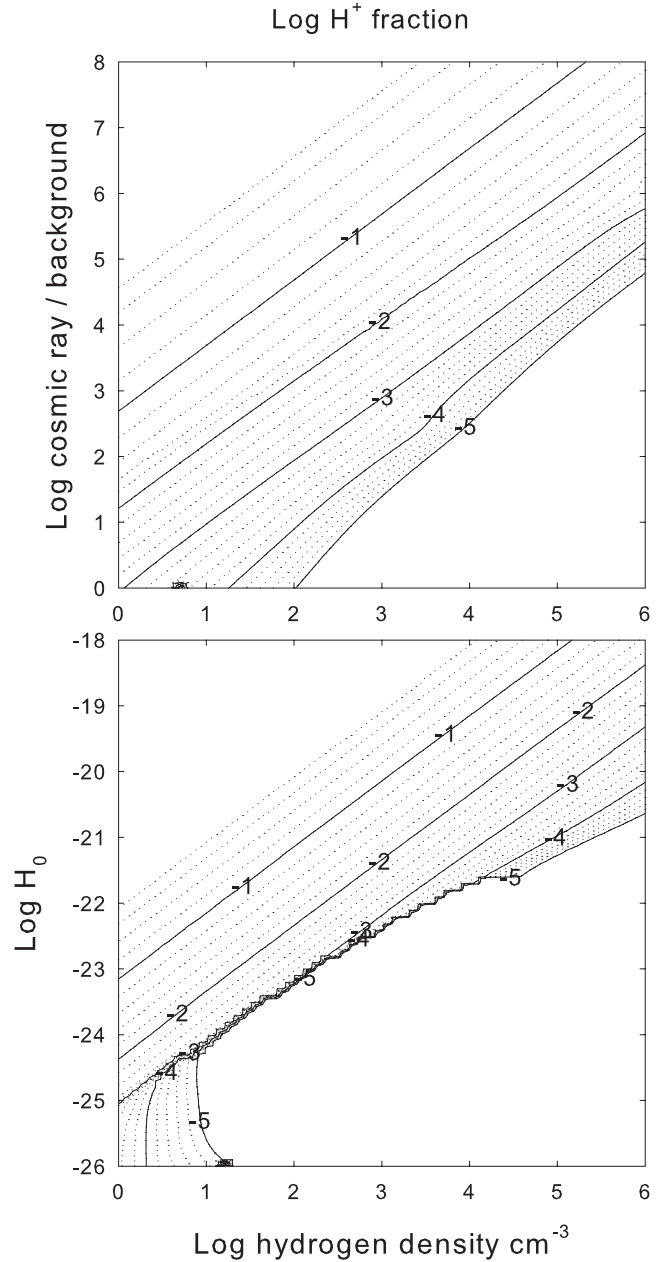}
\end{center}
\caption{The log of the H$^+$ fraction is shown as a function of the hydrogen
density and (top) the cosmic-ray
and (bottom) the extra-heating cases.
Gas is ionized in the upper left corner of the diagram,
where the density is low and the non-radiative heating rates are high,
and molecular in the lower right corner.
}
\label{fig:HydroIonFrac}
\end{figure}

Figure~\ref{fig:HeatCrTempA} shows the gas kinetic temperature,
Figure~\ref{fig:HydroMoleFrac} shows the hydrogen molecular fraction,
and Figure~\ref{fig:HydroIonFrac} shows the fraction of H in H$^+$.
In all pairs of figures the upper and lower panels show the
cosmic ray and extra-heating cases respectively.

% sims are
% C:\projects\papers\CoolFlow\LOCSpectrum\CR\varyCrHden
% C:\projects\papers\CoolFlow\LOCSpectrum\Heat\varyHeatHden
% fig HeatCrTemp in varyCrHeatHden.JNB

% sims are
% C:\projects\papers\CoolFlow\LOCSpectrum\CR\varyCrHden
% C:\projects\papers\CoolFlow\LOCSpectrum\Heat\varyHeatHden
% fig in varyCrHeatHden.JNB in C:\projects\papers\CoolFlow\LOCSpectrum\figures

% sims are
% C:\projects\papers\CoolFlow\LOCSpectrum\CR\varyCrHden
% C:\projects\papers\CoolFlow\LOCSpectrum\Heat\varyHeatHden
% fig in varyCrHeatHden.JNB in C:\projects\papers\CoolFlow\LOCSpectrum\figures

An `effective' ionization parameter, given by the ratio of the non-radiative
heating to the gas density $u_{nr} = r_{nr}/n(_{\rm{H}})$,
characterizes physical conditions in these environments.
The non-radiative heating acts to heat and ionize the gas,
while cooling and recombination processes
are often two-body processes which increase with density.
The ionization and temperature tend to increase with $u_{nr}$
along a diagonal running from low heating and high density,
the lower right corner,
to high heating and low density,
the upper left corner.
The temperature and molecular fractions tend to change along
diagonals from the lower left to upper right corners.
These are along lines of roughly constant $u_{nr}$.

The lower right and upper left corners of the plane have the most extreme conditions.
Gas is cold and molecular in the lower left high density, low heating, corner.
The kinetic temperature has
fallen to nearly that of the CMB in this region.
A small degree of ionization is maintained
by background light or cosmic rays.
This is to be contrasted with
gas in the upper left high-heating, low-density, corner.
That gas is hot, with $T \gg 10^4$~K,
and highly ionized.
Gas in either region would emit little visible/IR light
and so would be difficult to detect.

\subsection{The \htwo\ emission spectrum}

Paper 1 described the \htwo\ emission spectrum in some detail.
Although the detailed predictions have changed due to the improved collisional rate
coefficients used in this paper,
the overall trends of \htwo\ emission shown in that paper
have remained the same.

\begin{figure}
\begin{center}
\includegraphics[clip=on,width=\columnwidth,height=0.8\textheight,keepaspectratio]{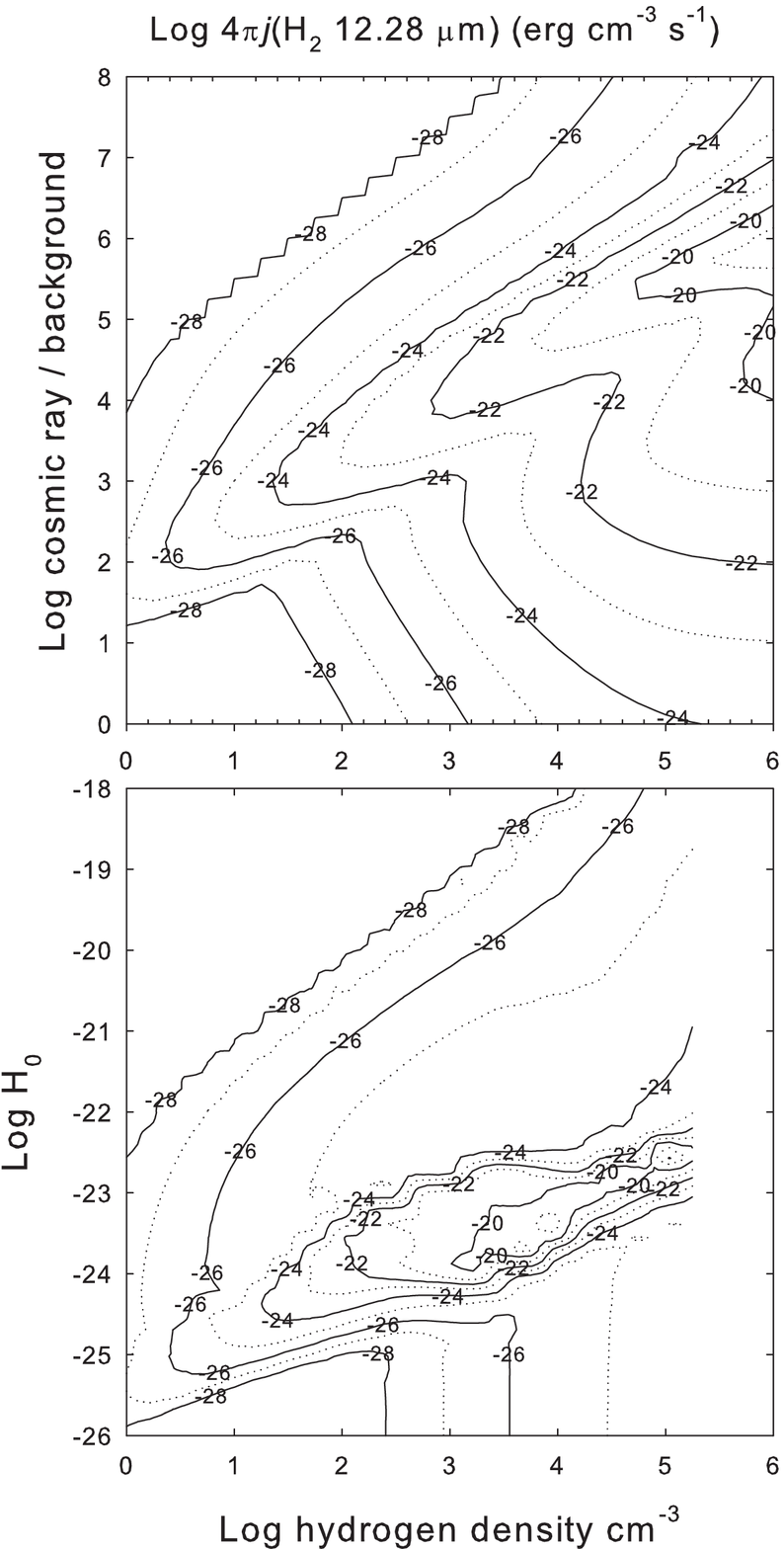}
\end{center}
\caption{The log of the emissivity $4 \pi j$ (erg $\pcc \ps $)
of the \htwo\ 0-0 S(2) 12.28 \micron\ line is shown as
a function of cloud parameters.
This figure is to be contrasted with the next figure
showing the \hi\ line emissivity over a similar range of parameters.
}
\label{fig:H2_emissitivty_contourA}
\end{figure}

The full \htwo\ spectrum is presented below.
Here we concentrate on the \htwo\ 0-0 S(2) $\lambda$12.28 \micron\ line
which we will combine with optical \hi\ lines to set constraints
on our simulations.
The log of the emissivity $4 \pi j$ (erg cm$^{-3}$ s$^{-1}$)
of the \htwo\ line is shown in Figure \ref{fig:H2_emissitivty_contourA}.
\htwo\ emission occurs within regions
where the \htwo\ molecular fraction,
shown in Figure \ref{fig:HydroMoleFrac}, is significant.
Other factors besides the \htwo\ density
determine the line's brightness.

% the graph is Hb_emissitivty_contour in H_lines_countours.jnb

The emissivity of an \htwo\ line that is excited by thermal collisions
is proportional to
\begin{equation}
4\pi j\left( {{\rm{H}}_2 } \right) \propto n\left( {{\rm{H}}_2 } \right)n_S \exp \left( { - \chi /kT} \right)
\label{eq:H2Emissivity}
\end{equation}
where $n_S$ is the density of all colliders and $\chi$
is the excitation potential of the upper level of the transition.
Higher temperatures and \htwo\ densities tend to make the line more emissive.
Figures \ref{fig:HeatCrTempA} and \ref{fig:HydroMoleFrac} show that
$ n\left( {{\rm{H}}_2 } \right)$ and $T$ are anti-correlated.
From this result, together with the form of equation \ref{eq:H2Emissivity},
it follows that there will be a band of peak emission that runs from the
upper right to lower left in Figure \ref{fig:H2_emissitivty_contourA}.
This is parallel to contours of constant temperature and
molecular fraction shown in Figures \ref{fig:HeatCrTempA}
and \ref{fig:HydroMoleFrac}.

The $\lambda 12.28$ \micron\ line is a pure rotational transition
with an level excitation energy of 1681~K.
The line is excited by two processes in our simulations.
The line is emitted efficiently when the gas becomes warm enough
for thermal particles to collisionally excite the upper level
but not so warm as to dissociate the \htwo .
This produces the region of peak emission that tracks the contour giving
$T \sim 2000 \K$ in Figure \ref{fig:HeatCrTempA}.

The \htwo\ line remains emissive in much cooler regions in the cosmic-ray case.
Here suprathermal electrons excite the line by a two-step mechanism
that is analogous to the Solomon process in starlight-excited PDRs.
The first step is a suprathermal excitation from the ground electronic state
X to one of the excited-electronic states.
A minority of these return to rotation-vibration levels within X.
Many of these eventually decay to the upper level of the
$\lambda 12.28$ \micron\ line.
Other electronic excitations lead to dissociation which are
then followed by formation by the grain-surface or \hminus\ routes.
These formation processes populate excited levels within X
which then lead to emission.
Details of our implementation of these \htwo\ excitation processes
are given in \citet{ShawEtal05} and \citet{ShawEtal08}.

\subsection{The \hi\ emission spectrum}

We consider the emissivity of H$\beta \ \lambda 4861$\AA\ in detail
since we will use the
\hi\ $\lambda 4861$\AA\ / \htwo\ $\lambda 12.28$ \micron\ line ratio to constrain our simulations.
We also consider two \hi\ line ratios that
can be used to indicate whether the spectrum has been
affected by interstellar reddening.
This is important because the dust content of the
filaments is currently unknown.

\begin{figure}
\begin{center}
\includegraphics[clip=on,width=\columnwidth,height=0.8\textheight,keepaspectratio]{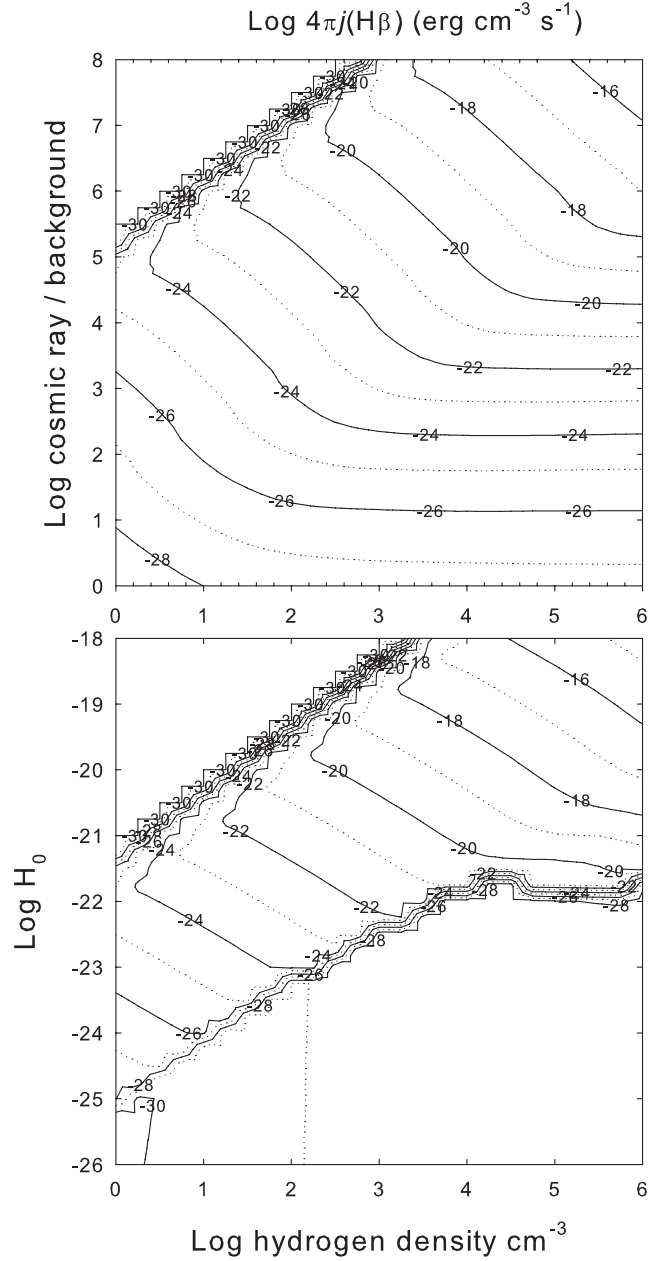}
\end{center}
\caption{The log of the emissivity $4 \pi j$ (erg $\pcc \ps $)
of H$\beta$ $\lambda 4861$\AA\ is shown as
a function of cloud parameters.
This figure is to be contrasted with those in Paper 1 which
show the \htwo\ line emissivity over a similar range of parameters.
}
\label{fig:Hb_emissitivty_contourA}
\end{figure}

Figure \ref{fig:Hb_emissitivty_contourA}
shows the log of the emissivity $4 \pi j$ of H$\beta \ \lambda 4861$\AA,
one of the brightest lines in the optical spectrum.
In a photoionized environment
\hi\ lines form by radiative recombination of
$e^- + \hplus $.
\hi\ lines also form by recombination if the gas is collisionally heated
to a temperature so warm that nearly all H is ionized.
The emissivity of an \hi\ recombination line is proportional to
\begin{equation}
4\pi j\left( \mbox{\rm H\,\sc i} \right) \propto n\left( {{\rm{H}}^{\rm{ + }} } \right)n_e T^{ - 0.8}
\label{eq:HIEmissivity}
\end{equation}
where $T$ is the gas kinetic temperature and the inverse
temperature dependence is a fit to recombination coefficients in
the neighborhood of 10$^4$~K.
\hi\ recombination lines are emitted most efficiently by gas that is dense,
ionized, and cool.

% the graph is Hb_emissitivty_contour in H_lines_countours.jnb

\hi\ lines can also form by collisional excitation
in regions where H is mostly atomic.
The gas must be warm enough to excite the upper levels of the optical
or IR \hi\ lines, requiring a temperature $T\geq 4000 \K$,
although collisional excitation is important at low temperatures in the
cosmic ray case when suprathermal secondary electrons are present.
The emissivity of a collisionally excited \hi\ line is proportional
to $n(\h0 ) n_x f(T)$ where $n(\rm{H}^0)$ is the atomic hydrogen density,
mostly in the ground state, and $n_x$ is the density of the colliding species.
The function $f(T)$ is the Boltzmann factor of the upper level of the \hi\ line
for the case of excitation by thermal particles
and is a constant for excitation by secondary electrons.
Collisionally excited \hi\ lines are emitted most efficiently by
gas that is dense, atomic, and warm.

The extra heat case is relatively simple since only thermal processes
affect the excitation and ionization of the gas.
The lower panel of Figure \ref{fig:Hb_emissitivty_contourA} shows that
the line has a ridge of relatively high emissivity which corresponds to moderate
ionization and temperature.
The \hi\ lines along this ridge are mainly produced by collisional excitation
with a contribution from recombination following
collisional ionization.
The emission falls off dramatically when the gas becomes molecular in the
lower right corner or very hot in the upper left corner.

Figure \ref{fig:Hb_emissitivty_contourA} shows the H$\beta$
emissivity in the cosmic-ray case.
Collisional excitation is also a dominant contributor to the line.
The ridge of peak emission occurs for the same reason as in the
extra heating case.
Significant \hi\ emission occurs in the lower right corner since the
dissociative cosmic rays prevent the gas from becoming fully molecular
even when it is quite cold.
The population of secondary electrons produces a significant collisional excitation
contribution to the \hi\ lines across most of the lower right half of
Figure \ref{fig:Hb_emissitivty_contourA}.

In both heating cases the emissivity tends to rise along a diagonal from
the lower left to upper right.
This corresponds to rising $n_e n_p$ at nearly constant $T$.
This figure clearly indicates that powerful selection effects
operate in this environment.
As we show below, the emissivity of the \hi\ lines is substantially
higher than produced in the pure-recombination case due to the collisional
contribution.
This affects the use of \hi\ lines as indicators or the gas of ionized gas.

The \hi\ lines are often used to measure the amount of interstellar extinction.
It is unusual for the \hi\ recombination spectrum to deviate very far from
Case~B for moderate densities and optical depths in
the Lyman continuum in a photoionized cloud \citep{AGN3}.
This is because the relative \hi\ line intensities are
mainly determined by the transition probabilities.
These determine which lines are emitted as the
electrons cascade to ground after capture from the continuum.
The result is that the spectrum depends mainly on atomic constants and less
so on the physical conditions in the gas.

\begin{figure}
\begin{center}
\includegraphics[clip=on,width=\columnwidth,height=0.8\textheight,keepaspectratio]{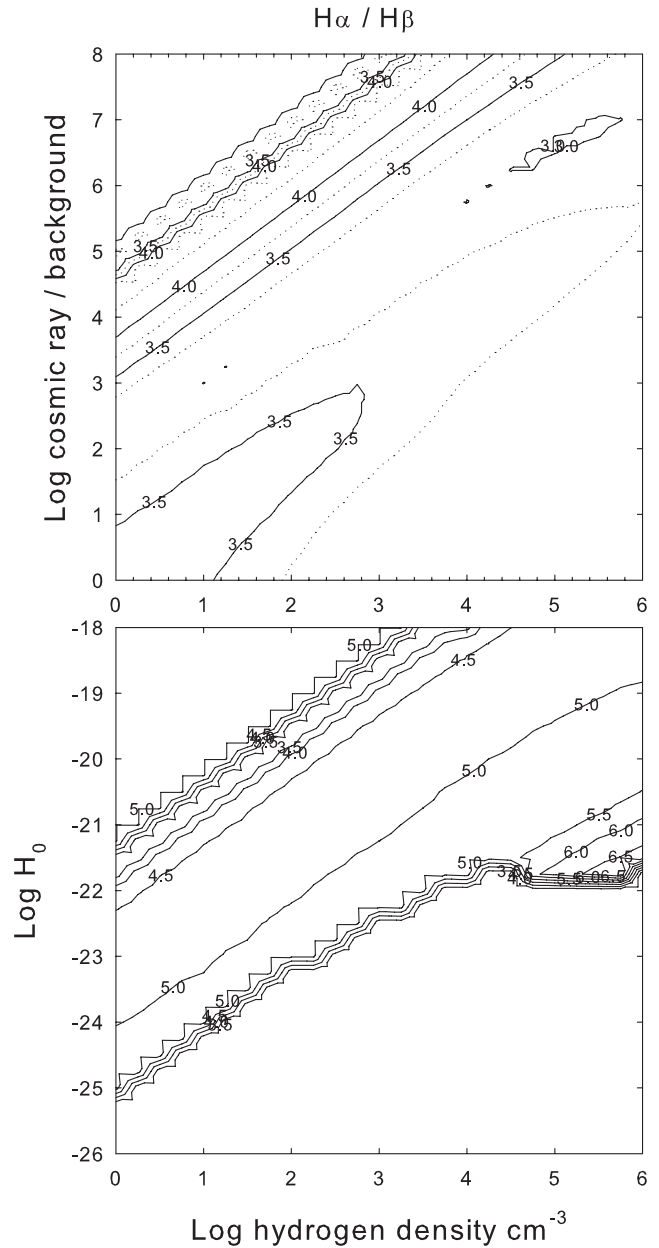}
\end{center}
\caption{The ratio of emissivities of \hi\
H$\alpha\ \lambda 6563$ to H$\beta\ \lambda 4861$ is shown as
a function of cloud parameters.
The ratio is $\sim 2.8$ under the Case~B conditions
expected for photoionized nebulae.
It is significantly larger for most parameters shown here
because of collisional contributions to the line.
}
\label{fig:Ha2HbcontourA}
\end{figure}

The ratio of the intensity of H$\alpha \ \lambda 6563$\AA\
relative to H$\beta \ \lambda 4861$\AA,
two of the strongest lines in the optical spectrum and
a possible reddening indicator,
is shown in Figure \ref{fig:Ha2HbcontourA}.
The H$\alpha$/H$\beta$ intensity ratio is $\sim 2.8$
for Case B in low-density photoionized gas \citep{AGN3}.
The predicted ratio is significantly larger than this in most regions where
the \hi\ lines have a large emissivity.

% the graph is Ha2Hbcontour in H_lines_countours.jnb

It is possible to measure intensities of \hi\ lines that originate
in a common upper atomic level with
spectra that cover both the optical and infrared.
The P$\alpha \ \lambda 1.87 \micron$ and H$\beta \ \lambda 4861$\AA\
lines, with a common $n = 4$ upper configuration, is an example.
Each of these lines is actually a multiplet with 4 $nl$ terms
within the upper level.
At conventional spectroscopic resolution the lines within the
multiplet appears as a single \hi\ line.
The transition probability of this multiplet
depends on the distribution of $nl$ populations with the upper terms,
which in turn depend on the gas density \citep{Pengelly64}.
Such line ratios are not expected to depend on the physical conditions
as much as lines that originate in different configurations.

\begin{figure}
\begin{center}
\includegraphics[clip=on,width=\columnwidth,height=0.8\textheight,keepaspectratio]{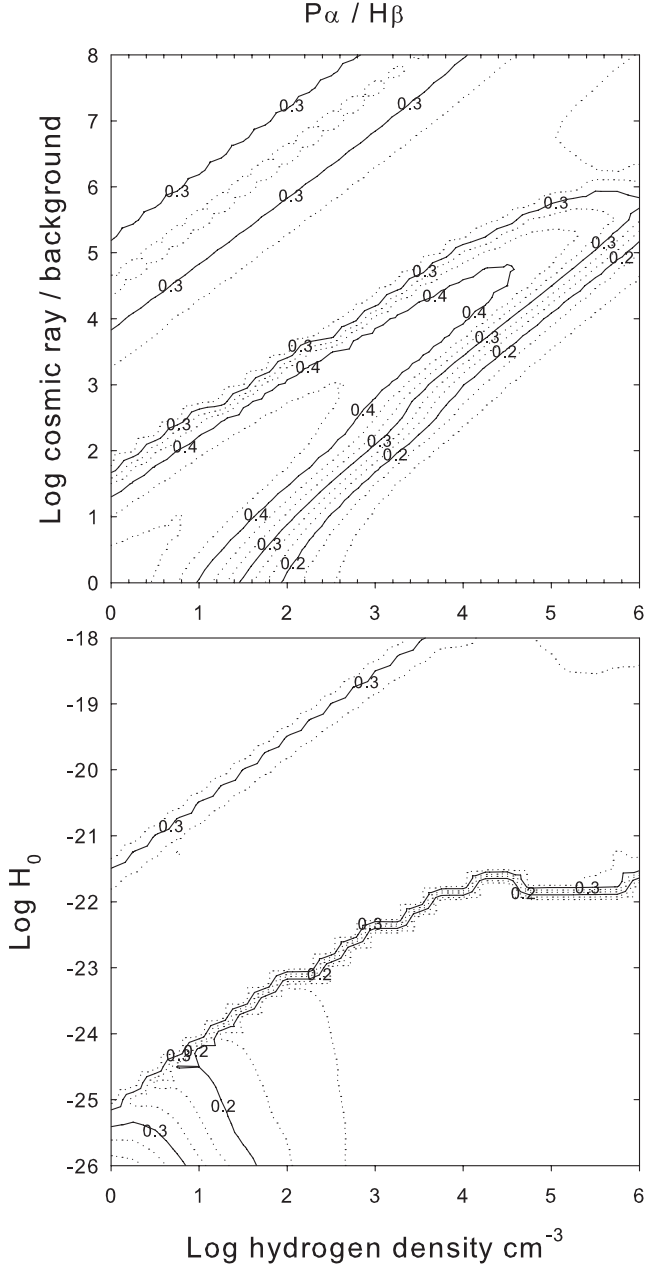}
\end{center}
\caption{The \hi\
P$\alpha\ \lambda 1.87 \micron $ to H$\beta\ \lambda 4861$
emissivity ratio is shown as
a function of cloud parameters.
These lines have a common upper $n=4$ configuration
so changes in the relative intensities
are due to changes in the $nl$ populations within
the $n = 4$ configuration.
The Case B value is 0.34.
}
\label{fig:Pa2HbcontourA}
\end{figure}

Figure \ref{fig:Pa2HbcontourA} shows the computed
P$\alpha \ \lambda 1.87 \micron$ / H$\beta \ \lambda 4861$\AA\ intensity ratio.
It is not constant because of the changing populations of $nl$ terms within
the $n = 4$ configuration of \h0.
At high densities the \emph{l} terms will be populated
according to their statistical weight and
the level populations are said to be well \emph{l}-mixed.
The line intensity ratio is then a constant that is determined by the
ratio of transition probabilities and photon energies.
Cloudy treated $n > 2$ configurations in
this well \emph{l}-mixed limit in versions before C08,
the version described in this paper.
As described above, these calculations use an improved \h0\ model
which fully resolves the $nl$ terms within a configuration.

% the graph is Pa2Hbcontour in H_lines_countours.jnb

When the density is too low to collisionally mix
the \emph{nl} terms their populations tend to peak
at smaller $l$ since recombinations from the continuum
and suprathermal excitation from ground tend to populate
low-$l$ levels.
It is only when the density becomes large enough for \emph{l}-changing collisions
to mix the $l$ levels that the well \emph{l}-mixed limit is reached.
This makes the
P$\alpha \ \lambda 1.87 \micron$ / H$\beta \ \lambda 4861$\AA\ intensity ratio
depend on density.

Figure \ref{fig:Pa2HbcontourA} shows that the
P$\alpha$ / H$\beta$ intensity ratio
does not vary over a broad range.
The extra-heating case, shown in the lower panel, has a P$\alpha$ / H$\beta$
ratio in the range of $0.2 - 0.3$, below the Case B ratio of 0.34
at $10^4 \K$ and low densities.
The cosmic ray case has a P$\alpha$ / H$\beta$ intensity ratio that
varies between $\sim 0.2 - 0.4$,
within a factor of two of the Case B ratio.
The range is larger because suprathermal
electrons excite the atom for nearly all temperatures.
These non-thermal excitations are mainly to $np$ terms and create a
distribution of $nl$ populations that differs from the recombination case.
This line ratio remains a good reddening indicator because of
the relatively modest range in the predicted values when combined
with the very wide wavelength separation of the two lines.

\subsection{Chemical and ionization state of the gas}

\begin{figure*}
\begin{center}
\includegraphics[clip=on,width=2\columnwidth,height=0.8\textheight,keepaspectratio]{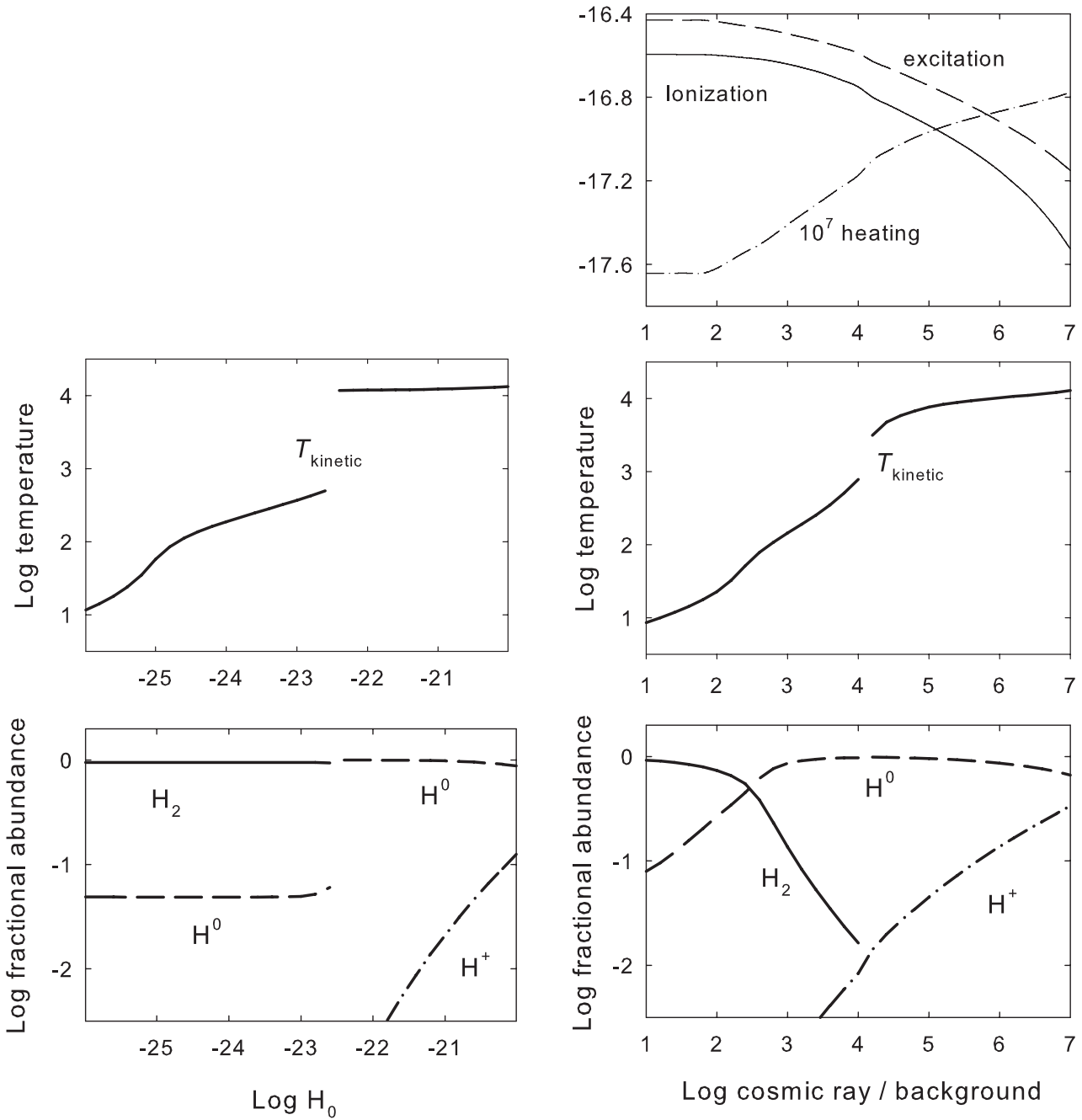}
\end{center}
\caption{
This shows the thermal and physical state
along a vertical line at $n_{\rm{H}} = 10^3 \pcc $ in the previous contour plots.
The left panels are the extra-heating case while the right panels are
the cosmic-ray case.
The non-radiative
rates are the independent axis in each panel.
The range in both was adjusted so that the phase transition, where many
physical quantities change abruptly, occurs near the middle of the independent
axis.
The ionization and temperature increase to the right as the non-radiative
heating rates increase.
The top right panel shows the cosmic ray heating, ionization,
and line-excitation efficiencies.
Cosmic rays ionize and excite predominantly neutral gas and heat ionized gas.
The abrupt change in temperature, shown in the second to bottom
pair of panels, is more extreme in the extra-heating case.
The two lower panels show the physical state of hydrogen.
There is a mix
of atomic and molecular gas in the cosmic-ray case due to the ionization and
dissociation that they produce in cold neutral gas.
}
\label{fig:Den3VarHeatCR}
\end{figure*}

The two non-radiative heating cases,
while having similar overall trends, have very different detailed properties,
which we examine next.
Figure~\ref{fig:Den3VarHeatCR} shows predictions for
a vertical line in Figures
\ref{fig:HeatCrTempA} -- \ref{fig:HydroIonFrac}
corresponding to $n_{\rm{H}} = 10^3 \pcc$
and varying non-radiative heating rates.
The independent axis gives the
extra heating rate pre-coefficient
and the cosmic-ray density relative to the Galactic background.
These parameters were chosen to cross the region where
the physical conditions in Figures \ref{fig:HeatCrTempA},
\ref{fig:HydroMoleFrac} and \ref{fig:HydroIonFrac} change abruptly.
The left panels show the extra heating case and the right panels the
cosmic ray case.
The second from the bottom panels of Figure~\ref{fig:Den3VarHeatCR}
show the kinetic temperature.
The discontinuous jump in temperature, the regions where the
contouring in Figures \ref{fig:HeatCrTempA} -- \ref{fig:HydroIonFrac}
become ragged, is clear.
The lowest panels show the molecular, atomic, and ionized
fractions of hydrogen.
The upper right panel of Figure~\ref{fig:Den3VarHeatCR} shows some
heating and excitation efficiencies for the cosmic-ray case.
In all cases the temperature and ionization of the gas increases to the right
as non-radiative energy sources become more important.

% sims are
% C:\projects\papers\CoolFlow\LOCSpectrum\CR\varyCR
% C:\projects\papers\CoolFlow\LOCSpectrum\Heat\varyHeat
% fig in Den3VarHeatCR.JNB in C:\projects\papers\CoolFlow\LOCSpectrum\figures

Both non-radiative heating cases have discontinuous jumps in temperature.
These are due to
inflection points in the gas cooling function that occur when
more than one solution is possible.
These stability issues are discussed further below.
We first focus on the changes in the ionization of the gas
shown in the lower two panels.

Table \ref{tab:Den3VarHeatCR} compares physical conditions
at three temperatures.
The low and high temperatures of 100~K and $10^4$~K are near the extremes
of the regions which produce the molecular and low-ionization emission.
The mid-temperature was evaluated as close to $10^3 \K$ as possible.
This is a typical temperature for \htwo -- emitting gas,
as shown in Paper 1.
No stable thermal phase exists at exactly this temperature so
the conditions at 700~K, the warmest stable region
below $10^3 \K$, is given.

% calculations are in LOCSpectrum\heat\varyHeat\varHeat.out and
%
\begin{table}
\caption{Physical conditions for the two cases at a range of
kinetic temperatures.}
\begin{center}
\begin{tabular}{llll}
\hline
Species & $T = 10^2 \K$ & $T \sim 700 \K $ & $T = 10^4 \K $ \\
\hline
$n_e$(Heat, $\pcc$)        & 0.013& 0.0089& 0.59 \\
$n_e$(CR, $\pcc$)          & 0.92& 9.3& 155\\
\hline
$n( \htwo )$(Heat, $\pcc$) & 952& 940& -\\
$n( \htwo )$(CR, $\pcc$)   & 232&  16& -\\
\hline
$n( \h0 )$(Heat, $\pcc$)   & 48 & 60 & 999\\
$n( \h0 )$(CR, $\pcc$)     & 768& 975& 863\\
\hline
$n( \hplus )$(Heat, $\pcc$)& -   & - & 0.0004\\
$n( \hplus )$(CR, $\pcc$)  & 0.80& 8.4& 137\\
\hline
\end{tabular}
\end{center}
\label{tab:Den3VarHeatCR}
\end{table}

In the extra-heating case thermal collisions are the only ionization source.
The result is that at low temperatures
the gas is predominantly molecular with a trace of H$^0$.
As the heating rate and temperature increase
there is an \emph{abrupt}\/ change in the chemical state when the kinetic temperature
approaches the dissociation potential of \htwo .
The dissociation of \htwo\ causes the particle density to increase and the
mean molecular weight to decrease.
Both cause the collisional rates, excitations, cooling, and ionization, to increase.
The result is a positive feed-back process that causes an abrupt phase transition
from \htwo\ to \h0 .
This is the reason that contours overlap in the lower panel of Figures
\ref{fig:HeatCrTempA} -- \ref{fig:HydroIonFrac}.

This behavior is in sharp contrast with the cosmic-ray case.
The effects of relativistic particles on a predominantly thermal gas
has been well documented in a number of papers starting with \citet{SpitzerTomasko68}
and most recently by \citet{XuMcCray91} and \citet{DalgarnoEtAl99}.
As shown in the upper right panel of Figure~\ref{fig:Den3VarHeatCR},
cosmic rays excite, heat, and ionize the gas.
The curves marked `Ionization' and `excitation' show the ionization and
excitation rate ($\ps$) but have been divided by the cosmic-ray-to-background ratio
to remove the effects of increasing cosmic-ray densities.
The curve marked `heating' has been similarly scaled and further
multiplied by 10$^7$ to facilitate plotting.
The fraction of the cosmic-ray energy that goes into each process
depends mainly on the electron fraction, $n_e / n_{\rm{H}}$,
the ratio of the thermal electron to total hydrogen densities.
This electron fraction increases as the cosmic-ray rate increases.

For low electron fractions a cosmic ray produces a first generation secondary
electron that is more likely to strike atoms or molecules causing
secondary excitations or ionizations.
Radiation produced by line excitation or recombination following ionization
will be absorbed by dust which then reradiates the energy in the FIR.
Relatively little of the cosmic-ray energy goes into directly heating the gas.
As the electron fraction increases, moving to the right in Figure~\ref{fig:Den3VarHeatCR},
more of the cosmic-ray energy goes into heating rather than exciting or ionizing the gas.
This is because for larger electron fractions the secondary electrons
have a greater probability of undergoing an elastic collision with a thermal electron.
This adds to the thermal energy of the free electrons and so heats the gas.

The effect is to produce a more gentle change in the ionization of the
gas as the cosmic-ray rate is increased.
Ionization and dissociation are mainly caused by non-thermal secondary particles.
Significant levels of dissociation or ionization exist at temperatures
that are too low to produce these effects by thermal collisions.
The lower two panels of Figure~\ref{fig:Den3VarHeatCR} show
how the \htwo\ and H$^0$ fractions change .
In the cosmic-ray case they change continuously
until the thermal instability point is reached.
Although a jump still occurs, the change in temperature is mitigated by
the more continuous change in ionization and molecular fractions.
This is to be contrasted with the extra-heating case where the gas is almost
entirely molecular or atomic with little mixing of the two phases.

Some specific values are given in Table~\ref{tab:Den3VarHeatCR}.
At the lowest temperature, 100~K, nearly all H is molecular in the
extra-heating case, while in the cosmic-ray case substantial amounts
of \h0\ are present.
There are even significant amounts of \hplus\ at this temperature because
of the ionization produced by the cosmic ray secondaries.
At the highest temperature shown, $10^4$~K, nearly all H is atomic in the
extra heating case while a substantial amount of \hplus\ is present
in the cosmic-ray case.

A substantial population of H$^0$ and H$^+$ is mixed with \htwo\
in the cosmic-ray heated gas.
Molecular regions become far warmer and have far fewer
H$^+$ ions mixed with \htwo\ in the extra-heating case.
This leads to the most important distinctions between the two cases.
Ions have larger collision cross sections so are more active in altering the
level populations of \htwo , as shown in Table~\ref{tab:CriticalDensities}.
The \htwo\ populations will be different as a result.
Further, H$^0$ and H$^+$ can underdo orth-para exchange collisions with \htwo .
This process will be far more rapid in the cosmic-ray case and
will be another distinction between the two cases.

\subsection{Thermal state of the gas}
\label{sec:ThermalStateAndHistoryOfTheGas}

Figure~\ref{fig:Den3VarHeatCR} shows that both cases have
temperatures that change discontinuously.
The origin of these jumps is described here.
We concentrate on the extra-heating case where the effects are the largest.

\begin{figure}
\begin{center}
\includegraphics[clip=on,width=\columnwidth,height=0.8\textheight,keepaspectratio]{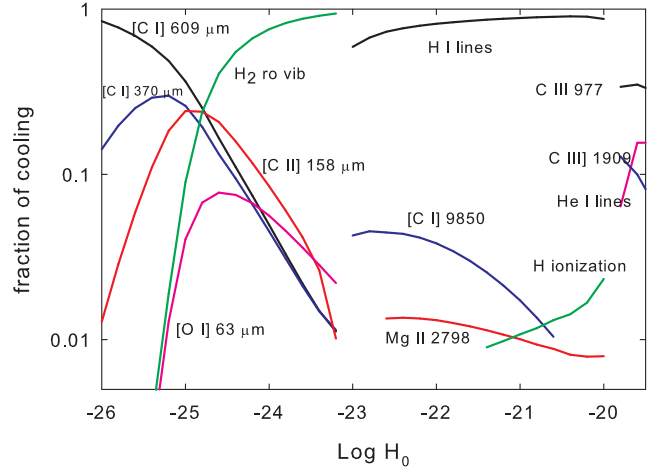}
\end{center}
\caption{This identifies the most important cooling transitions for the
extra-heating case shown in Figure~\ref{fig:Den3VarHeatCR}.
The cooling shifts from the IR when the gas is cold and molecular,
regions with $\log H_0 < -23$,
into the optical and UV when the gas is warm and ionized.}
\label{fig:IonizationCooling}
\end{figure}

Figure~\ref{fig:IonizationCooling} shows gas coolants
as the extra heating rate is varied across Figure \ref{fig:Den3VarHeatCR}.
The coolants leftward of the discontinuous break are classical
atomic and molecular `PDR' coolants.
A Galactic PDR is the H$^0$ region adjacent to galactic regions of star formation
\citep{Tielens05}.
These lines can be detected by current and planned IR instrumentation.

% the sims are in C:\projects\papers\CoolFlow\LOCSpectrum\Heat\varyHeat
% the graph is one of three in varyCrHeatHden.jnb

The gas abruptly changes from H$^0$ to H$^+$ at the discontinuity
where the temperature
jumps from `warm' ($\sim 10^3$~K) to `hot' ($\sim 10^4$~K).
The coolants on the hot side are mainly emission lines of atoms and ions of
the more common elements.
The strongest coolant, marked `H~{\sc i}~lines', is the set of Lyman lines that are
collisionally excited from the ground state.
Although their intrinsic intensity is large they do not escape the cloud because
of the large H~{\sc i} line optical depths and the presence of dust.
These lines are absorbed by dust as the photons undergo multiple scattering.
This is an example of a process that cools the gas by initially converting
free electron kinetic energy into line emission which is then absorbed by,
and heats, the dust.
The energy eventually escapes as reprocessed FIR emission.
Note that the gas and dust temperatures are not well coupled for the low
densities found in these filaments.
The dust is generally considerably cooler than the gas.

\begin{figure}
\begin{center}
\includegraphics[clip=on,width=\columnwidth,height=0.8\textheight,keepaspectratio]{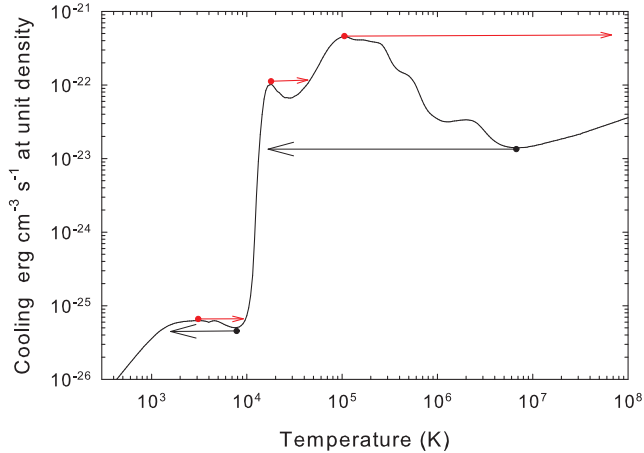}
\end{center}
\caption{The cooling function for gas with unit density and
a range of extra-heating rates.
The heating is varied and the equilibrium temperature determined.
The derived gas kinetic temperature is used as the independent variable
to better compare with previous calculations and
so that thermal stability can be judged.
The arrows indicate regions where the gas will undergo a discontinuous jump
in temperature to avoid thermally instable regions.
A gas that was originally cold and molecular would follow the curve moving from left to the right while initially hot gas would move from right to left.}
\label{fig:CoolingCurve}
\end{figure}

The behavior shown in the two previous figures,
with discontinuous jumps in the physical conditions, is due to
well-known thermal instabilities in interstellar clouds.
Figure~\ref{fig:CoolingCurve} shows the calculated cooling rate for a unit
density and, again, the extra heating case.
The heating rate is varied and the temperature determined by balancing
heating and cooling.
We plot the cooling as a function of temperature rather than
the heating rate to make it
easier to compare these results with previous calculations.

% in figures CoolCurve.jnb

The overall shape of the cooling function is discussed, for instance, in the
review by \citet{DalragnoMcCray72}.
At low temperatures cooling is mainly by molecules and low-lying levels
within ground terms of atoms and first ions of the abundant elements.
At low temperatures cooling usually involves changes in levels
which have excitation energies of $\leq 10^3$~K.
The cooling increases as the Boltzmann factor for excitation increases
and reaches a peak at $\sim 10^3$~K where the Boltzmann factor reaches unity.
The cooling then declines as $T$ increases further
due to the $T^{-1/2}$ dependence of the Coulomb focusing factor.
Starting at roughly 5000~K lines involving changes in term become energetically
accessible, increasing the cooling, and producing a peak at $\sim 10^4$~K.
The decline after the peak is again due to the $T^{-1/2}$ dependence
of the collision rate coefficient when the Boltzmann factor has reached unity.
The similar peak at $\sim 10^5$~K is due to transitions involving changes in
electronic configuration in ions of second and third-row elements.

The form of the cooling curve is affected by the presence of dust.
Many of the most important coolants in a dust-free mixture, such as iron,
calcium, silicon, and others, are strongly depleted when dust forms.
The result is that cooling is less efficient due to the loss of these
gas-phase coolants.
This is the major reason that the cooling curve differs from
the solar-abundance case.

The condition for thermal stability of a constant-density gas is
\begin{equation}
\left[ {\frac{{\partial \left( {C - H} \right)}}{{\partial T}}} \right]_\rho   > 0
\end{equation}
\citep{Field65}.
Here $C$ and $H$ are the cooling and heating rates,
erg cm$^{-3}$ s$^{-1}$,
and $\rho$ indicates that the derivative is at constant density.
Small perturbations in the temperature cause the temperature to continue to
change in the direction of the perturbation if the derivative is negative.
Our hypothesized non-radiative heating processes have no explicit
temperature dependence so the partial derivative with respect to temperature
simplifies to the derivative of the cooling.
The portions of the cooling curve shown in
Figure~\ref{fig:CoolingCurve} with negative slope are thermally unstable.

Thermal instabilities cause the gas to have a memory of its history.
The present state
of the gas will depend on whether it approaches thermal equilibrium
from the high or low-temperature state.
One possibility is that the filaments cooled down from the surrounding
hot-ionized plasma \citep{RevazEtAl08}.
They would have reached their current state
after moving from right to left in Figure~\ref{fig:CoolingCurve}.
In this case the gas would follow the leftward-pointing arrows when it
passed the unstable regions.
Filaments originating as cold molecular gas,
perhaps in the interstellar medium of the central galaxy,
would move from left to right and would follow the rightward-pointing arrows.
Different regions of the cooling curve are reached by a heating or cooling gas.
This difference could provide a signature of the history of the gas.

\subsection{The grain/molecule inventory and the history of the gas}
\label{sec:GrainMoleculeInventoryHistoryOfTheGas}

The dust content and molecular inventory provides a constraint
on the history of the filaments.
If the filaments formed from the surrounding hot X-ray plasma
they would likely be dust free.
Properties of dust-free clouds within a galaxy cluster were
examined by \citet{FerlandFabianJohnstone94} and
\citet{FerlandFabianJohnstone02}.
If the gas originated in the interstellar medium of the central
galaxy it would most likely contain dust, as we have assumed so far.

The timescales required to form the observed molecules provide a clue to the
origin of the gas.
The slowest step is the formation of \htwo .
There are no direct formation processes because the homonuclear molecule has no
permanent dipole moment.
In dusty environments \htwo\ forms by catalysis on grain surfaces.
We have adopted the rate derived by \citet{Jura75} for the galactic ISM.
The \htwo -formation timescale is then
\begin{equation}
\tau _{grain}  \sim 10^6 \left[ {\frac{{n\left( {{\rm{H}}^0 } \right)}}{{{\rm{10}}^{\rm{3}} \,{\rm{cm}}^{{\rm{ - 3}}} }}}  \frac{{A_{dust} }}{{A_{ISM} }}\right]^{ - 1}\quad {\rm{years}}
\end{equation}
where $n(\rm{H}^0)$ is the atomic hydrogen density
and $A_{dust}/A_{ISM}$ is the filament dust to gas ratio relative to the
ISM value.

If grains are not present then \htwo\ will mainly form by associative
detachment of H$^-$ \citep{FerlandFabianJohnstone94}.
This process is limited by the rate of the slowest step,
radiative association to form H$^-$.
This has a rate coefficient given by
$
{r = \rm{8}} \times {\rm{10}}^{{\rm{ - 16}}} t_3^{0.87} n_e \;{\rm{s}}^{{\rm{ - 1}}}
$
in the neighborhood of 10$^3$~K.
Carbon is the most abundant electron donor in \h0\ regions
where this process is fast.
The timescale can be posed in terms of the hydrogen density by assuming
a solar C/H ratio and that all C is singly ionized.
The \htwo\ formation timescale in a dust-free environment is then
\begin{equation}
\tau _e  \sim 2 \times 10^8 \left[ {\frac{{n\left( {{\rm{H}}^0 } \right)}}{{{\rm{10}}^{\rm{3}} \,{\rm{cm}}^{{\rm{ - 3}}} }}} \right]^{ - 1} t_3^{ - 0.87} \left( {\frac{{{\rm{C}}/{\rm{H}}}}{{{\rm{C}}/{\rm{H}}_ \odot  }}} \right)^{ - 1} \quad {\rm{years}}
\end{equation}
where the density, temperature, and C/H abundance are given in units of
10$^3 \pcc $, 10$^3$~K, and solar, respectively.

The dust-free \htwo\ formation timescales are $\sim 2.5$ dex longer than those in a dusty
environment.
Unfortunately the ages of the filaments are poorly constrained.
\citet{HatchEtAl06} find that a filament in the Perseus cluster with a projected
length of $\sim 25 \kpc $ and a velocity spread of $\sim 400 \kmps $.
The corresponding expansion age is $\geq 7 \times 10^7 \yr $.
This is comfortably longer than the formation time for a dusty gas.
\htwo\ could only form in the dust-free case if the density is
considerably higher than expected from typical pressures.
This is possible but unlikely.
We return to this question below.

The stable thermal solutions presented in the remainder of this paper
are limited to those on the rising-temperature branch of the cooling curve.
This, together with our assumed dust content, implicitly assumes that
the gas was originally cold and dusty.
In practice this means that the temperature solver determines
the initial temperature with a search that starts
at low temperatures and increases $T$ to reach thermal equilibrium.

\section{Properties of constant-pressure clouds}

\subsection{Equation of state}
The equation of state, the relationship between gas density and the temperature
or pressure, is completely unknown for the filaments.
Possible contributors to the total pressure include gas, turbulent,
magnetic, cosmic ray, and radiation pressures.
The gross structure of the filaments, forming large lines or arcs,
is reminiscent of magnetic phenomena like coronal loops.
This suggests that magnetic fields guide the gas morphology with
the field lines lying along the arc.
Gas would then be free to flow along field lines,
the long axis of the filaments, but not across them.

We assume that the gas pressure within the filaments is the same as the gas
pressure in the surrounding hot gas, $nT = 10^{6.5} \pcc\ \rm{K}$,
found by \citet{SandersFabian07} in the Perseus cluster.
The magnetic field must then be strong enough to guide
the cool material and so maintain
the linear geometry despite any motion with respect to the surrounding hot gas.
A magnetic field of $\sim 100 \mu $G has an energy density
equivalent to a thermal gas with $nT = 10^{6.5} \pcc\ \rm{K}$
so a field stronger than this is needed.
Little is known about the filament geometry at the sub-arcsec
($\delta r \leq 35 \pc$ at Perseus) level
and nothing is known about the magnetic field strength within the filaments.
Magnetic confinement does provide a plausible explanation
for the observed geometry however.

\begin{figure}
\begin{center}
\includegraphics[clip=on,width=\columnwidth,height=0.8\textheight,keepaspectratio]{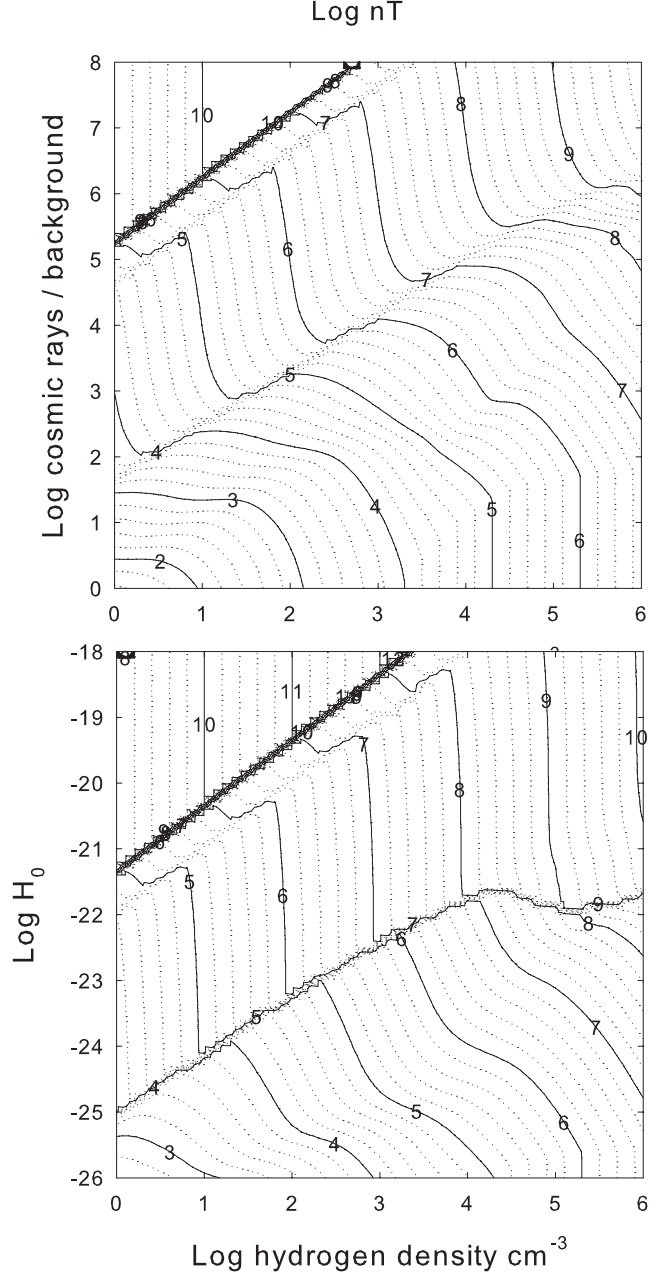}
\end{center}
\caption{The log of the gas pressure, expressed as $P/k = nT (\K \pcc)$,
is shown as a function of the hydrogen
density and (top panel) the cosmic-ray rate relative to the Galactic background
and (bottom panel) the extra-heating rate.
For comparison inner regions of the Perseus cluster
have $nT \approx 10^{6.5}$ cm$^{-3}$ K.}
\label{fig:pressure}
\end{figure}

Figure~\ref{fig:pressure} shows the gas pressure in
conventional ISM unit ($P/k = nT \pcc \K$) for the two cases.
If individual sub-components that make up the filaments have constant pressure
then they will lie along one of these contour lines.

% figure pressure is in varyCrHeatHden.jnb

\begin{figure}
\begin{center}
\includegraphics[clip=on,width=\columnwidth,height=0.8\textheight,keepaspectratio]{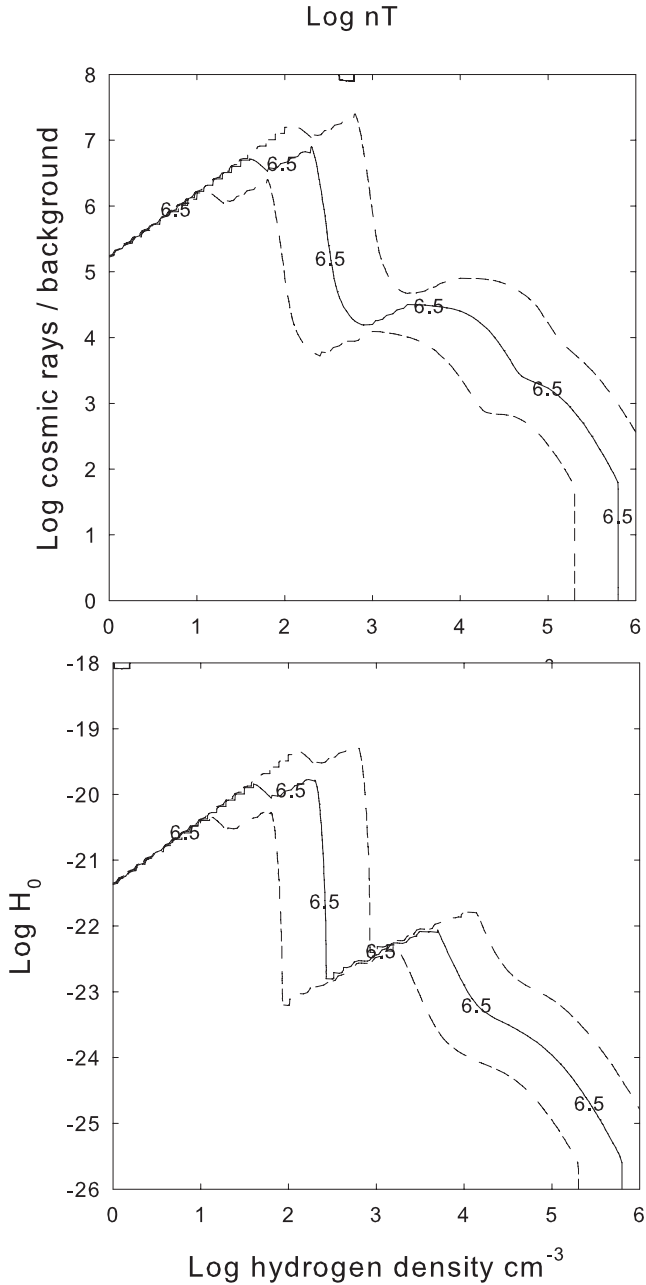}
\end{center}
\caption{The solid contour corresponds to the preferred
gas pressure $P/k = nT = 10^{6.5} \pcc $ K.
The dotted lines correspond to gas pressures 0.5 dex above and below this value.
Regions where the contours have a steeper than 45 degree slope, where the
contours nearly overlap, are thermally unstable.}
\label{fig:PressurePreferred}
\end{figure}

The dynamic range in Figure~\ref{fig:pressure} is large.
The solid line in Figure~\ref{fig:PressurePreferred}
corresponds to our preferred pressure of $nT = 10^{6.5} \pcc $~K.
The dashed lines in the figure show pressures 0.5 dex to either side of
this value.
As we show next,
regions where the contours exceed a 45 degree angle, and where the $\pm 0.5$ dex
contours nearly overlap, are thermally unstable.

% figure PressurePreferred is in varyCrHeatHden.jnb

\subsection{Thermal stability}
\label{sec:ThermalStability}

The criterion for thermal stability in a constant-pressure gas is
\begin{equation}
\left[ {\frac{{\partial \left( {C - H} \right)}}{{\partial T}}} \right]_P  = \left[ {\frac{{\partial \left( {C - H} \right)}}{{\partial T}}} \right]_\rho   - \frac{{\rho _0 }}{{T_0 }}\left[ {\frac{{\partial \left( {C - H} \right)}}{{\partial \rho }}} \right]_T  > 0
\label{eq:StabilityConstantPressure}
\end{equation}
\citep{Field65}, where the equality assumes constant mean molecular mass.
The first term on the right
is the same as that considered in the constant-density case discussed
in the Section \ref{sec:ThermalStateAndHistoryOfTheGas}.
Both of the heating sources considered here have linear density dependencies
and no explicit dependence on temperature.
At constant pressure this will carry over into a heating rate that is inversely
proportional to the temperature so the second term is non-zero.

\begin{figure}
\begin{center}
\includegraphics[clip=on,width=\columnwidth,height=0.8\textheight,keepaspectratio]{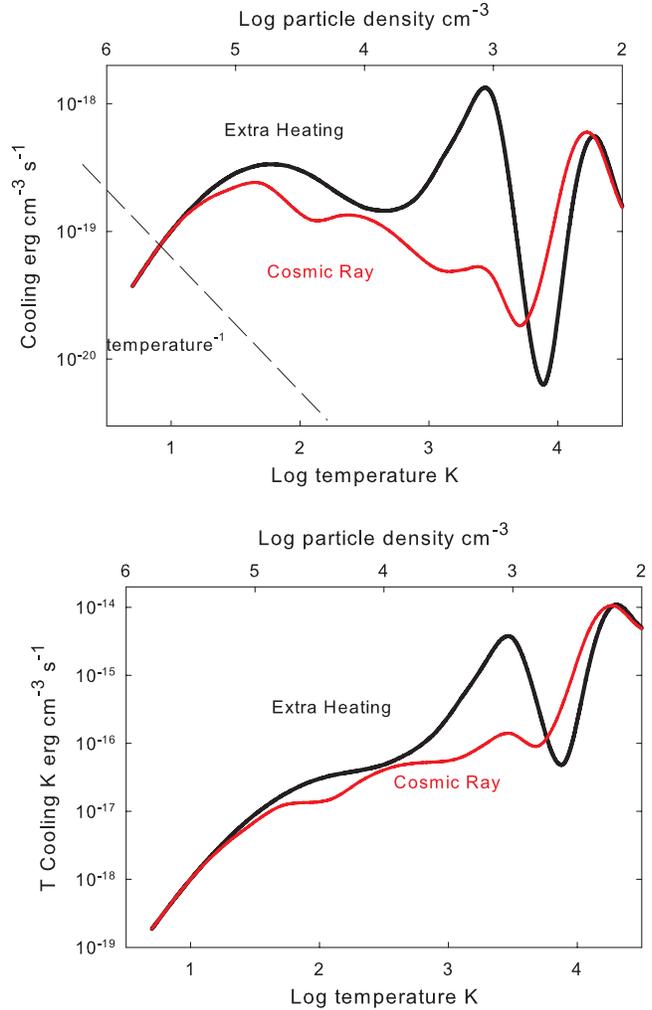}
\end{center}
\caption{The solid lines in the upper panel give the volume cooling rates
for the two cases as a function of the temperature
(the lower independent axis)
and particle density (he upper axis)
along the $P/k = nT = 10^{6.5} \pcc $ K isobaric contour of the
previous figure.
Regions where the slope of the cooling curve is steeper than the dashed line
marked temperature$^{-1}$ in the upper panel are thermally unstable.
The lower panel shows the product of the cooling and the temperature.
Regions where this product has a negative slope are unstable.
}
\label{fig:CP_stabilityA}
\end{figure}

The upper panel of Figure~\ref{fig:CP_stabilityA}
shows the gas cooling rate along the
isobaric log $nT$ = 6.5 contour in Figure~\ref{fig:PressurePreferred}.
Both heating cases are drawn in each panel.
The figure shows the cooling as a function of temperature
rather than the heating rate so that
the gas stability can be more easily judged.
The detailed cooling properties of the two sources of non-radiative
heating are quite different, as can be seen from the physical properties
shown in Figure~\ref{fig:Den3VarHeatCR} and listed in
Table \ref{tab:Den3VarHeatCR}.
The line marked `temperature$^{-1}$'
shows the approximate form of the second term
in equation \ref{eq:StabilityConstantPressure}.
Regions where the slope of the cooling function is steeper than this are unstable.
The lower panel of Figure~\ref{fig:CP_stabilityA} shows the
product of the cooling and the temperature.
Expressed this way, regions with negative slope are unstable.
Two distinct phases, corresponding to cold molecular
and warm atomic/ionized, exist.

% the graph is CP_stability in CP_stability.jnb

The upper axis in Figure~\ref{fig:CP_stabilityA},
and the figures that follow, gives the
particle density along the isobaric line
while the lower axis gives the kinetic temperature.
Temperature and density are related
since the product $nT$ is constant.
Note that the density $n$ is the \emph{total}\/
particle density not the hydrogen density.
For molecular regions the total particle density is
about half the hydrogen density
while in ionized regions it is about twice
due to the presence of free electrons.

\begin{figure}
\begin{center}
\includegraphics[clip=on,width=\columnwidth,height=0.8\textheight,keepaspectratio]{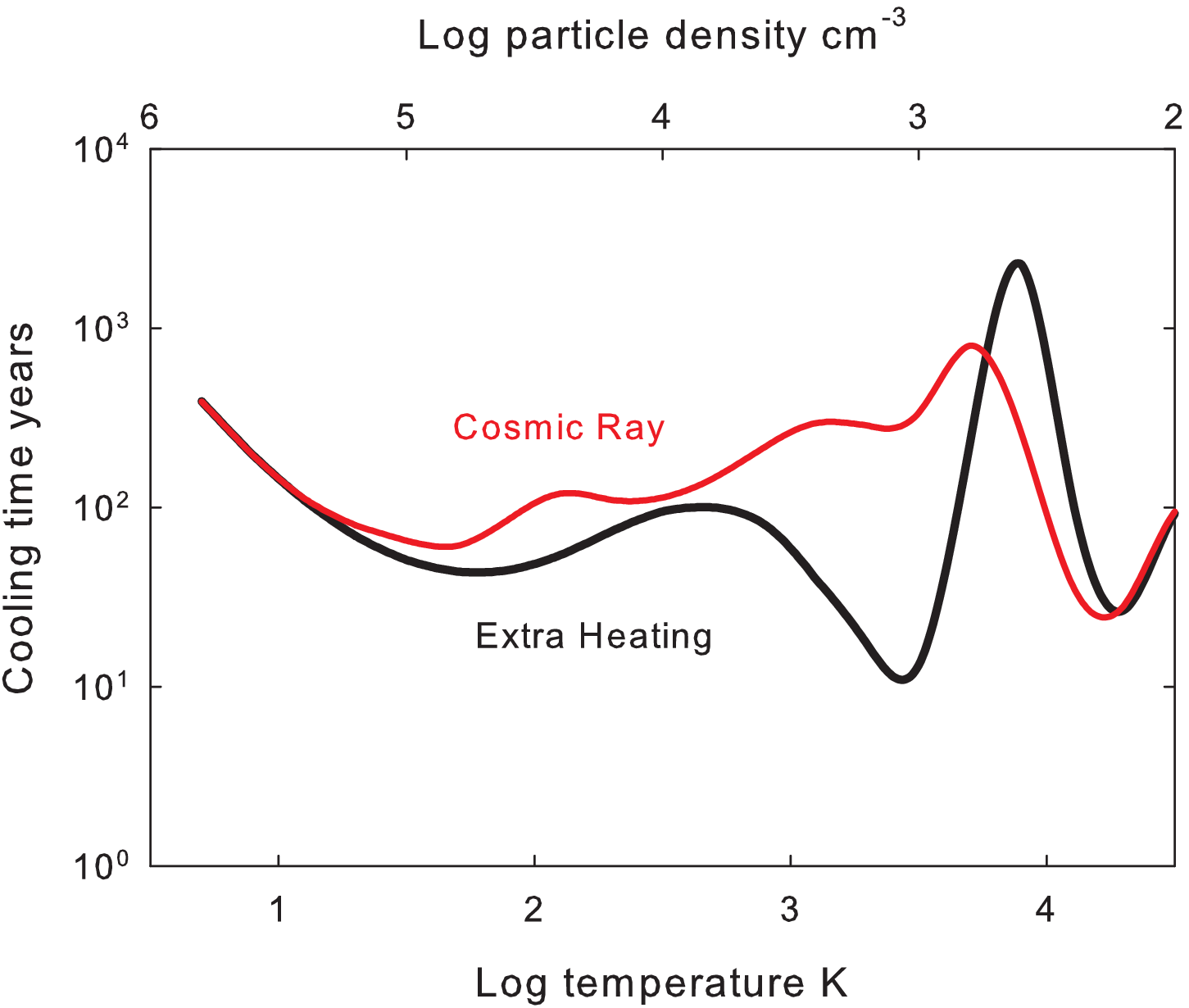}
\end{center}
\caption{The cooling timescales for the two cases shown in the previous
figure is shown.
Gas in unstable regions of the previous figure will move to a stable portion
of the cooling curve on this timescale.
}
\label{fig:CP_cooling_timeA}
\end{figure}

Figure~\ref{fig:CP_cooling_timeA} shows the cooling time for the
cases shown in Figure~\ref{fig:CP_stabilityA}
The cooling time will determine how quickly the gas can respond to changes
in the environment.
It also determines how quickly thermally unstable gas will heat or cool
and reach stable regions of the cooling curve.
These times are short compared with timescales over which
the galaxy cluster changes and are far shorter than the
\htwo\ formation timescales mentioned above.

% the graph is one of three in CP_stability.jnb

\subsection{Spectrum emitted by a homogeneous cloud}

\begin{figure}
\begin{center}
\includegraphics[clip=on,width=\columnwidth,height=0.8\textheight,keepaspectratio]{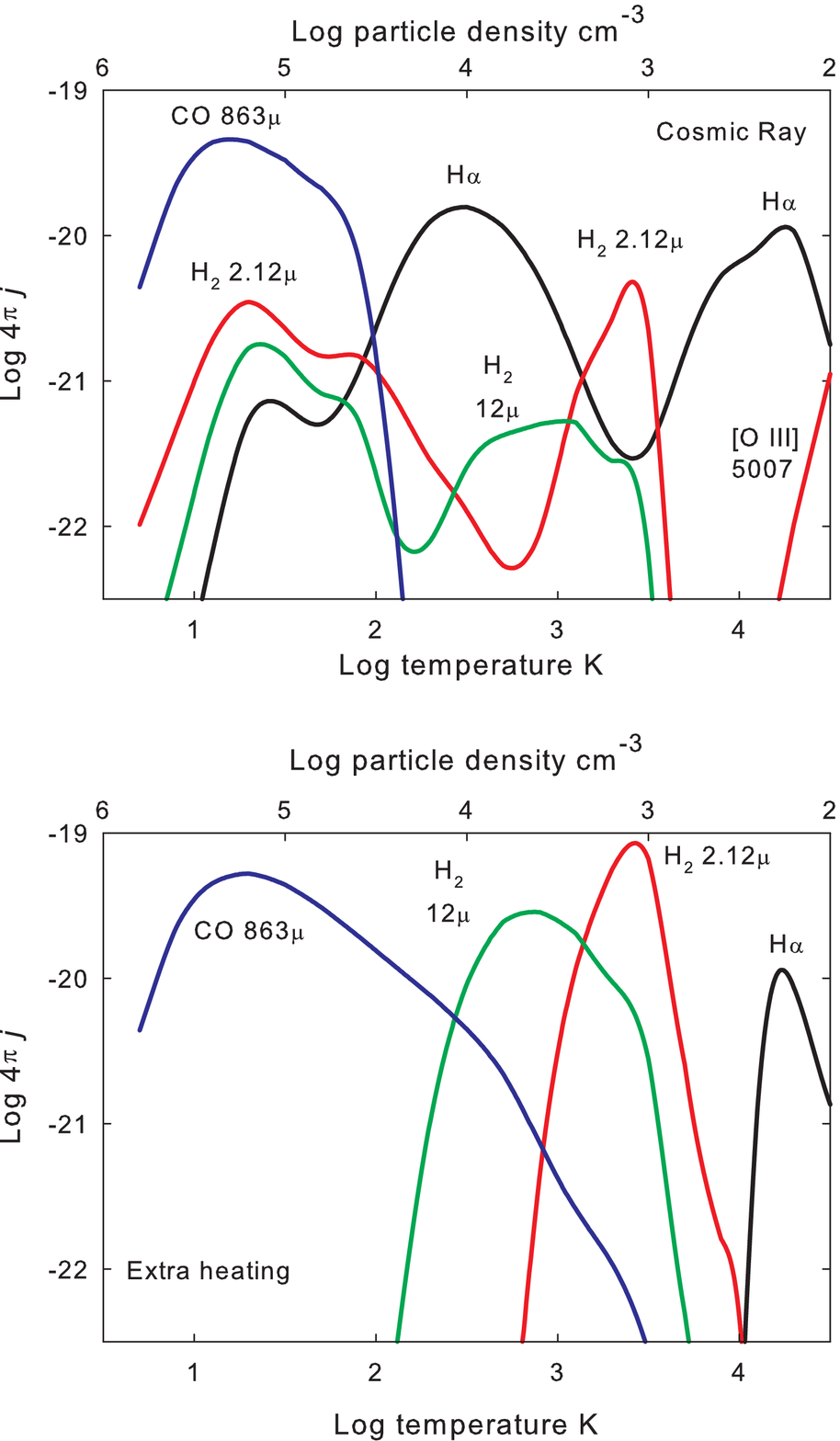}
\end{center}
\caption{The emissivities, the emission per unit volume,
are shown for several emission lines along the isobaric line
corresponding to $P/k = nT = 10^{6.5} \pcc \K $.
The total particle density and temperature are indicated as the independent axes.
The upper panel is the cosmic ray case and the lower panel is for extra heating.
}
\label{fig:CP_spectrum}
\end{figure}

The full emitted spectrum is computed for each point along the isobaric line
in Figure \ref{fig:PressurePreferred}.
Emissivities $4\pi j$ of a few representative lines
are shown in Figure~\ref{fig:CP_spectrum}.
The luminosity of a line will be the integral of $4\pi j$
over the volume of the emitting region.

% the graph is in CP_spectrum.jnb

Most of the lines shown in Figure~\ref{fig:CP_spectrum} will be optically thin
for any reasonable column density.
In this case their emissivity
has no explicit dependence on column density.
If we can neglect internal reddening,
a good approximation for the IR and radio lines
but an additional uncertainty for optical or UV transitions,
then the line luminosity is the integral of the emissivity
over the volume containing gas.
This is simple since the luminosity of an optically thin line
is simply proportional to the total volume of
emitting gas rather than on details of how the gas is arranged.

The CO lines are more problematic since lower-$J$ transitions
are generally optically thick in the Galactic ISM.
The emissivity of the CO $J=3 \rightarrow 2\ \lambda 863 \micron$
transition is shown
rather than the lower-$J$ lines normally observed \citep{SalomeEtAl06},
since it is more likely to be optically thin,
if the column density of CO is large enough to
thermalize an optically-thick line then the luminosity will be
set by the gas kinetic temperature and the cloud area
rather than on the emissivity.
It is by no means certain that the $\lambda 863 \micron$
will be optically thin in the filaments and we shall
return to the CO line spectrum below.

Figure~\ref{fig:CP_spectrum}
shows that different lines form at different temperatures and densities because,
as Figure~\ref{fig:Den3VarHeatCR} shows,
atoms and molecules exist at different
regions with little overlap.

Paper 1 showed that \htwo\ lines in the
filaments are much stronger relative to \hi\
recombination lines than is found in normal star-forming regions like Orion.
\hi\ recombination lines will have a peak emissivity that is weighted towards the
coolest and densest regions where H$^+$ is present,
as shown in Equation \ref{eq:HIEmissivity} and
Figure \ref{fig:Hb_emissitivty_contourA}.
This is indeed the case in the extra-heating case as shown in the lower panel of
Figure~\ref{fig:CP_spectrum},
where we also see that the \hi\ and \htwo\ lines form in gas with distinctly
different densities and temperatures.

The extra-heating case, shown in the lower panel of Figure~\ref{fig:CP_spectrum},
is simplest and we consider it first.
Lines excited by thermal collisions tend to form in warmer gas
because of the exponential Boltzmann factor.
Low-excitation \htwo\ lines such as $\lambda 12.28 \micron$
form in cooler gas than the \htwo\ $\lambda 2.12 \micron $,
which has an upper level with an excitation potential of $\sim 7000\K$,
in the extra-heating case.

The emissivities of the H$\alpha$ and \htwo\ lines have two local peaks
in the cosmic-ray case,
as shown in the upper panel of Figure~\ref{fig:CP_spectrum}.
The lower-temperature peak is due to the direct
excitation of \hi\ and \htwo\ lines by
secondary electrons.
Excitation to \htwo\ electronic levels that
then decay into excited states of the \htwo\ ground state,
a process analogous to the photon Solomon process \citep{ShawEtal05},
contributes to the \htwo\ emission.
The higher-temperature peak occurs when cosmic rays heat the gas
to a warm enough temperature to excite \htwo\ transitions with thermal collisions.
Eventually, in the highest-$T$ regions of the figure, the gas becomes warm enough to
be predominantly ionized and the \hi\ lines begin to form by recombination.

\begin{figure}
\begin{center}
\includegraphics[clip=on,width=\columnwidth,height=0.8\textheight,keepaspectratio]{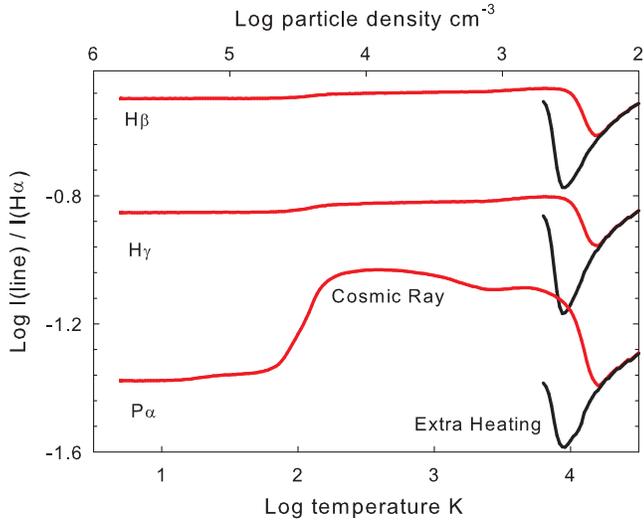}
\end{center}
\caption{The intensities of several strong optical and IR H~{\sc i} lines
are shown for points along the isobaric line
corresponding to $\log nT = 6.5$.
The total particle density and temperature are indicated as the independent axes.
}
\label{fig:CP_Hlines}
\end{figure}

The hydrogen emission-line spectrum is predicted to be different from
Case B relative intensities in regions where the lines form by
collisional excitation of \h0\ rather than by recombination of \hplus.
Figure~\ref{fig:CP_Hlines} shows the predicted intensities
of the brighter optical and
IR lines relative to H$\alpha$.
There are two reasons for differences from simple Case B, the first being
the large range in kinetic temperature along the isobaric line.
The standard Case B spectrum is most often quoted for $T = 10^4$~K, appropriate
for a photoionized cloud.
The fact that collisional processes excite the lines in this environment
cause further deviations from Case B.

% the graph is in CP_spectrum.jnb

Table \ref{tab:HI_emission_spectrum} compares the \hi\ spectrum
for two points along the constant-$nT$ line with Case B
predictions.
Little \hi\ emission occurs for the extra-heating case at $T \approx 300 \K$
so only the cosmic-ray case is given.
The extra heating and cosmic ray predictions agree at
$T \approx 2\times 10^4 \K$ because the gas is warm enough to collisionally
excite the \hi\ lines.
The emissivity $4 \pi j$ is $\sim 1$~dex brighter
than a pure recombination \hi\ spectrum
because of the contribution of collisional excitation of \h0 .
Note that the quantities given in Table \ref{tab:HI_emission_spectrum}
are the emissivity $4 \pi j$ and not the emission coefficient
 $4 \pi j / n_e n_p$ given in, for instance, \citet{AGN3}
The relative intensities of the \hi\ lines are also different in the two cases.
The \hi\ lines at $T \approx 300 \K$ are mainly excited
by non-thermal secondary electrons
with collision rates that are proportional
to the oscillator strength of the Lyman line.
The emission at $T \approx 2\times 10^4 \K$ is mainly produced by collisions
with thermal electrons whose rates are dominated by the
near-threshold cross section.
Case B applies when the lines form by recombination, which is not the case here.

Table \ref{tab:HI_emission_spectrum} shows that, although the \hi\ lines
have a far greater emissivity than is given by Case B, the relative
intensities are not too dissimilar.
This is because the relative intensities of the \hi\ spectrum are
most strongly affected by the transition probabilities that determine
how a highly-excited electron decays to ground rather than detail detains
of how the electron became excited.

\begin{table}
\centering
\caption{Predicted \hi\ emission spectrum}
\begin{tabular}{ c c c c c c }
\hline

&CR&CR,H&Ca B&Heat&CR\\
log $n_{\rm{H}}$&4&2.2&2& integrated& integrated\\
log $T$ &2.5&4.3&4.3&&\\
\hline
$4\pi j(\rm{H}\alpha )$&-19.85  &-19.96  &-21.01 & -20.029 & -20.171\\
\hi\   3798\AA         & 0.0132 &0.00076 &0.0197 &6.41(-3) & 0.0109\\
\hi\   3835\AA         & 0.0181 &0.0112  &0.0272 &9.38(-3) & 0.0152\\
\hi\   3889\AA         & 0.0261 &0.0172  &0.0391 &0.0145   & 0.0224\\
\hi\   3970\AA         & 0.0400 &0.0286  &0.0591 &0.0244   & 0.0352\\
\hi\   4102\AA         & 0.0665 &0.0531  &0.0962 &0.0459   & 0.0608\\
\hi\   4340\AA         & 0.125  &0.116   &0.173  &0.102    & 0.120\\
\hi\   4861\AA         & 0.288  &0.254   &0.364  &0.227    & 0.271\\
\hi\   6563\AA         & 1.000  &1.000   &1.000  &1.000    & 1.000\\
\hi\  1.875\micron     & 0.121  &0.0815  &0.104  &0.0839   & 0.0992\\
\hi\  2.166\micron     & 0.0082 &0.0053  &0.085  &4.84(-3) & 6.35(-3)\\

\hline
\end{tabular}
\label{tab:HI_emission_spectrum}
\end{table}

\subsection{Allowed range of cloud density and temperature}
\label{sec:DensityTemperatureRangeOfClouds}

The observed spectrum has a wide range of molecular and atomic emission,
requiring that cloudlets with a range of densities and temperatures occur.
Only some of the solutions shown in Figure~\ref{fig:CP_spectrum} will exist
in a constant-pressure filament.
The lowest temperature is set by the CMB temperature,
the lowest temperature that molecules and grains will have.
This low-temperature limit sets a high-density limit
since the product $nT = 10^{6.5} \K \pcc$ is constant.
Gas denser than $n_{\rm CMB} \sim 10^6 \K$ will remain at the CMB temperature,
be over-pressurized relative to its environment, and would expand.
We do not consider gas denser than $n_{\rm CMB}$.

Thermal stability (Section \ref{sec:ThermalStability} and Figure~\ref{fig:CP_stabilityA})
sets a low-density, high-temperature, limit if the system is time steady.
Gas on unstable parts of the cooling curve will move to regions of
thermal stability on the cooling timescale of the region.
We only consider thermally stable solutions here and come back to discuss
this point below.

\begin{figure}
\begin{center}
\includegraphics[clip=on,width=\columnwidth,height=0.8\textheight,keepaspectratio]{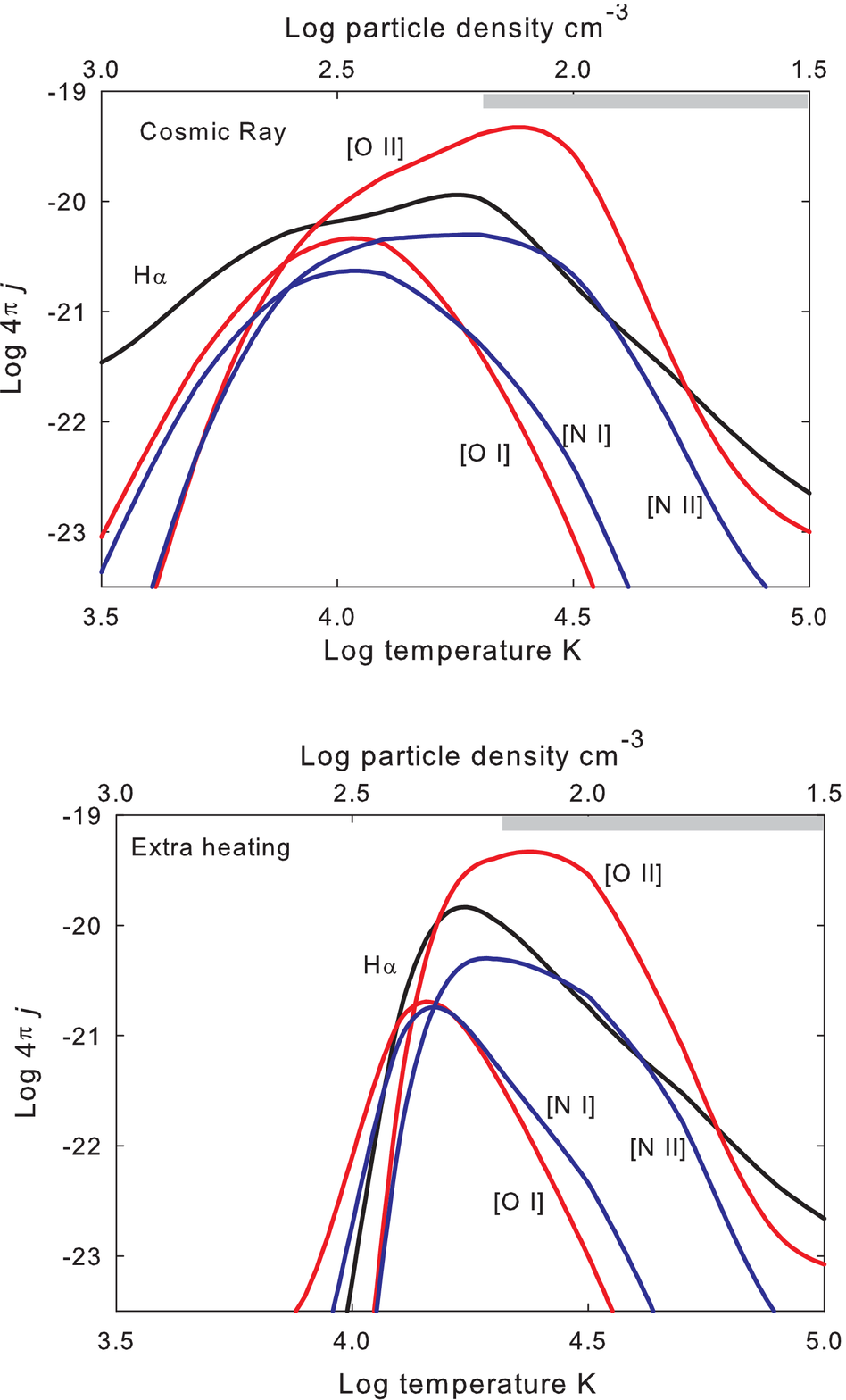}
\end{center}
\caption{The emissivities of several of the prominent optical
atomic and low-ionization lines seen in the filaments are shown.
The lines include [\oi ] $\lambda$6300,
[\oii ] $\lambda$3727, [\ni ] $\lambda$5199, and [\nii] $\lambda$6584.
The shaded rectangle indicates thermally unstable regions.
}
\label{fig:CP_spectrum_Hi_T_end_strongA}
\end{figure}

Figure~\ref{fig:CP_spectrum_Hi_T_end_strongA} shows the emissivities of
some strong lines that are produced near the low-density, high-temperature,
end of the cloud distribution.
We concentrate on the optical spectrum in this Figure
since these lines form in warm gas and because
their relative intensities can be measured with greater precision
owing to the fact
that this spectral range can be observed with a single entrance slit.
The grey band in the upper right of each panel indicates regions
which are thermally unstable.
Most of the lines shown in the Figure have peak emissivities that
occur in stable regions.
The predicted intensities of these lines will not depend
greatly on the precise high-$T$
cutoff of the cloud distribution.

% the graph is in CP_spectrum.jnb

\begin{figure}
\begin{center}
\includegraphics[clip=on,width=\columnwidth,height=0.8\textheight,keepaspectratio]{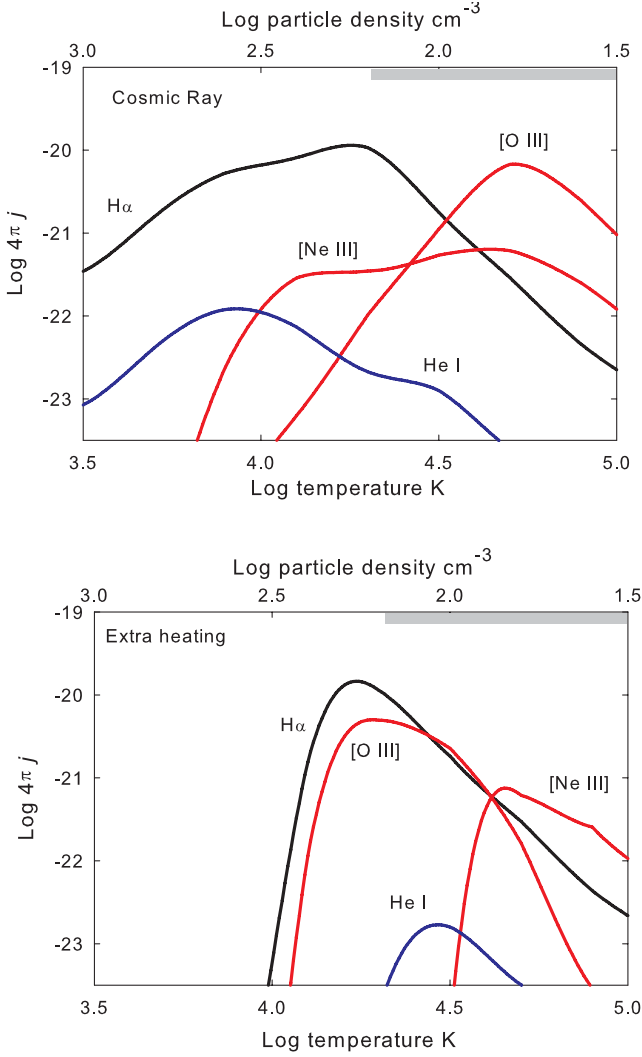}
\end{center}
\caption{The emissivities of several optical
atomic and low-ionization lines that are either sensitive to the
thermal stability cutoff or which might discriminate between the
two non-radiative heating models we discuss.
The lines include [\oiii ] $\lambda$5007, [\neiii ] $\lambda$3869,
and \hei\ $\lambda$5876.
H$\alpha$ is shown for reference.
The shaded rectangle indicates thermally unstable regions.}
\label{fig:CP_spectrum_Hi_T_end_faintA}
\end{figure}

Figure \ref{fig:CP_spectrum_Hi_T_end_faintA} shows some fainter lines which are
currently unobserved in the Horseshoe filament in the Perseus cluster
which we model in detail below.
Some, for instance [\oiii ] $\lambda$5007\AA, will be very sensitive to the
thermal stability cutoff since the peak emission occurs in unstable regions.
We show below that these lines can
discriminate between the two non-radiative heating processes.

% the graph is in CP_spectrum.jnb

\section{Emission from a distribution of Clouds}

\subsection{The need for clouds with a range of properties}

The spectrum shows that emission from both molecular and ionized gas comes
from gas that is nearly spatially coincident
\citep{SalomeEtAl08b}.
Physically this requires that gas with a variety of densities contributes
to the net emission detected by the entrance aperture of a spectrometer.

No one choice of cloud parameters can reproduce the full range of molecular, atom,
and ionic emission.
Figures \ref{fig:H2_emissitivty_contourA} and \ref{fig:Hb_emissitivty_contourA}
show that \hi\ and \htwo\ emission originate in warm and hot regions respectively.
The observed [\nii ] or [\oii ] emission cannot originate in the \htwo\ region.
Gas with a range of temperature and ionization
must contribute to the observed emission.

The Galactic ISM has two or more thermally distinct phases
that are in pressure equilibrium \citep{FieldGoldsmithHabing69}.
Figure \ref{fig:PressurePreferred} shows that gas with two temperatures
but the same pressure exist
for some values of the non-radiative heating rates.
This occurs for cosmic ray fluxes $\approx 4.2$ dex times the Galactic background
and for heating rate pre-coefficients of log $H_o \approx -22.5$.
The range in gas properties for these two points is not broad enough to
produce the observed range of emission however.
For instance, in the cosmic-ray case the two hydrogen densities are
$\approx 10^3 \pcc$ and $\approx 10^{4.3} \pcc$, corresponding
to kinetic temperatures of $T \approx 10^{3.5} \K$ and $T \approx 10^{2.2} \K$.
Gas with this pair of densities,
when co-added in the appropriate mix of
filling factors, can account for some of the \hi\ and \htwo\ emission
but none of the optical forbidden lines,
as shown in Figure \ref{fig:CP_spectrum_Hi_T_end_strongA}.

The full range of thermally stable clouds shown in Figure \ref{fig:CP_stabilityA}
must exist and contribute to the spectrum.
It is then necessary to adopt a weighting function, as described next,
to decide how to co-add clouds with different conditions.
This weighting function is not a physical model for why this mix of clouds exist.
Rather it represents a parametric way of solving the spectroscopic problem
by specifying how the volume of clouds depends on density.
We will end up with an empirical fit to the spectrum and finally discuss
various scenarios that might establish this range of conditions.

\subsection{Formalism}
\label{sec:FormalismLOC}

We assume that the filaments are composed of cloudlets with a range of
density $n$.  The cumulative filling factor $f(n)$ of material with
density below $n$ (i.e.\ fractional volume filled by material with
this density or less) is
\begin{equation}
f( n ) = \left\{
\begin{array}{cl}
0 & n \le n_{\rm low} \\
k \left\vert n^{\alpha}-n_{\rm low}^{\alpha} \right\vert &
n_{\rm low} < n \le n_{\rm high}, \alpha \ne 0 \\
k \log \left(n/n_{\rm low}\right) &
n_{\rm low} < n \le n_{\rm high}, \alpha \equiv 0 \\
1 & n > n_{\rm high},
\label{eqn:FillingFactorDefinition}
\end{array}
\right.
\end{equation}
where $n$ is the total hydrogen density, $\alpha$ is a parameter to be
determined by fitting the observed spectrum, and $k$ is a
normalization constant.  The constant value $1$
for densities above $n_{\rm high}$
corresponds to the observation that the whole volume is filled by
material of lower densities.  The  more commonly-used filling factor,
described in Section 5.9 of \citet{AGN3},
is the dimensionless fraction of space that is
filled with a particular phase of material.  For a continuous density
distribution, which we consider here, it is more useful to define
a differential filling factor which will
be a function of density.
This is most easily defined in this cumulative form.

The physical picture is that the aperture of the spectrometer takes in
emitting material with a range of densities and corresponding
temperatures but a single pressure.  This might occur, for instance,
if the non-radiative heating rate is not uniform in space or time.
Gas would heat up or cool down, on the cooling timescale given in
Figure~\ref{fig:CP_cooling_timeA}, in response to changes in the
heating.  The density and temperature would change to maintain
constant pressure but clouds with a variety of densities and
temperature would contribute to the emission from a region on the sky.

With these assumptions the line emissivity
$4 \pi \bar j$ (erg $\pcc \ps$) integrated over a distribution
of clouds is given by
\begin{eqnarray}
4\pi \bar j &=&
  \int_0^1 {4\pi j\left( n \right)} \,\,df\\
	&=&
	\int_{n_{\rm low}}^{n_{\rm high}} {4\pi j\left( n \right)} \,\,\, n^{\alpha  - 1} \,dn\left/
	\int_{n_{\rm low} }^{n_{\rm high} } { n^{\alpha  - 1} \,dn} \right.,
\label{eqn:EmissivityDefinition}
\end{eqnarray}
where the integration is over the physical limits to the filling
factor, $0 \leq f(n) \leq 1$, and the integration over filling factor
is rewritten in terms of an integral over density.
Here $n_{\rm low}$ and $n_{\rm
high}$ are the lower and upper density bounds to the integration.  In
practice we simulate physical conditions and emission from clouds for
a wide range of specific densities $n$ and replace the integral with
the sum
\begin{equation}
4 \pi \bar j \simeq
\sum\limits_{i=0 }^{i=j }
  {4\pi j_i\left( n_i \right)\,\, n_i^{\alpha  - 1} \,\Delta n_i} \left/
\sum\limits_{i=0}^{i=j }
  {\,\, n_i^{\alpha  - 1} \,\Delta n_i}
 \right.,
 \label{eqn:EmissivitySumOverDensity}
\end{equation}
where the density bins are separated by $\delta n/n\sim 0.1$ dex.

The hydrogen density averaged over the cloud distribution, $\bar n$ ($\pcc$),
is similarly defined as
\begin{equation}
\bar n =
  \int_0^1 {n} \,\,df.
  \label{eqn:DensityAveragedDistribution}
\end{equation}
This becomes
\begin{equation}
\bar n = \frac{\alpha }{{\alpha  + 1}}\,\,\frac{{n_{high}^{\alpha  + 1}  - n_{low}^{\alpha  + 1} }}{{n_{high}^\alpha   - n_{low}^\alpha  }}
\label{eqn:DensityAverageExplicit}
\end{equation}

With these definitions the luminosity of a line $L_{line}$ ($\ergps$) is related to
the emissivity $4 \pi \bar j$ and the total volume containing gas $V$ ($\cm^3$) by
\begin{equation}
L_{line} = V 4 \pi \bar j
\label{eqn:LuminosityVolumeEmissivityRelationship}
\end{equation}
The mass contained in this volume is given by
\begin{equation}
M = V \bar n \mu
\label{eqn:MassVolumeRelationship}
\end{equation}
where $\mu$ is the mean mass per proton
which we take to be $1.4 \, u$.
The line surface brightness is then
\begin{equation}
S_{line} = 1.8704 \times 10^{-12} \times 4 \pi \bar j \ dl
\label{eqn:SurfaceBrightnessThicknessEmissivityRelationship}
\end{equation}

The physical meaning of these quantities is as follows.
The volume $V$ is the total volume containing gas.
The gas within this volume has a range of density and temperature and
may be molecular, atomic, or ionized.
The line emissivity $4 \pi \bar j$ takes into account both
the atomic physics of the emission process and the fact
that a line is only emitted for certain combinations of density and temperature.
The weighting parameter $\alpha$ determines
the amounts of gas at various densities.
The density $\bar n$ depends on $\alpha$ since this sets how much
weight is given to gas at various densities.
It further depends on thermal stability criterion which sets $n_{low}$.

There are three free parameters,
the lower and upper density bounds $n_{\rm low}$ and $n_{\rm high}$
and the power-law index $\alpha$.
Physically motivated limits are used to set the density bounds
in Equation \ref{eqn:EmissivitySumOverDensity}.
As described in Section \ref{sec:DensityTemperatureRangeOfClouds},
thermally unstable densities are excluded from the integrals.
This is equivalent
to assuming that gas that lies on unstable portions of the cooling curve,
shown in Figure~\ref{fig:CP_stabilityA}, will quickly move
to stable regions so that the volume is fully filled by stable gas.
The CMB temperature is assumed to set a lower limit to the kinetic temperature
which then sets the upper limit to the density since the product $nT$ is held fixed.
This leaves only $\alpha$ as a free parameter.

\subsection{Dependence on $\alpha$ }

The full emission-line spectrum, with many hundreds of thousands of lines,
is computed for each point along the isobaric curve illustrated by
Figure \ref{fig:CP_spectrum}.
The emissivity of each line is stored and integrated over density
using Equation \ref{eqn:EmissivitySumOverDensity}.
Integrated predictions are presented relative to H$\alpha$
since it is easily observed and,
of the bright optical lines, is the least affected by reddening.

Figure~\ref{fig:CP_spectrum} shows that different lines have emissivities that
peak in different regions along the isobaric line.
Lines that are formed in warm or hot ionized gas,
such as the [S II] or [O II] optical lines,
will form in nearly the same region as H$\alpha$ itself as shown
in Figures \ref{fig:CP_spectrum_Hi_T_end_strongA} and \ref{fig:CP_spectrum_Hi_T_end_faintA}.
We expect that the intensities of such lines relative to H$\alpha$ will
have little dependence on $\alpha$ but will depend
on the chemical composition and the energy source.
This is in contrast with molecular lines, such as the CO or \htwo\ lines shown
in Figure \ref{fig:CP_spectrum},
which form in gas that is much colder and denser that the gas producing
the \hi\ emission.
Because of the form of the filling factor function we expect that
molecular lines will be stronger relative to H$\alpha$
for $\alpha$ close to 0 and that
the intensity ratio decreases
as $\alpha$ becomes more negative.

\begin{figure}
\begin{center}
\includegraphics[clip=on,width=\columnwidth,height=0.8\textheight,keepaspectratio]{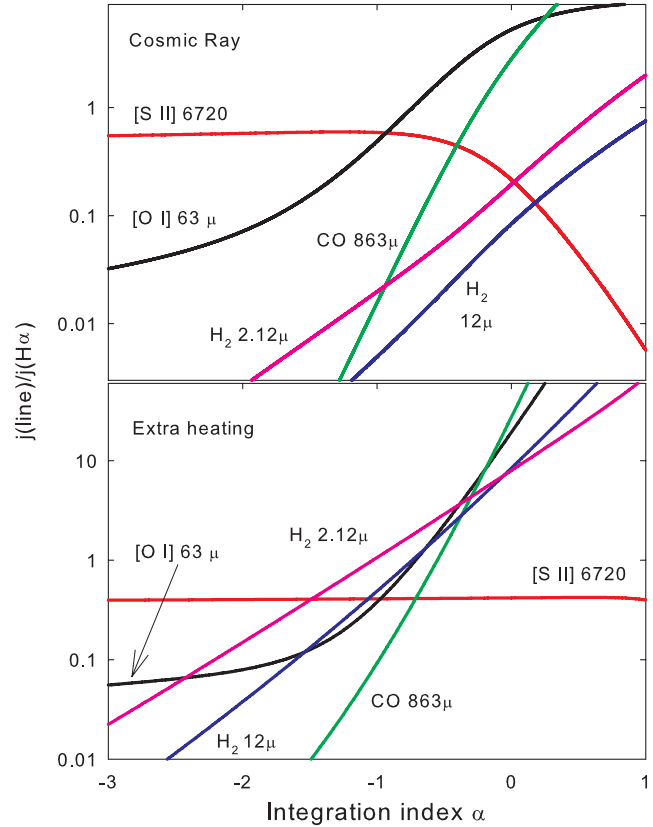}
\end{center}
\caption{
The integrated intensities of some strong lines
are shown as a function of the power-law index $\alpha$.
The upper panel shows the \hi\ H$\alpha$ emissivity $4 \pi \bar j$.
The middle and lower panels show the predicted intensities of other lines relative
to H$\alpha$ for the cosmic ray (middle panel) and extra heating (lower panel) cases.
}
\label{fig:CP_spectrum_alpha}
\end{figure}

Figure~\ref{fig:CP_spectrum_alpha} shows the predicted line
intensities as a function of $\alpha$.
Lines that form along with H$\alpha$,
such as [\sii ] $\lambda \lambda 6716$,
are not strongly sensitive to the value of $\alpha$.
The line ratio is constant in the extra-heating case because H$\alpha$ forms only
in ionized gas along with the [\sii ] lines.
The ratio decreases for large values of $\alpha$ in the cosmic-ray case
because this gives strongest weight to the molecular regions where the \hi\ lines
form by suprathermal excitation of \h0 .
For the regions where the line ratio is a constant the intensities
reflect the abundances of the elements
and the underlying heating process rather than the integration process.

% the graph is in CP_spectrum_alpha.jnb

The intensities of molecular lines relative to H$\alpha$ have a strong
dependence on $\alpha$.
The power-law index $\alpha$ affects how the dense and cold molecular regions
are added to the hotter ionized gas.
Figure~\ref{fig:CP_spectrum_alpha} shows that such lines as
[\oi ] $\lambda 63 \micron$,
\htwo\ 2.12$\micron$, and CO $\lambda 863 \micron$ have powerful dependencies on $\alpha$.
We will use these lines to determine $\alpha$ in the following section.

\begin{figure}
\begin{center}
\includegraphics[clip=on,width=\columnwidth,height=0.8\textheight,keepaspectratio]{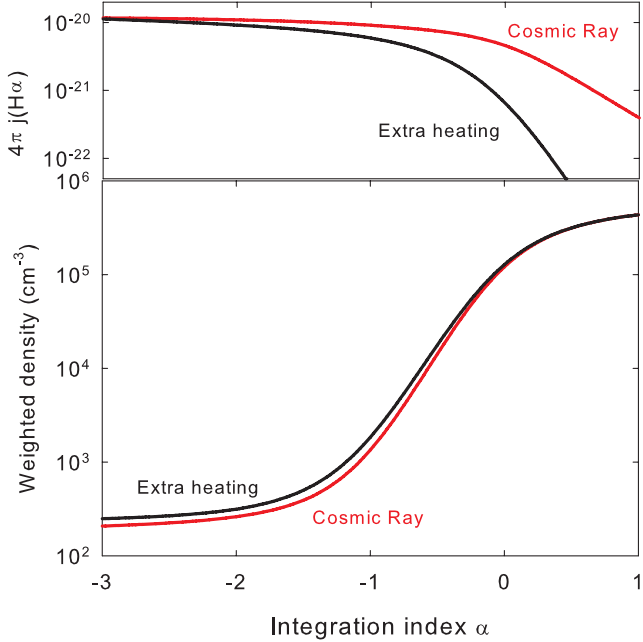}
\end{center}
\caption{
The upper panel shows the \hi\ H$\alpha$ emissivity $4 \pi \bar j$
as a function of the power-law index $\alpha$.
The lower panel shows the integrated density defined by
Equation \ref{eqn:DensityAveragedDistribution}.
}
\label{fig:CP_EmissionDensity_alpha}
\end{figure}

Figure \ref{fig:CP_EmissionDensity_alpha} shows the H$\alpha$ emissivity and
mean density as a function of $\alpha$.
The upper panel shows the emissivity $4\pi \bar j$ of \hi\ H$\alpha$
as defined by Equation \ref{eqn:EmissivityDefinition}.
The emissivities are quite similar for when $\alpha$ is very small
since this give greatest weight to the warm ionized regions where the
emissivities are similar.
They differ for large $\alpha$ since this weights the cold dense
regions where suprathermal excitation of \hi\ is important in the cosmic-ray case.
The total volume of gas in all forms,
molecule, atomic, and ionized, will be the observed luminosity
of the H$\alpha$ line divided by this emissivity, as given by
Equation \ref{eqn:LuminosityVolumeEmissivityRelationship}.
The lower panel of Figure \ref{fig:CP_EmissionDensity_alpha} shows the
mean density defined by Equation \ref{eqn:DensityAveragedDistribution}.
The two cases differ slightly at low densities because the low-density
cutoff given in Equation \ref{eqn:DensityAverageExplicit} is set by
the thermal stability requirement.
The total mass of gas in the filaments can be derived from the
H$\alpha$ luminosity and the information given in this Figure
together with Equation \ref{eqn:MassVolumeRelationship}.
Note that although we give these results for H$\alpha$ there is enough
information in the Tables and Figures to do this with any emission line.

% the graph is in CP_spectrum_alpha.jnb

\subsection{The observed spectrum}
\label{sec:ObservedSpectrum}

The next step is to use the observed intensities of
molecular lines relative to H$\alpha$ to set the power-law index $\alpha$.
This requires an observed spectrum that extends across the optical
into the far-infrared.
We take the Horseshoe region of the Perseus cluster as representative.
It is relatively isolated, at a projected distance of 25 \kpc\
from the centre of NGC 1275, so confusion
with other sources is minimized.
Its spectrum does not have underlying stellar absorption lines
so optical \hi\ lines can be measured without stellar contamination.

The upper section of Table~\ref{tab:ObservedPredicted} gives the line
ratios observed in the Horseshoe region together with predictions
from the two models.
The optical (Gemini GMOS) and near IR (UKIRT CGS4)
data are from \citet{HatchEtAl06}
and \citet{HatchEtAl05} respectively and the
mid-IR data are from \citet{JohnstoneEtAl07}.
All of the observed quantites in Table~\ref{tab:ObservedPredicted} have been
corrected for an assumed Galactic extinction of E(B-V) = 0.163
\citep{SchlegelEtAl98}.

The lower section of Table~\ref{tab:ObservedPredicted} gives the
observed H$\alpha$ surface brightness, $S(\rm{H}\alpha)$, and
luminosity $L(\rm{H}\alpha$) corresponding to the same region
that the optical line ratios were extracted from. The averaged line
emissivity, $4\pi \bar j(\rm{H}\alpha )$, and averaged density, $\bar n$, from
the wave heated and cosmic-ray heated models are also listed.

\begin{table}
\centering
\caption{
Upper: Observed and predicted line strengths near the
Horseshoe Position 11 region
of \protect{\citet{ConseliceEtAl02}}. Column 1 lists the emitting species, column 2
is the wavelength in microns, column 3 is the observed, extinction corrected flux ratio of
each line with respect to H$\alpha$, column 4 is the line flux ratio with respect to
H$\alpha$ predicted by the extra-heating model, column 5 is the line flux ratio with respect to
H$\alpha$ predicted by the cosmic-ray heating model.
Note that the optical, NIR and MIR
lines are each observed though a different aperture.
See Figure \ref{fig:apertures} for a visualization of each aperture.
Lower: Observed H$\alpha$ surface brightness, $S(\rm{H}\alpha$) ($\ergps \pscm \rm{arcsec}^{-2}$)
and luminosity, $L(\rm{H}\alpha$) ($\ergps$) corresponding to the same region
that the optical line ratios were extracted from. The
integrated line emissivity averaged over the distribution of clouds,
$4\pi \bar j(\rm{H}\alpha )$, ($\ergps \pcc$) defined by equation \ref{eqn:EmissivityDefinition},
and the density averaged over the distribution of clouds, $\bar n$, ($\pcc$) defined by
equation \ref{eqn:DensityAveragedDistribution} in
the wave heated and cosmic-ray heated models are also tabulated.}

\label{tab:ObservedPredicted}
\begin{tabular}{ c c c c c c }
\hline
Species&$\lambda (\micron )$&F(line)/F(H$\alpha$)&Pred./Obs.& Pred./Obs.\\
       &                    &Observed&Extra heating&Cosmic ray \\
\hline
$\alpha$   &       &        &-2.10  &-0.35 \\
\hline
H~{\sc i}  &0.4861 &0.24    &0.87   &1.04 \\
$[\oiii ]$ &0.5007 &$<0.03$ &       &     \\
$[\ni ]$   &0.5199 &0.06    &1.88   &1.66 \\
$\hei $    &0.5876 &0.04    &$6.4\times 10^{-5}$&0.37 \\
$[\oi]$    &0.6300 &0.19    &1.00   &0.93 \\
H~{\sc i}  &0.6563 &1.00    &1.00   &1.00 \\
$[\nii ]$  &0.6584 &0.71    &0.29   &0.34 \\
$[\sii ]$  &0.6716 &0.23    &1.05   &1.40 \\
$[\sii ]$  &0.6731 &0.17    &1.15   &1.08 \\
\htwo      &1.957  &0.09    &1.28   &0.81 \\
\htwo      &2.033  &0.03    &1.44   &1.41 \\
\htwo      &2.121  &0.08    &1.54   &1.02 \\
\htwo      &2.223  &0.02    &1.38   &0.77 \\
\htwo      &12.28  &0.03    &1.00   &1.00 \\
$[\neii ]$ &12.81  &0.06    &$4.6\times 10^{-3}$&0.46 \\
$[\neiii ]$&15.55  &$<0.02$ &       &     \\
\htwo      &17.03  &0.06    &0.29   &0.53 \\
\hline
$S(\rm{H}\alpha$)&&$1\times 10^{-14}$&&\\
$L(\rm{H}\alpha$)&&$7\times 10^{39}$&&\\
$4\pi \bar j(\rm{H}\alpha )$&&&$9.40\times 10^{-21}$ &$6.75\times 10^{-21}$\\
$\bar n$  &&        &300    &$2.63\times 10^4$ \\
\hline
\end{tabular}
\end{table}

\begin{figure}
\centering
\includegraphics[clip=on,width=\columnwidth,clip]{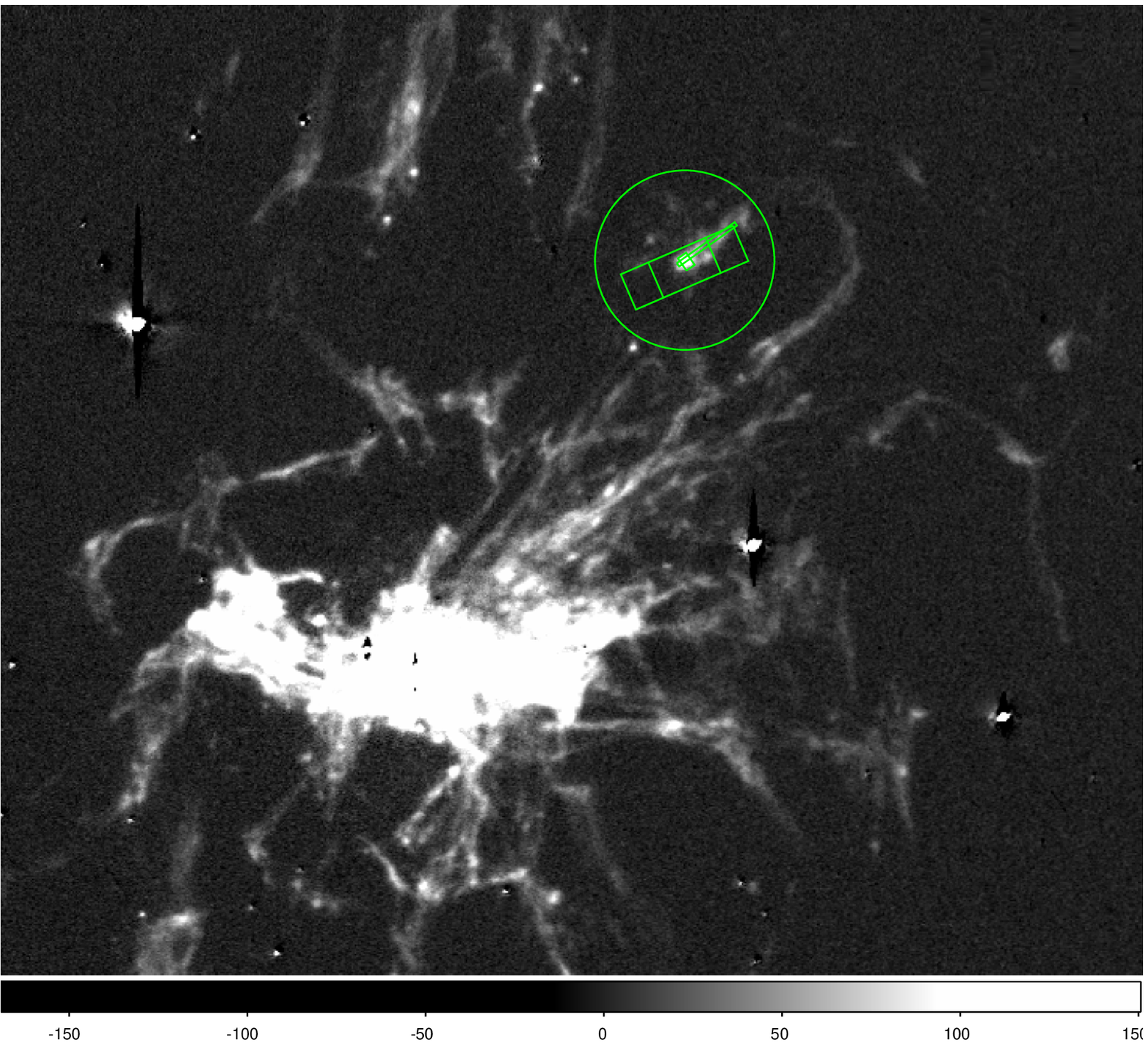}
\hfill
\includegraphics[clip=on,width=\columnwidth,clip]{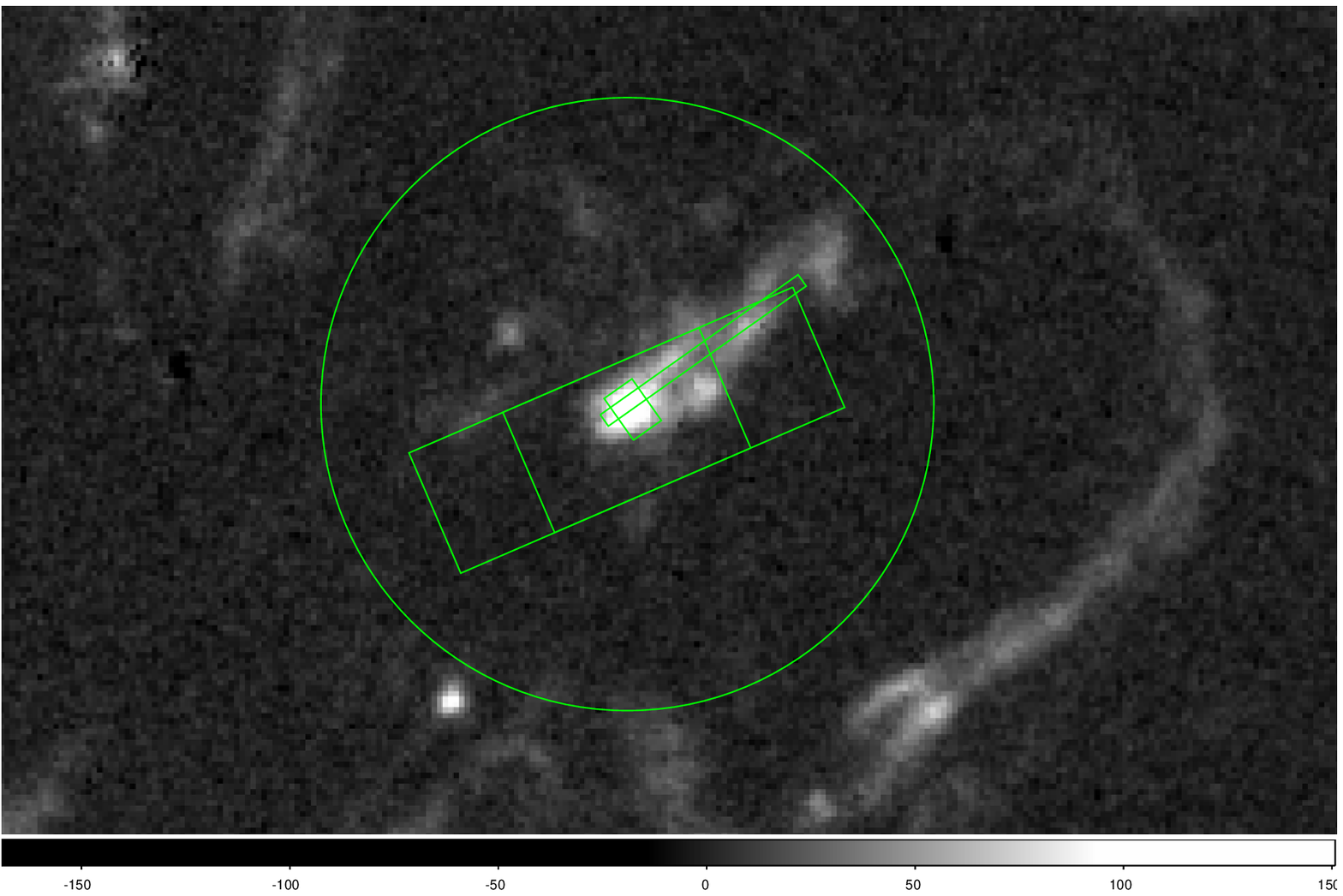}\\
\caption{Entrance apertures of the various instruments used to
determine the line ratios presented in
Table~\ref{tab:ObservedPredicted} and in the text.  Gemini GMOS: long
thin rectangle, UKIRT CGS4: small rectangle centred on brightest
knot. The two larger overlapping rectangles are the positions of the
Spitzer IRS SH aperture. Line fluxes were averaged between these two
nod positions. The large circle is the beam of the IRAM (CO 1--0)
observations.
}
\label{fig:apertures}
\end{figure}

The observational data presented in Table~\ref{tab:ObservedPredicted}
were measured from a variety of ground and space based telescopes. These facilities
each have different entrance apertures, which we show overlaid on the
H$\alpha$ emission map of \citet{ConseliceEtAl02} in Fig.\ref{fig:apertures}.
As stressed by \citet{JohnstoneEtAl07}, these
aperture corrections make it difficult to combine
observations taken with different instrumentation.
The 2$\micron$ NIR and optical spectra \citep{HatchEtAl05,HatchEtAl06}
were combined by scaled them to the same observed H$\alpha$ flux.
The H$\alpha$ flux in the NIR data was
estimated from the observed P$\alpha$ line flux assuming that the P$\alpha$ / H$\alpha$
intensity ratio has the Case B value (8.45) for $T = 10^4 \K$.
The MIR Spitzer observations of the low-$J$ \htwo\ lines
were placed on the same flux scale as
the optical/NIR lines by another method that did not assume Case B.
H$\alpha$ images of the region covered by the Spitzer entrance aperture
were used to find the area covered by emitting gas and maps of the optical
\hi\ and MIR \htwo\ emission were compared to introduce
aperture corrections and derive flux ratios \citep{JohnstoneEtAl07}.

There are substantial uncertainties in the derived spectrum.
The calculations presented above show that the P$\alpha$/H$\alpha$ ratio
is not expected to be given by Case B although the differences are not great
(Table \ref{tab:HI_emission_spectrum}).
The extinction within the filaments is unknown so no correction is made.
Renormalizing the optical / NIR \hi\ spectrum to agree with Case B intensities
largely removes the effects of reddening.
Given the uncertainties in both the observed
and theoretical predictions
our goal is to reproduce the spectrum to within a factor of two.

\citet{SalomeEtAl08b} detect the CO 1-0 line in the Horseshoe
and note that the line is optically thick.
This is in contrast to the optically thin optical and IR lines that
are the focus of this paper.
The observed flux of an optically thick line depends on the geometry
while that of an optically thin line depends only on the volume of
emitting gas.

\citet{SalomeEtAl08b} derive a mass of $10^8 \Msun$ for the region of the
Horseshoe included in their 11\,arcsec radius beam (see  Fig.\ref{fig:apertures}).
We can derive the total mass from the luminosity of any emission
line using Equation \ref{eqn:MassVolumeRelationship}.
We measured the H$\alpha$ luminosity in the IRAM beam from the
data of \citep{ConseliceEtAl02}
as $L(\rm{H}\alpha) = 5.5 \times 10^{40} \ergps$.

\subsection{The predicted spectrum}

We vary $\alpha$ to match the observed \htwo\ 12.28$\micron$/H$\alpha$
intensity ratio of 0.03 (Table~\ref{tab:ObservedPredicted}).
This line ratio has several advantages.
It has one of the widest ranges in excitation and ionization
in the spectrum and so is a strong function of $\alpha$.
The IR \htwo\ lines are, like the optical \hi\ lines,
optically thin so geometric effects do not enter.
Finally, by comparing emission from an ion and molecule of the same element
there are no additional uncertainties introduced by that fact
that the composition is unknown.
The uncertainty in the ratio is likely to be dominated by
systematic errors which we cannot quantify.

This procedure gives $\alpha = -0.35$ for the cosmic ray and
$\alpha = -2.10$ for the extra heating cases.
Both cases have negative $\alpha$ which,
as Equation \ref{eqn:FillingFactorDefinition} shows,
means that the lowest-density gas has the largest volume.
We are led to a picture in which the molecular gas lies
in cold and dense cores while the low-ionization emission
originates in extended low-density regions.
The ionized material may be located in extended envelopes
surrounding the molecular cores.

The last two columns of Table~\ref{tab:ObservedPredicted}
give the predicted line intensities relative to the observed values.
A more complete set of predicted intensities is given in Appendix \ref{sec:AppendixFullEmissionLineSpectrum}.
Our goal is to match
the spectrum to within a factor of two.
Both heating cases do this for the majority of lines.
The discriminants between the two cases are described in the next section.

The last two columns of Table \ref{tab:HI_emission_spectrum} give the
integrated \hi\ spectra for the two cases.
The predicted P$\alpha$/H$\alpha$ ratio is 0.0839 and 0.0992
for the extra heating and cosmic ray cases, not greatly different from
the Case B ratio of 0.104.
The P$\alpha$/H$\beta$ is 0.37 for both heating cases and is 0.29 for Case B.
These differences are significant but the wide wavelength separation of
the line pairs make them useful reddening indicators.

\subsection{The need for ionizing particles}

Several emission lines indicate that the cosmic-ray case best describes the
heating and ionization of the gas.
Most lines are matched by both cases to within our desired factor of two.
It is not possible to formally quantify a goodness of fit in Table~\ref{tab:ObservedPredicted}
because the uncertainties are dominated by systematic errors
which we cannot quantify.
Rather we point to a few lines which clearly indicate
that the cosmic-ray case applies.

Figures \ref{fig:CP_spectrum_Hi_T_end_strongA}
and \ref{fig:CP_spectrum_Hi_T_end_faintA} show the critical distinction
between the two cases.
The extra-heating case behaves as expected for a collisionally-excited thermal gas.
As the temperature increases, going from left to right in the figures, we see
lines of [\oi ], [\ni ], [\nii ], [\oii ], [\oiii ], \hei ,
and eventually [\neiii ], increasing in intensity.
This corresponds to increasing ionization and excitation potentials
as expected for a collisionally-ionized gas.
As stressed in the discussion of Figure \ref{fig:Den3VarHeatCR}
the distribution of ions and molecules in the extra-heating case
is sharply defined by temperature.
At a given temperature hydrogen will be nearly all ionized, atomic,
or molecular with the change from one species to the next occurring
over a relatively small range in temperature.

This behavior is to be contrasted with the cosmic-ray case.
Here a certain level of ionization and excitation is present at all gas kinetic
temperatures because the cosmic-ray ionization rates are only indirectly related
to temperature.
The result is that some \h0\ is present in regions that
are predominantly \htwo\ or \hplus .
We find the somewhat paradoxical result that
the higher-ionization (3.00 Ryd to create)
[\neiii ] $\lambda 3869$\AA\ line becomes strong
at lower temperatures than the lower-ionization (2.58 Ryd)
[\oiii ] $\lambda \lambda 5007, 4959$\AA\ lines.
This is because of the importance of charge-transfer recombination
in setting the ionization distribution of O.
The process $\rm{O}^{+n} + \h0 \rightarrow \rm{O}^{+n-1} + \hplus $
is very fast for both O$^+$ and O$^{+2}$
but is slow for He$^+$, Ne$^+$, and Ne$^{+2}$.
Larger amounts of \h0\ are present in the cosmic-ray case because of the
mixing of different ionization species in the same region.
The result is that the cosmic-ray case produces significant
[\neii ] $\lambda 12.81$\micron\ and [\neiii ] $\lambda 3869$\AA\
emission without exceeding the upper limit to [\oiii ] $\lambda 5007$\AA .

\hei\ $\lambda 5876$\AA\ provides another distinction between the cosmic
ray and extra-heat cases.
He$^0$ is more efficiently ionized by suprathermal than by lower-energy
thermal particles due to its large ionization potential.
The $\lambda 5876$\AA\ line is in fair agreement with the observed value
in the cosmic ray case while it is underpredicted by more than 4 dex
in the extra heating case.

Figure \ref{fig:CP_spectrum_Hi_T_end_faintA} shows the fundamental
tension experienced by the two cases.
We do not include the thermally unstable temperatures indicated by the
grey bar in the Figure.
The extra-heating case could only produce significant \hei\ and [\neiii ] emission
by including large amounts of unstable gas.
This would overpredict
the unobserved [\oiii ] $\lambda 5007$\AA\ line by large factors.
The cosmic-ray case produces strong He and Ne emission while not
overproducing [\oiii ].

\citet{JohnstoneEtAl07} discuss the puzzle posed by the observed
[\neiii ] emission from filaments that do not show significant [\oiii ] $\lambda 5007$\AA.
Neon is special because of the high ionization potentials compared with other
second row elements like N or O.
[\neiii ] was not detected in the Horseshoe
but was seen in another region where [\oiii ] was weak.
\citet{JohnstoneEtAl07} suggested that a low kinetic temperature in the
ionized gas, perhaps the result of enhanced abundances and cooling efficiency,
might be the cause.
Cosmic-ray ionization is another possibility.

These calculations demonstrate that cosmic rays are needed
to account for the spectrum.
Suprathermal particles such as those produced by cosmic rays
create a gas with a broad
range of ionization at any one point.
Charge transfer recombination then produces the observed unusual
ionization distribution of the heavy elements.
Although we have explicitly considered cosmic rays,
which are known to exist in the cluster environment,
any source of ionizing particles would work as well.
Although gentle heating by MHD waves is ruled out,
MHD waves could still be the fundamental source of the ionization since
high-energy particles may be produced {\em in situ}\/ by MHD phenomena
such as magnetic reconnection \citep{Lazarian05}.

\citet{CrawfordFabian92} showed that photoionization by a very
hard continuum source can produce the optical spectrum.
High-energy photons produce primary and Auger electrons
that have very high energies and so behave as ionizing particles.
Their model worked for the same reasons that the cosmic-ray model
discussed above works.
The ionizing particles generate a gas with a mix of molecules, atoms and ions,
charge transfer effects are important, and a
peculiar low-ionization spectrum is produced.
Photoionization by very hot objects is another viable source of the
deduced population of suprathermal particles.

A form of mixing layers might also produce the ionizing particles.
If hot ionized gas were to mix with cold molecular gas the resulting
physical state would be determined by the microphysics of the collisional
interactions rather than by the ratio of kinetic temperatures.
Any energy source that deposits its energy as
ionizing particles rather than as heat can produce the observed spectrum.
Our predictions do not depend on the fundamental energy injection,
only on the eventual production of a population of suprathermal particles.
We shall refer to this as ionizing particle heating in the remainder of this paper
to reflect the fact that there are many possible sources of these particles.

\section{Discussion}
\label{sec:Discussion}

\subsection{The geometry}

Our deduced value of $\alpha = -0.35$ corresponds to the majority of the volume
being filled by low-density ionized gas.
The density of the molecular regions is typically
$\bar n_{mole} / \bar n_{ion} \sim 10^3$ times higher than the
density of the ionized gas (Figure \ref{fig:CP_spectrum}).
The ratio of volumes is then
$V_{mole}/V_{ion} \sim 10^{3 \alpha} \sim 0.09$.
By contrast, most of the mass is in dense molecular cores.
The mass depends on $n^{\alpha + 1}$ so the ratio of masses is
$M_{mole}/M_{ion} \sim 10^{3 (\alpha +1)} \sim 90$.
The small dense molecular cores may provide the reservoir of material
for an extended halo of low-density ionized gas.
If the gas is constrained to follow magnetic field lines then the
individual clouds may resemble comets with a cold dense core and
extended warm low-density tail.

The observed surface brightness in H$\alpha$
and the predicted $4 \pi \bar j(\rm{H}\alpha )$,
both listed in Table \ref{tab:ObservedPredicted},
can be combined with Equation \ref{eqn:SurfaceBrightnessThicknessEmissivityRelationship}
to find the line-of-sight thickness of emitting material.
The deduced thickness is $dl = 1.0\times 10^{18} \cm = 0.3 \pc$.
Note that this is the thickness of the entire cloud, not just the H$^+$ region,
although most of this is filled by ionized gas.

This depth is similar to that found by \citet{FabianEtAl03} from fundamentally
different assumptions.
They assumed that the optical \hi\ lines are formed by recombination
with Case B intensities, that the pressure of the \hi -emitting gas
was equal to that of the surrounding hot gas, and that the gas was fully ionized
with a temperature of $\sim 10^4 \K$.
If the gas is excited by ionizing particles then
\hi\ is far more emissive than
predicted by Case B due to the enhancement by collisional excitation
(Table \ref{tab:HI_emission_spectrum}).
This large enhancement makes up for the fact that hydrogen is not
fully ionized across the volume so the \hi\ emission measure,
the product $n_e n_p \, dl$, is much smaller.
These two factors nearly cancel leading to a similar depth.

For this surface brightness argument to work the entire entrance
aperture of the spectrometer must be covered by emitting gas.
A slit width of 1 \arcsec\ corresponds to a projected width
of $\sim 350 \pc$ at the distance of Perseus.
Each cloud is $\sim 0.3 \pc$ thick along the line of sight but
must cover a region of the sky $\sim 350 \pc$ across.
The aspect ratio of the width to depth of the emitting gas is
$> 10^3:1$ since the region of the
Horseshoe we observed is well resolved on the sky.
This aspect ratio, although surprising, is similar to other
geometric quantities in the system.
HST observations show that the projected width of the narrowest parts of the
filaments is $\sim 70 \pc$ \citep{FabianEtAl08} while
they have lengths of tens of \kpc .
The filaments are composed of remarkably thin and long clouds.

The filaments are most likely composed of a network of
much smaller threads.
In the magnetic-field dominated
constant-pressure case assumed here gas is free to move along
but not across field lines.
The line surface brightness suggests that individual threads
have a width of a fraction of a \pc .
We view this geometry from a direction that is nearly orthogonal
to the field lines.
This is an observational selection effect introduced by the
way that isolated filaments were chosen for study.
The narrowest filaments, with a width of 70 \pc ,
must be composed of $\sim 10^2 - 10^3$ of these thin threads.
If the threads are uniformly distributed across the filaments,
and the filaments have a depth equal to their observed width,
then the mean separation
between threads is $\sim 10 \pc$, about thirty times their thickness.

There are two possible physical interpretations for our $\alpha$ power-law index.
It could represent a radial gradient in density across a thread
normal to the field lines.
In this case the dense molecular regions would form a very small linear
core with the lower-density ionized gas surrounding it in a much larger cylinder.
We consider this unlikely since gas could not move across field lines
to change its density in response to changes in the heating.
The more likely possibility is that the $\alpha$ index represents
the variation of the density along magnetic field lines.
In this picture the ionizing-particle density would vary with position or time,
the temperature of the gas would adjust itself in response,
and then expand or contract to maintain constant pressure.
A magnetic field line would then have most of its volume occupied by
lower-density ionized gas intermixed with dense molecular regions.
The cold dense molecular regions could become warm and ionized, and return
to being cold and molecular, as the heating rate changes.

\subsection{Internal extinction across a thread}

The hydrogen column density through a thread normal to its
surface is $N(\rm{H}) \sim \bar n \, dl \approx 10^{22.5} \pscm$.
% dl is 0.3 pc from second paragraph this section, 1.0e18 cm
% \hat n is 2.63e4 cm-3 from Table 5
% N(H) is 2.63e22 from product
If the gas has a Galactic dust to gas ratio, as we have assumed,
then the total visual extinction across a thread will be
$A_v \sim 17$ mag.
The effects of this extinction depends on the geometry.
The dense molecular regions, having most of the column density,
would also be most extinguished.
This region detected emits MIR and sub-mm emission
where the extinction will have little effect.
The column density through the partially ionized gas which
produces most of the optical emission is small so this
emission is not directly affected by internal extinction.
We could not detect optical emission from ionized regions
which happened to lie on the far side of the molecular cores of course.

\subsection{The total mass at position 11}

The total mass can be derived from the luminosity of any emission line using
the formalism derived in Section \ref{sec:FormalismLOC}.
Section \ref{sec:ObservedSpectrum} derived an
$L(\rm{H}\alpha)$ of $5.5 \times 10^{40} \ergps$ for position 11
and the beam used by \citet{SalomeEtAl08b}.
Equation \ref{eqn:MassVolumeRelationship} gives a total mass of
$2.1 \times 10^8 \Msun$.
This is in reasonable agreement with the mass of $10^8 \Msun$ derived by
\citet{SalomeEtAl08b} from observations of the CO 1-0 transition.
% mass is 5.5e40 / 6.75e-21 * 2.63e4 * mu = 2.14e65 mu
% mu is 1.2*m_p = 1.2*1.66e-24 = 1.99e-24 gm
% is 4.26e41 gm = 4.26e41/1.989e33 = 2.14e8 M_sun

The mass of $2.1 \times 10^8 \Msun$ derived from emission lines
counts only the spectroscopically active gas.
As Figure \ref{fig:CP_spectrum} shows, little emission is
produced by cold gas at the high-density end of the distribution.
Gas denser than our high-density limit cannot exist and be in pressure
equilibrium because the CMB sets the lowest possible temperature.
However there is no observational constraint that actually limits the
amount of material that could be at this highest density
and the CMB temperature.
A large reservoir of very cold high density gas could exist
and remain undetected.
Our derived mass is a lower limit as a result.
Gas at the very coldest limit of the distribution
could serve as the source of fresh material
for the spectroscopically active regions.

\subsection{The pressure of the molecular cores}

A number of studies have estimated the pressure of
the molecular gas.
Next we compare these with our derived temperature and density which assume a
pressure equal to that of the surrounding hot gas.

\citet{JaffeEtAl01} found a density $n_{\rm{H}} > 10^5 \pcc$
and $T \sim 10^{3.2} \K$ from
the distribution of \htwo\ level populations.
The gas pressure, $nT > 10^{8.2} \K$,
was substantially higher than the pressure of the surrounding hot gas.
For our best fitting ionizing particle model most of the
$v=0$, low-$J$, \htwo\ emission originates in gas with
a temperature of about $T \sim 10^{1.3} \K$
and a density $n_{\rm{H}} \sim 10^{5.2} \pcc $,
as shown in Figure \ref{fig:CP_spectrum}.
Excited-$v$ emission occurs at a higher temperature,
$T \sim 10^{3.2} \K$, and lower density,
$n_{\rm{H}} \sim 10^{3.3} \pcc $.
We do not present level-excitation diagrams in this paper but note that
we have fitted the observed \htwo\ spectrum,
from which the \htwo\ level-excitation populations were derived,
to within a factor of two.
Our model produces a population distribution
that is consistent with these observations.

\citet{SalomeEtAl08a} argue that the filaments are composed of
an ensemble of Galactic Giant Molecular Cloud-like components.
By analogy with GMCs they argue that the density in the \CO\ region is
several hundred particles per cubic centimeter and that
the gas has a kinetic temperature of
$T \approx 20 - 40 \K$.
The resulting pressure, $nT \approx 10^{3.3} \pcc \K$,
is 3.2 dex smaller than the pressure of the surrounding hot gas.

We avoid geometrical complexities
by focusing on optically thin emission lines.
Figure \ref{fig:CP_spectrum} shows that the optically thin emissivity of the
\CO\ $\lambda 863$ \micron\ line
peaks at $T_{k,\CO} \approx 20 \K$
and $n \approx 10^{4.2} \pcc$.
This temperature is similar to the \CO\ temperature
quoted by \citet{SalomeEtAl08a} but the density is
considerably higher in keeping with our assumed high gas pressure.
The difference is due to the higher cosmic ray density in our simulations.
Well-shielded regions of Galactic GMCs are similarly heated by cosmic rays
but with a considerably lower intensity equal to the Galactic background.
The cosmic ray density that produced the peak in our simulations
is $\sim 10^4$ times the Galactic background.
This accounts for the GMC-like temperatures at such a high pressure.
It is not possible to make a specific prediction of the \CO\ 1-0 brightness
temperature without first deriving an explicit geometric model
of the cloud, a task beyond the scope of this paper.

We conclude that both the \htwo\ spectrum and \CO\ brightness
temperature are consistent with
a gas pressure of $nT = 10^{6.5} \K \pcc $ and excitation
by ionizing particles.

\subsection{Gravitational stability}

The stability of the material in the filaments to gravitational
collapse may be influenced by a number of factors: basic geometry,
turbulence and magnetic fields.

If the support of the filaments is purely by thermal pressure, then
they will be liable to collapse when their size becomes greater than
the Jeans length,
\begin{equation}
R_{\rm J} \simeq 10^{18} c_5n_5^{-1/2}{\rm\,cm},
\end{equation}
where $c_5$ is the sound speed (or effective virial velocity) in $\rm
km\,s^{-1}$ and the density is $10^5n_5{\rm\,cm^{-3}}$.  For $c_5=1$,
$n_5=1$ the thickness of the molecular material will be
$0.3{\rm\,pc}$, a similar scale to that inferred for the H$\alpha$
emitting region.  We infer above that the molecular material has a
volume filling fraction about one tenth that of the ionized gas, so
the size scale of the molecular material will be comparable to this.
This suggests that the gravitational stability of the molecular
component may have a significant role in determining the structure of
the filaments.

Magnetic fields and turbulence may play a significant role in
supporting the molecular material against gravitational collapse.  The
standard criterion for support against collapse by uniform magnetic
fields is to compare the column density along a flux line to the
critical value $B/2\pi G^{1/2}$.  If the dense material is threaded
along flux tubes, this is not a significant constraint as is stands.
Likely more important is the coupled effects of turbulence and
randomly directed internal magnetic fields.  These will act to
increase the effective virial velocity of the material, and so support
it against collapse.

However, both turbulence and random magnetic fields are inherently
transient, and so must be driven in some fashion to maintain their
role in supporting the material.  In the local interstellar medium,
this driving is achieved by a number of processes, including star
formation.  While there is some evidence of star formation in the
filaments, this appears to be a relatively localised phenomenon.

What will happen if part of the molecular material in a filament
starts to undergo a gravitational collapse?  We find that the cooling
timescale ($\sim 10^2{\rm\,yr}$) is significantly smaller than the
dynamical timescale ($\sim 3\times10^5{\rm\,yr}$), so the material may
initially fragment.  However, the enhanced density will mean that the
net tidal force between the material and the surrounding diffuse
envelope will increase, both allowing the filament to break up, and
enhancing the net heating.  This process may thus naturally maintain
the filaments with marginal gravitational stability, in a similar
fashion to that in which galactic disks are maintained with a Toomre Q
parameter around unity.  Note that if this is the case, it requires
that the heating rate is determined by {\em local}\/ processes, such
as drag-driven reconnection.

\subsection{The energy budget}

The energy dissipation timescale for the line-emitting material is
short, requiring that the energy be replenished {\em in situ}.  While
it is clear from Figure~\ref{fig:CP_spectrum} that below
$10^4{\rm\,K}$ the detailed cooling mechanisms depend on the
temperature, from Figure~\ref{fig:CP_stabilityA} the overall cooling
rates do not depend strongly on these details.  The heating need to
power the emission regions in steady state ranges from $\Lambda \sim
10^{-19 \pm 1} \erg \pcc \ps$ while the thermal timescale of the gas
is $\sim 10^2 \yr$ (Figure \ref{fig:CP_cooling_timeA}).

A lower limit to the energy density of MHD waves in equipartition with
a magnetic field of 100 $\mu$G, the weakest B capable of confining the
clouds, is
\[
U_{\rm wave}  = 4.4 \times 10^{ - 10} \left( B/100\,\mu {\rm G} \right)^2
\erg \pcc .
\]
For the quoted heating rate this energy will be dissipated in a timescale
\[
\tau_{\rm wave}  = U_{\rm wave} /\Lambda
= 10^{2 \pm 1} \left( B / 100\,\mu {\rm G} \right)^2  \;{\rm{yr}}
\]
over which the wave energy has to be replenished.

The timescales for the cosmic-ray heated case are somewhat longer.
The Galactic background cosmic-ray density corresponds to an energy
density of $\sim 1.8 \eV \pcc$ \citep{Webber98}, mostly in high-energy
nuclei.  The cosmic-ray intensity required to maintain the material at
$nT\sim 10^{6.5}\pcc\,{\rm K} $ ranges between $10^2 \mbox{--} 10^6$
times the Galactic background value, depending on the filament density
(Figure~\ref{fig:PressurePreferred}).  Molecular regions are predicted
to have a density of $n_{\rm{H}} \sim 10^5 \pcc$ and require a
cosmic-ray $\sim 10^3$ times the background.  The low-ionization gas
requires a rate $\sim 10^6$ times the background with a density of
$n_{\rm{H}} \sim 10^{2.5} \pcc$.  For these values the energy
dissipation time for the molecular and low-ionization regions are
$\sim 10^3 \yr$ and $\sim 5 \times 10^5 \yr$ respectively.

These replenishment timescales are generally of order or longer than
the cooling time (Figure~\ref{fig:CP_cooling_timeA}), the time
required for the gas to adjust itself to a changing heating rate.
This means that non-equilibrium effects, in which the ionization,
heating, and cooling decouple from one another, may be important.  The
short heating / cooling times suggests that the gas pressure can
fluctuate about its equilibrium value in response to variations in the
environmental heating, and may drive sound waves.  The resulting
structures may be intrinsically quite dynamic and disequilibrium.

The size scale of a thread appears to be $\sim 0.3{\rm\,pc}$.
Assuming that they are cylindrical, then if they are heated through
their surfaces the energy flux required to keep them in equilibrium is
\begin{equation}
	F \simeq {r_{\rm fil}\over 2}\Lambda \simeq 0.05{\rm\,erg\,cm^{-2}\,s^{-1}}.
\end{equation}
If the field which provides this energy has an energy density $X$
times the thermal energy density inferred for the surrounding X-ray
bright gas, the net speed of energy transport through the surface of
the filaments must be
\begin{equation}
	v \simeq 10^{8} {r_{\rm fil}\over X 0.3{\rm\,pc}} {\rm\,cm\,s^{-1}}.
\end{equation}
While this suggests that it is easier to heat more narrow filaments,
the same energetic constraints will apply to an assembly of filaments
if the observed surface brightness is to be produced.  For comparison,
the Alfv\'en velocity is
\begin{eqnarray}
	v_{\rm A} &=& {B\over (4\pi \rho)^{1/2}}\\
	&\simeq& 7\times10^4 n_5^{-1/2}\left( B/100\,\mu {\rm G} \right)
	{\rm\,cm\,s^{-1}},
\end{eqnarray}
and the sound speed will also be of order $1{\rm\,km\,s^{-1}}$.  This
suggests that the energy required to heat the filaments can only be
transported through the molecular material by non-thermal particles or
radiation.  However, this does not rule out it being converted into
this form locally to the filaments, as energy transport is easier
through the more diffuse components with higher characteristic
transport speeds.

The filament material does not appear to be supported against gravity
by lateral kinetic energy, as the observed speeds are below the local
Keplerian velocity, and the filaments are being dragged away from the
stellar populations which are believed to have formed within them.  If
this is the case, the net rate of drag heating of the material per
unit mass will be approximately
\begin{eqnarray}
\Gamma_{\rm tidal} &\simeq& g v_{\rm drag} \simeq {v_{\rm Kep}^2
v_{\rm drag} \over r}\\
& \simeq & 0.3 {\rm\,erg\,g^{-1}\,s^{-1}}
\simeq 5\times10^{-20} n_5 {\rm\,erg\,cm^{-3}\,s^{-1}}
\end{eqnarray}
assuming $v_{\rm Kep} \simeq 10^3{\rm\,km\,s^{-1}}$, $v_{\rm drag}
\simeq 10^2{\rm\,km\,s^{-1}}$ and that the material has a density of
$10^5n_5{\rm\,cm^{-1}}$.  This energy input rate is in the range of
interest for heating the filaments, as already suggested from the
overall energy budget by \citet{PopeEtAl08}.  Note that the spatial
distribution of this energy input will be determined by the details of
the drag processes between the phases, so that the resulting energy
input may appear either in the dense material or within the halo of
the material.

Note that these scaling values can be formed into a simple estimate of
the rate at which mass passes through the system of filaments, by
dividing the total mass inferred for the molecular filaments by the
dynamical timescale of the material, i.e.\
\begin{equation}
% (4e10 Msum * 1e7 cm/s) / (25 kpc) =
\simeq {M_{\rm CO} v_{\rm drag}/r} \simeq 50 {\rm\,M_\odot\,yr^{-1}},
\end{equation}
which is similar to the mass processing rate inferred for the
radiative cooling of the hot material in these regions.

At high Reynolds number, the drag force is roughly equal to the ram
pressure between the phases.  At the relative density of the X-ray hot
medium and the molecular gas, this would suggest that to prevent the
dense material moving at free fall velocities, the relative motion of
the hot material must be at $(n_{\rm mol}/n_{\rm hot})^{1/2} v_{\rm
ff}$, which are not likely to be the case.  The filament material may be
transitory on a timescale $\sim v_{\rm fil}/g \simeq
3\times10^6{\rm\,yr}$, in which time the filament material will travel
$\sim {1/2}(v/v_{\rm Kep})^2$ of the distance to the core of the
cluster.  Alternatively, their effective cross section may be
increased either by an effective viscosity or by magnetic fields.

The interphase drag will lead to magnetic field lines in the diffuse
phases becoming wrapped around the dense material.  This suggests that
the magnetic structure may be somewhat like the bubble in the solar
wind maintained by the earth's magnetosphere.  The resulting magnetic
segregation would limit high energy particle transport into the
molecular material.  It is likely that some high energy particles will
be able to leak in, either through magnetic neutral points, as a
result of turbulent motions, or by diffusion resulting from field
inhomogeneities.  Alternatively, the drag-driven turbulence may itself
be the source of the high-energy particle flux, as a result of
reconnection events.

Overall, it is clear that the dissipation of the energy resulting from
buoyancy has potential as a source of heating for the filaments.
However, if this is the source of the heating, the means by which the
energy is transformed into the non-thermal ionizations which we infer
are the immediate means by which the molecular material is heated must
be complex.

\subsection{What is the source of the ionizing particles?}

The chemical-ionization state of the gas strongly suggests that
the the heating and ionization is produced by ionizing particles.
Several considerations suggest that the population of ionizing
particles is not produced by a strongly enhanced version of Galactic
cosmic rays.

The energy density Galactic cosmic rays is dominated by particles with
energies around $1{\rm\,GeV}$ in the solar neighborhood (where at
higher energies the energy spectrum is $\propto E^{-2.5}$,
\citet{SpitzerTomasko68}).  Most of the energy is in high-energy
protons which have a long range and do not interact strongly with
matter.  The CR particles lose energy by ionizing the material they
pass through, primarily by producing secondary electrons with energies
around $50 {\rm\,eV}$.  These secondary electrons further ionize and
heat the gas, but they will be transported by relatively small
distances.

As a rough estimate of their range, the velocity of CR protons at
$1{\rm\,GeV}$ is $v = \sqrt{3}/2 c$, so from equation (3) of
\citet{SpitzerTomasko68} the interaction cross section with atomic
hydrogen is $\sim 10^{-19}{\rm\,\cm^{2}}$.  Each interaction causes
the particle to lose $\sim 50{\rm\,eV}$ of its energy.  The range of
the CR heating particles is therefore
\begin{equation}
L_{\rm CR} \simeq {E\over n\sigma\Delta E} \simeq 700
n_5^{-1}{\rm\,pc}.
\end{equation}
However, this range may be significantly reduced by the magnetic
field.  The gyroradius of the protons in a $100{\rm\,\mu G} = 10
{\rm nT}$ field is
\begin{equation}
r_{\rm g} = {\gamma m v\over eB} = 3\times10^{10}\gamma\beta{\rm\,cm}
\end{equation}
so the protons will be tightly coupled to the magnetic field lines,
but (at least at first order) free to travel along them.
Non-uniformities in the field geometry will allow particles to diffuse
across the field lines.

The distance which the particles can travel in a dissipation time is
$\sim 30{\rm\,pc}$ in the molecular gas and $\sim 150{\rm\,kpc}$ in
the atomic gas.  This suggests that the cosmic rays can travel with
little reduction in intensity across the size scale of an individual
thread.  It is difficult to see how the flux of cosmic rays could be
modulated to the extent that we need to produce the range in density
shown in Figure \ref{fig:PressurePreferred}.

Even if we were to assume that the large flux of CR required to heat
the ionized material could be excluded from the molecular core, this
would result on a pressure acting on the molecular material which
would be inconsistent with our models.  The Galactic background cosmic
ray energy density of $\sim 1.8 \eV \pcc$ corresponds to a pressure of
$\sim 2\times10^4\pcc\,{\rm K}$.  Our spectral model requires a range
of ionization rates between $10^2 - 10^6$ times the background.  The
corresponding pressure is between $10^{6.3} - 10^{10.3}\pcc\,{\rm K}$,
with the top of this range being far larger than that inferred for the
molecular material.  The logical conclusion is that the inferred
suprathermal ionization is not associated with a population of
relativistic nuclei that is similar to Galactic cosmic rays.

As described above, other possible sources of suprathermal electrons
include magnetic reconnection, mixing layers, and photoionization by
very hard radiation fields.

\subsection{The history of the gas}

The origin of the filaments is unknown.  Two possibilities are that
molecular gas was ejected from the ISM of the central galaxy or that
they formed from the surrounding hot gas which fills the cluster
\citep{RevazEtAl08}.

We have shown that the filaments are energized by a population of
ionizing particles.
High-energy particles destroy molecules in addition to heating the gas.
Figure \ref{fig:CP_spectrum} shows the the peak CO and \htwo\ line
emission occurs at $n \sim 10^{5.3} \pcc $ while
Figure \ref{fig:PressurePreferred} shows that
this occurs at a cosmic ray ionization rate of $\sim 10^{2.5}$ times
the Galactic background, which \citet{WilliamsElAl98} give as
$5 \times 10^{-17} \ps $.
The corresponding \htwo\ dissociation rate of $\sim 1.6 \times 10^{14} \ps $
corresponds to an \htwo\ survival time of only $ \sim 2 \times 10^6 \yr $.
This is very close to the \htwo\ formation timescale for the dusty case,
as give in Section \ref{sec:ThermalStateAndHistoryOfTheGas} above,
but far shorter than the formation timescale in the dust-free case.
We conclude that that gas must be dusty to support the observed
molecular inventory against rapid destruction by ionizing particles.

Perhaps surprisingly,
the abundances are fairly close to that of the ISM of our galaxy.
In photoionization
equilibrium the `thermostat effect' governs the strength of forbidden lines relative
to hydrogen lines.
\textbf{Their} intensity ratio has no direct dependence on the abundances
\citep{FerlandARAA}.
In the non-radiatively heated cases considered here the strength of the forbidden
lines is a product of both the temperature of the gas and the abundance of the element.
In this case the intensities of forbidden lines relative to hydrogen lines depends linearly
on the abundance of the heavy element relative to hydrogen.
This suggests that the metallicity of the filaments is within a factor of two of solar.

The [Ca~II] doublet near $\lambda 7320$\AA\ is strong if grain are not
present \citep{KingdonEtAl95}.
The absence of this line shows that Ca is depleted by at least an order
of magnitude, also suggesting that dust is present.

\section{Conclusions}

\begin{enumerate}
\renewcommand{\theenumi}{(\arabic{enumi})}

\item
We propose a model in which the filaments have a constant
gas pressure that is set by the pressure of the surrounding hot gas.
The overall geometry is set by magnetic field lines.
Gas is freed to move along field lines but not across them.

\item
We consider gas heating by dissipation of
MHD wave energy and by cosmic rays.  If a range of
heating rates occur then the gas will expand or contract along
field lines to maintain constant pressure.  Low heating
rates result in dense molecular cores while high heating rates
produce an extended warm ionized gas.

\item
The relative proportions of molecular and low-ionization emission
is set by the proportion of high and low
density gas that is present.  In our final best-fitting
model most of the volume is filled
by low-density ionized gas while most of the mass is in dense
cold molecular cores.

\item
Both wave and cosmic-ray heating can match
most emission line intensities within a factor of two.
The distribution of low-ionization emission lines shows
that charge transfer strongly affects the ionization.
This is only produced in the cosmic-ray case, where a population of
suprathermal electrons produces a gas with a wide mix of ionization.
Although we explicitly consider cosmic rays any source of ionizing
particles would work as well.

\item
The emission measure shows that the filaments are composed
of a network of much smaller threads, each
$\sim 0.3 \pc$ in radius and separated by $\sim 10 \pc$.
A single filament is composed of hundreds to thousands of these threads.

\item
The mass we derive from fitting the entire spectrum is in reasonable
agreement with mass estimates from optically thick CO lines.
There may be an even larger reservoir of dense molecular gas at
the CMB temperature which would not be spectroscopically active.

\item
The observed molecular inventory requires that dust be present to
sustain the needed \htwo\ formation rate.
The presence of dust suggests that the filaments have not been
strongly shocked and did not form out of the surrounding hot and presumably
dust-free gas.

\item
The metallicity is within a factor of two of solar.  Both this and
the presence of dust suggests that the gas originated in the
ISM of the central galaxy.

\item
The calculations presented here use the development version of Cloudy,
which we have expanded to include new \htwo\ collision rates and resolve
\h0\ configurations into $nl$ terms.  The critical densities
of the \htwo\ levels which produce the observed emission are considerably
lower than previously estimated.  The ionizing particle-excited \hi\ spectrum has relative intensities that are not too different from Case B
but has emissivities which are far higher.

\end{enumerate}

\section{Acknowledgments}
We thank Steven A. Wrathmall and David Flower for providing analytical fits to their
new H -- \htwo\ collision rate coefficients
and Philippe Salom{\'e} for comments on the manuscript.
ACF acknowledges support by the Royal Society.  RMJ
acknowledges support by the Science and Technology Facilities
Council. GJF thanks the NSF (AST 0607028), NASA (NNG05GD81G), STScI
(HST-AR-10653) and the Spitzer Science Center (20343) for support.
PvH acknowledges support from the Belgian Science Policy Office (grant MO/33/017).

\bsp

\appendix

\section{The emission-line spectrum}
\label{sec:AppendixFullEmissionLineSpectrum}

\begin{table}
\centering
\caption{\label{tab:Predicted_UV_Optical}  Predicted UV and optical line ratios.}
\begin{tabular}{ c c c c }
\hline
Line&$\lambda (\micron )$&Extra Heat&Cosmic Ray \\
\hline
$\mgii $ & 0.2798 & 1.64E-01 & 1.04E-01 \\
$[\oii ]$ & 0.3727 & 1.18E+00 & 1.27E+00 \\
$\hi $ & 0.3798 & 6.41E-03 & 1.09E-02 \\
$\hi $ & 0.3835 & 9.38E-03 & 1.52E-02 \\
$[\neiii ]$ & 0.3869 & 1.28E-10 & 1.32E-02 \\
$\hei $ & 0.3889 & 7.45E-06 & 9.01E-03 \\
$\hi $ & 0.3889 & 1.45E-02 & 2.24E-02 \\
$\hi $ & 0.3970 & 2.44E-02 & 3.52E-02 \\
$[\sii ]$ & 0.4074 & 4.78E-02 & 4.16E-02 \\
$\hi $ & 0.4102 & 4.59E-02 & 6.08E-02 \\
$\hi $ & 0.4340 & 1.02E-01 & 1.20E-01 \\
$\hei $ & 0.4471 & 8.51E-07 & 5.39E-03 \\
$\heii $ & 0.4686 & 1.40E-13 & 4.36E-03 \\
$\hi $ & 0.4861 & 2.27E-01 & 2.71E-01 \\
$\nu F_{\nu}$ & 0.4885 & 2.39E+00 & 1.43E+00 \\
$[\oiii ]$ & 0.5007 & 9.73E-10 & 2.01E-03 \\
$[\ni ]$ & 0.5199 & 1.13E-01 & 1.02E-01 \\
$[\nii ]$ & 0.5755 & 7.15E-03 & 7.24E-03 \\
$\hei $ & 0.5876 & 3.19E-06 & 1.83E-02 \\
$[\oi ]$ & 0.6300 & 1.91E-01 & 1.76E-01 \\
$\hi $ & 0.6563 & 1.00E+00 & 1.00E+00 \\
$[\nii ]$ & 0.6584 & 2.15E-01 & 2.49E-01 \\
$\hei $ & 0.6678 & 1.84E-06 & 5.41E-03 \\
$[\sii ]$& 0.6716 & 2.30E-01 & 2.37E-01 \\
$[\sii ]$& 0.6731 & 1.72E-01 & 1.78E-01 \\
$\hei $ & 0.7065 & 8.36E-06 & 1.75E-03 \\
$[\ariii ]$ & 0.7135 & 1.77E-06 & 8.57E-03 \\
$[\caii ]$ & 0.7306 & 9.39E-04 & 2.17E-03 \\
$[\oii ]$  & 0.7325 & 4.48E-02 & 4.95E-02 \\
$[\siii ]$& 0.9532 & 4.90E-05 & 8.43E-03 \\

\hline
\end{tabular}
\end{table}

Table \ref{tab:Predicted_UV_Optical} lists the intensities of some of the
brighter optical emission lines
while Tables \ref{tab:Predicted_NIR} and \ref{tab:Predicted_MID_IR}
give the predicted lines in the near IR through sub-mm regions.

\begin{table}
\centering
\caption{\label{tab:Predicted_NIR}  Predicted near IR line ratios.}
\begin{tabular}{ c c c c }
\hline
Line&$\lambda (\micron )$&Extra Heat&Cosmic Ray \\
\hline
$\hei $ & 1.083 & 7.72E-05 & 2.37E-02 \\
$\htwo $ & 1.631 & 4.45E-06 & 2.64E-04 \\
$\htwo $ & 1.636 & 5.55E-07 & 1.63E-04 \\
$\htwo $ & 1.639 & 9.00E-06 & 2.00E-04 \\
$\htwo $ & 1.645 & 1.62E-06 & 2.58E-04 \\
$\htwo $ & 1.650 & 2.05E-04 & 4.82E-03 \\
$\htwo $ & 1.658 & 1.83E-07 & 2.29E-04 \\
$\htwo $ & 1.666 & 2.80E-04 & 4.77E-03 \\
$\htwo $ & 1.675 & 6.16E-07 & 1.61E-03 \\
$\htwo $ & 1.687 & 2.58E-03 & 3.13E-02 \\
$\htwo $ & 1.695 & 2.28E-07 & 5.78E-04 \\
$\htwo $ & 1.714 & 2.32E-03 & 1.68E-02 \\
$\htwo $ & 1.719 & 6.73E-07 & 6.18E-04 \\
$\htwo $ & 1.747 & 2.43E-07 & 3.69E-04 \\
$\htwo $ & 1.748 & 1.58E-02 & 6.32E-02 \\
$\htwo $ & 1.748 & 1.58E-02 & 6.32E-02 \\
$\htwo $ & 1.780 & 5.95E-07 & 3.60E-04 \\
$\htwo $ & 1.787 & 1.05E-02 & 4.35E-02 \\
$\htwo $ & 1.835 & 5.57E-02 & 6.37E-02 \\
$\hi $ & 1.875 & 8.39E-02 & 9.92E-02 \\
$\htwo $ & 1.891 & 2.83E-02 & 6.50E-02 \\
$\htwo $ & 1.957 & 1.15E-01 & 7.26E-02 \\
$\htwo $ & 2.033 & 4.33E-02 & 4.24E-02 \\
$\htwo $ & 2.121 & 1.23E-01 & 8.17E-02 \\
$\hi $ & 2.166 & 4.84E-03 & 6.35E-03 \\
$\htwo $ & 2.201 & 1.50E-03 & 3.16E-03 \\
$\htwo $ & 2.223 & 2.76E-02 & 6.17E-02 \\
$\htwo $ & 2.247 & 1.12E-02 & 2.46E-02 \\
$\htwo $ & 2.286 & 5.77E-04 & 6.93E-03 \\
$\htwo $ & 2.406 & 9.16E-02 & 7.88E-02 \\
$\htwo $ & 2.413 & 3.05E-02 & 6.81E-02 \\
$\htwo $ & 2.423 & 8.65E-02 & 5.73E-02 \\
$\htwo $ & 2.437 & 2.41E-02 & 2.35E-02 \\
$\htwo $ & 2.454 & 5.57E-02 & 3.50E-02 \\
$\htwo $ & 2.475 & 1.26E-02 & 2.90E-02 \\
$\htwo $ & 2.499 & 2.42E-02 & 2.77E-02 \\
$\htwo $ & 2.527 & 4.69E-03 & 1.94E-02 \\
$\htwo $ & 2.550 & 8.45E-03 & 2.52E-02 \\
$\htwo $ & 2.558 & 3.09E-03 & 9.46E-03 \\
$\htwo $ & 2.595 & 1.30E-03 & 9.44E-03 \\
$\htwo $ & 2.626 & 2.18E-02 & 4.93E-02 \\
$\htwo $ & 2.710 & 9.63E-04 & 2.06E-02 \\
$\htwo $ & 2.718 & 3.59E-04 & 5.05E-03 \\
$\htwo $ & 2.719 & 1.27E-03 & 8.02E-03 \\
$\htwo $ & 2.730 & 1.03E-03 & 7.14E-03 \\
$\htwo $ & 2.802 & 7.75E-02 & 6.66E-02 \\

\hline
\end{tabular}
\end{table}

\begin{table}
\centering
\caption{\label{tab:Predicted_MID_IR}  Predicted mid IR -- sub-mm line ratios.}
\begin{tabular}{ c c c c }
\hline
Line&$\lambda (\micron )$&Extra Heat&Cosmic Ray \\
\hline
$\htwo $ & 5.051 & 1.66E-02 & 3.70E-02 \\
$\htwo $ & 5.510 & 1.06E-01 & 1.57E-01 \\
$\htwo $ & 5.810 & 2.89E-03 & 1.16E-02 \\
$\htwo $ & 6.107 & 6.63E-02 & 7.30E-02 \\
$\htwo $ & 6.907 & 2.97E-01 & 2.47E-01 \\
$\htwo $ & 8.024 & 1.04E-01 & 1.04E-01 \\
$\htwo $ & 9.662 & 2.16E-01 & 1.83E-01 \\
$\htwo $ & 12.28 & 2.97E-02 & 3.15E-02 \\
$[\neii ]$ & 12.81 & 2.74E-04 & 2.78E-02 \\
$[\neiii ]$ & 15.55 & 3.17E-10 & 8.17E-03 \\
$\htwo $ & 17.03 & 1.79E-02 & 1.94E-02 \\
$[\siii ]$ & 18.67 & 1.48E-05 & 3.72E-03 \\
$\htwo $ & 28.21 & 3.12E-04 & 2.07E-03 \\
$[\siii ]$ & 33.47 & 2.89E-05 & 7.82E-03 \\
$[\rm{Si III} ]$ & 34.81 & 2.30E-02 & 8.98E-01 \\
$[\neiii ]$ & 36.01 & 2.76E-11 & 7.31E-04 \\
$[\oiii ]$ & 51.80 & 8.15E-11 & 1.29E-04 \\
$[\oi ]$ & 63.17 & 7.52E-02 & 2.75E+00 \\
$[\rm{Si I} ]$ & 68.40 & 1.43E-04 & 4.06E-03 \\
$[\oiii ]$ & 88.33 & 1.16E-10 & 1.81E-04 \\
$[\nii ]$ & 121.7 & 4.79E-03 & 7.83E-03 \\
$[\oi ]$ & 145.5 & 7.48E-03 & 1.43E-01 \\
$[\cii ]$ & 157.6 & 2.57E-02 & 1.31E-01 \\
CO & 323.6 & 4.17E-04 & 4.11E-02 \\
$[\ci ]$ & 609.2 & 2.75E-04 & 1.77E-02 \\
CO & 647.2 & 1.60E-03 & 7.05E-01 \\
CO & 863.0 & 1.07E-03 & 5.94E-01 \\

\hline
\end{tabular}
\end{table}

\label{lastpage}
\clearpage

\begin{thebibliography}{99}
\label{bibliography}

\bibitem[\protect\citeauthoryear{Abel et al}{2005}]{AbelEtal05}
Abel, N.P., Ferland, G.J., Shaw, G. \& van Hoof, P.A.M. 2005, ApJS, 161, 65

\bibitem[\protect\citeauthoryear{Adams}{1972}]{Adams72}
Adams, T. 1972, ApJ, 174, 439

\bibitem[\protect\citeauthoryear{Allers et al}{2005}]{AllersEtal05}
Allers, K. N., Jaffe, D. T., Lacy, J. H., Draine, B. T. \& Richter, M. J.
2005, ApJ, 630, 368

\bibitem[\protect\citeauthoryear{Anderson et al}{2000}]{Anderson00}
Anderson, H., Ballance, C.P., Badnell, N.R., \& Summers, H.P
2000, J Phys B, 33, 1255

\bibitem[\protect\citeauthoryear{Baker \& Menzel}{1938}]{BakerMenzel38}
Baker, J.G., \& Menzel, D.H. 1938, ApJ, 88, 52

\bibitem[\protect\citeauthoryear{Baldwin et al}{1995}]{BaldwinEtAl95}
Baldwin, J. A., Ferland, G. J., Korista K. T., \& Verner, D. 1995, ApJ, 455, L119

\bibitem[\protect\citeauthoryear{Black \& van Dishoeck}{1987}]{BlackvanDishoeck87}
Black, John H. \& van Dishoeck, E.F. 1987, ApJ, 322, 412

\bibitem[\protect\citeauthoryear{Cazaux \& Tielens}{2002}]{CazauxTielens02}
Cazaux, S., \& Tielens, A.G.G.M., 2002, ApJ, 575, L29-L32

\bibitem[\protect\citeauthoryear{Conselice et al}{2001}]{ConseliceEtAl02}
Conselice, C.J., Gallagher, J.S. \& Wyse, R.F.G. 2001, AJ, 122, 2281

\bibitem[\protect\citeauthoryear{Crawford \& Fabian}{1992}]{CrawfordFabian92}
Crawford, C.S., \& Fabian, A.C. 1992, MNRAS, 259, 265

\bibitem[\protect\citeauthoryear{Dalgarno}{1996}]{Dalragno96}
Dalgarno, A., 1996, PNAS 10312269

\bibitem[\protect\citeauthoryear{Dalgarno \& McCray}{1972}]{DalragnoMcCray72}
Dalgarno, A., \& McCray, R. A. 1972, ARA\&A, 10, 375

\bibitem[\protect\citeauthoryear{Dalgarno et al}{1973}]{DalgarnoEtAl73}
Dalgarno, A., Black, J.H., \& Weisheit, J.C. 1973, ApL, 14, 77

\bibitem[\protect\citeauthoryear{Dalgarno et al}{1999}]{DalgarnoEtAl99}
Dalgarno, A., Yan, Min, \& Liu, Weihong 1999, ApJS, 125, 237

\bibitem[\protect\citeauthoryear{Dyson \& Williams}{1997}]{DysonWilliams97}
Dyson, J.E., \& Williams, D.A. 1997, The Physics of the Interstellar Medium (Bristol; Institute of Physics Publishing)

\bibitem[\protect\citeauthoryear{Elitzur \& Ferland}{1986}]{ElitzurFerland86}
Elitzur, M., \& Ferland, G. J. 1986, ApJ, 305, 35

\bibitem[\protect\citeauthoryear{Fabian et al}{2008}]{FabianEtAl08}
Fabian, A.C., Johnstone, R.M., Sanders, J.S., Conselice, C.J., Crawford, C.S.,
Gallagher III, J.S. \& Zweibel, E., 2008, Nature submitted

\bibitem[\protect\citeauthoryear{Fabian et al}{2003}]{FabianEtAl03}
Fabian, A.C., Sanders, J.S., Crawford, C.S., Conselice, C.J.,
Gallagher III, J.S. \& Wyse, R.F.G., 2008, MNRAS 344, L48

\bibitem[\protect\citeauthoryear{Ferland}{2003}]{FerlandARAA}
Ferland, G.J., 2003, ARA\&A, 41, 517

\bibitem[\protect\citeauthoryear{Ferguson \& Ferland}{1997}]{FergusonFerland97}
Ferguson J.W., \& Ferland G.J., 1997, ApJ, 479, 363

\bibitem[\protect\citeauthoryear{Ferland et al}{1998}]{FerlandCloudy98}
Ferland G.J., Korista K.T., Verner D.A., Ferguson J.W., Kingdon J.B., Verner E.M., 1998, PASP, 110, 761

\bibitem[\protect\citeauthoryear{Ferland et al}{2008}]{FerlandEtAl08}
Ferland, G.J., Fabian, A.C., Hatch, N., Johnstone, R., Porter, R.L.,
van Hoof, P.A.M., \& Williams, R.J.R. 2008,
MNRAS 386, 72L

\bibitem[\protect\citeauthoryear{Ferland et al}{1994}]{FerlandFabianJohnstone94}
Ferland, G. J., Fabian, A. C., \& Johnstone, R.M. 1994, MNRAS, 266, 399

\bibitem[\protect\citeauthoryear{Ferland et al}{2002}]{FerlandFabianJohnstone02}
Ferland, G. J., Fabian, A. C., \& Johnstone, R.M. 2002, MNRAS, 333, 876

\bibitem[\protect\citeauthoryear{Ferland \& Mushotzky}{1984}]{FerlandMushotzkyCR84}
Ferland, G. J., \& Mushotzky, R. F. 1984, ApJ, 286, 42

\bibitem[\protect\citeauthoryear{Ferland \& Netzer}{1979}]{FerlandNetzer79}
Ferland, G. J., \& Netzer, H. 1979, ApJ, 229, 274

\bibitem[\protect\citeauthoryear{Ferland \& Rees}{1988}]{FerlandRees84}
Ferland, G. J., \& Rees, M. J. 1988, ApJ, 332, 141

\bibitem[\protect\citeauthoryear{Fernandes et al}{1997}]{FernandesEtAl07}
Fernandes, A.J.L., Brand, P.W.J.L. \& Burton, M.G. 1997, MNRAS, 290, 216

\bibitem[\protect\citeauthoryear{Field}{1965}]{Field65}
Field, G. B. 1965, ApJ, 142, 431

\bibitem[\protect\citeauthoryear{Field et al.}{1969}]{FieldGoldsmithHabing69}
Field, G.B., Goldsmith, D.W. \& Habing, H.J. 1969, ApJ, 155L, 149

\bibitem[\protect\citeauthoryear{Guidetti et al}{2008}]{GuidettiEtAl08}
Guidetti, D., Murgia, M., Govoni, F., Parma, P., Gregorini, L., de Ruiter, H.R., Cameron, R.A. \& Fanti, R. 2008, A\&A, 483, 699

\bibitem[\protect\citeauthoryear{Haardt \& Madau}{1996}]{HaardtMadau96}
Haardt, F., \& Madau, P., 1996, ApJ, 461, 20

\bibitem[\protect\citeauthoryear{Harrington}{1973}]{Harrington73}
Harrington, J. P. 1973, MNRAS, 162, 43

\bibitem[\protect\citeauthoryear{Hatch et al}{2005}]{HatchEtAl05}
Hatch, N.A., Crawford, C.S., Fabian, A.C., \& Johnstone, R.M. 2005,
MNRAS 358, 765

\bibitem[\protect\citeauthoryear{Hatch et al}{2006}]{HatchEtAl06}
Hatch, N.A., Crawford, C.S., Johnstone, R.M. \& Fabian, A.C., 2006,
MNRAS 367, 433

\bibitem[\protect\citeauthoryear{Heiles \& Crutcher}{2005}]{HeilesCrutcher05}
Heiles, C. \& Crutcher, R. 2005, Magnetic Fields in Diffuse H I and Molecular Clouds,
Chapter in Cosmic Magnetic Fields, astro-ph/0501550

\bibitem[\protect\citeauthoryear{Indriolo et al}{2007}]{IndrioloEtAl07}
Indriolo, N., Geballe, T.R., Oka, Takesi, \& McCall, B.J. 2007, ApJ, 671, 1736

\bibitem[\protect\citeauthoryear{Jaffe et al}{2005}]{JaffeEtAl05}
Jaffe, W., Bremer, M.N. \& Baker, K. 2005,
MNRAS 360, 748

\bibitem[\protect\citeauthoryear{Jaffe et al}{2001}]{JaffeEtAl01}
Jaffe, W., Bremer, M.N. \& van der Werf, P. P. 2001,
MNRAS 324, 443

\bibitem[\protect\citeauthoryear{Johnstone et al}{2007}]{JohnstoneEtAl07}
Johnstone, R., Hatch, N., Ferland, G.J., Fabian, A.C., Crawford, C., \& Wilman, R. 2007,
MNRAS 382, 1246

\bibitem[\protect\citeauthoryear{Jura}{1975}]{Jura75}
Jura, M., 1975, ApJ, 197, 575

\bibitem[\protect\citeauthoryear{Kingdon et al}{1995}]{KingdonEtAl95}
Kingdon, J., Ferland, G.J., \& Feibelman, W.A. 1995, ApJ, 439, 793

\bibitem[\protect\citeauthoryear{Lazarian}{2005}]{Lazarian05}
Lazarian, A., 2005, in Magnetic Fields in the Universe: From
Laboratory and Stars to Primordial Structures, edited by de Gouveia
dal Pino, E.~M., Lugones, G., Lazarian, A., AIP Conference
Proceedings, 784, pp. 42-53

\bibitem[\protect\citeauthoryear{Le Bourlot et al}{1999}]{LeBourletEtAlCollFits99}
Le Bourlot, J.; Pineau des Forêts, G.; Flower, D. R. 1999, MNRAS, 305, 802

\bibitem[\protect\citeauthoryear{Loewenstein \& Fabian}{1990}]{LoewensteinFabian90}
Loewenstein, M. \& Fabian, A.C. 1990, MNRAS, 242, 120

\bibitem[\protect\citeauthoryear{Lupu, France \& McCandliss}{2006}]{LupuEtAl06}
Lupu, R.E., France, K. \& McCandliss, S.R. 2006, ApJ, 644, 981L

\bibitem[\protect\citeauthoryear{Mandy \& Martin}{1993}]{MandyMartin93}
Mandy, M.E. \& Martin, P.G. 1993, ApJS, 86, 119

\bibitem[\protect\citeauthoryear{Myers \& Goodman}{1998}]{MyersGoodman98}
Myers, P.C. \& Goodman, A.A. 1998, ApJ, 326, L27

\bibitem[\protect\citeauthoryear{Osterbrock \& Ferland}{2006}]{AGN3}
Osterbrock D., Ferland G.J., 2006,
Astrophysics of gaseous nebulae and active galactic nuclei, 2nd.~ed
.~by D.E.~Osterbrock and G.J.~Ferland.~Sausalito, CA: University Science Books,

\bibitem[\protect\citeauthoryear{Pengelly}{1964}]{Pengelly64}
Pengelly, R. M. 1964, MNRAS, 127, 145

\bibitem[\protect\citeauthoryear{Pengelly \& Seaton}{1964}]{PengellySeaton64}
Pengelly, R. M. \& Seaton, M.J. 1964, MNRAS, 127, 165

\bibitem[\protect\citeauthoryear{Pope et al}{2008}]{PopeEtAl08}
Pope, E. C. D., Hartquist, T. W. \& Pittard, J. M., 2008.  MNRAS,
{\bf 389, 1259}

\bibitem[\protect\citeauthoryear{Porter et al}{2005}]{PorterEtAl05}
Porter, R.L., Bauman, R.P., Ferland, G.J., \& MacAdam, K. B. 2005, ApJ, 622, 73L (astro-ph/0502224)

\bibitem[\protect\citeauthoryear{Porter \& Ferland}{2007}]{PorterFerland07}
Porter, R.L., \& Ferland, G.J. 2007, ApJ, 664, 586 (astro-ph/0704.2642)

\bibitem[\protect\citeauthoryear{Revaz et al}{2008}]{RevazEtAl08}
Revaz, Y., Combes, F., \& Salom\'e, P. 2008, A\&A, 477, L33

\bibitem[\protect\citeauthoryear{R\"{o}llig et al}{2007}]{Rollig07}
R\"{o}llig, M., Abel, N. P. Bell, T. Bensch, F. Black, J. Ferland, G. J. Jonkheid, B. Kamp, I. Kaufman, M.J. Le Bourlot, J. Le Petit, F. Meijerink, R. Morata, O. Ossenkopf, V. Roueff, E. Shaw, G. Spaans, M. Sternberg, A. Stutzki , J. Thi, W.-F. van Dishoeck, E. F. van Hoof, P. A. M. Viti, S. \& Wolfire, M.G. A\&A,
2007, 467, 187

\bibitem[\protect\citeauthoryear{Salom\'e et al}{2006}]{SalomeEtAl06}
Salom\'e, P., Combes, F., Edge, A.C. et al 2006, A\&A, 454, 437

\bibitem[\protect\citeauthoryear{Salom\'e et al}{2008a}]{SalomeEtAl08a}
Salom\'e, P., Revaz, Y., Combes, F., Pety, J., Downes, D., Edge, A.C.,
\& Fabian, A.C., 2008a, A\&A, 483, 793

\bibitem[\protect\citeauthoryear{Salom\'e et al}{2008b}]{SalomeEtAl08b}
Salom\'e, P., Combes, F., Revaz, Y., Edge, A.C., Hatch, N.A.,
Fabian, A.C. \& Johnstone, R.M., 2008b, A\&A, 484, 317

\bibitem[\protect\citeauthoryear{Sanders \& Fabian}{2007}]{SandersFabian07}
Sanders, J.S. \& Fabian, A.C. 2007, MNRAS, 381, 1381 (astro-ph 0705.2712)

\bibitem[\protect\citeauthoryear{Schlegel et al}{1998}]{SchlegelEtAl98}
Schlegel, D.~J., Finkbeiner, D.~P. \& Davis, M. 1998, ApJ, 500, 525S

\bibitem[\protect\citeauthoryear{Sellgren et al}{1990}]{SellgrenEtal90}
Sellgren, K., Tokunaga, A. T., \& Nakada, Y. 1990, ApJ, 349, 120

\bibitem[\protect\citeauthoryear{Shaw et al}{2005}]{ShawEtal05}
Shaw, G.,  Ferland, G.J. Abel, N.P., Stancil, P.C. \& van Hoof, P.A.M. 2005, ApJ, 624, 794

\bibitem[\protect\citeauthoryear{Shaw et al}{2006}]{ShawEtal06}
Shaw, G., Ferland, G. J., Srianand, R., \& Abel, N. P. 2006, ApJ, 639, 941

\bibitem[\protect\citeauthoryear{Shaw et al}{2008}]{ShawEtal08}
Shaw, G., Ferland, G. J., Srianand, R., Abel, N. P., van Hoof , P. A. M. \&
Stancil P. C. 2008, ApJ 675, 405

\bibitem[\protect\citeauthoryear{Shemansky et al}{1985}]{ShemanskyEtal85}
Shemansky, D.E., Ajello, J.M., \& Hall, D. T. 1985, ApJ, 296,765

\bibitem[\protect\citeauthoryear{Spitzer \& Tomasko}{1968}]{SpitzerTomasko68}
Spitzer, L., \& Tomasko, M. G. 1968, ApJ, 152, 971

\bibitem[\protect\citeauthoryear{Sternberg}{2005}]{Sternberg05}
Sternberg, A., 2005, Astrochemistry:
Recent Successes and Current Challenges, Proceedings of the 231st Symposium of the International Astronomical Union,
Edited by Lis, Dariusz C.; Blake, Geoffrey A.; Herbst, Eric. Cambridge: Cambridge University Press,
2005., pp.141-152

\bibitem[\protect\citeauthoryear{Sternberg \& Dalgarno}{1989}]{SternbergDalgarno89}
Sternberg, A., \& Dalgarno, A. 1989, ApJ, 338, 197

\bibitem[\protect\citeauthoryear{Sternberg et al}{1987}]{SternbergEtAl87}
Sternberg, A., Dalgarno, A. \& Lepp, S. 1987, ApJ, 320, 676

\bibitem[\protect\citeauthoryear{Taylor et al}{2007}]{TaylorEtAl07}
Taylor, G.B., Fabian, A.C., Bentile, G., Allen, S.W., Crawford, C. \& Sanders, J.S. 2007,
MNRAS 382, 67

\bibitem[\protect\citeauthoryear{Tielens}{2005}]{Tielens05}
Tielens, A. G. G. M., 2005, The Physics and Chemistry of the Interstellar Medium, Cambridge, UK: Cambridge University Press

\bibitem[\protect\citeauthoryear{Tine et al}{1997}]{TineEtAl97}
Tine, S.,; Lepp, S., Gredel, R. \& Dalgarno, A., 1997, ApJ, 481, 282

\bibitem[\protect\citeauthoryear{Tytarenko et al}{2002}]{TytarenkoWilliamsFalle02}
Tytarenko, P. V., Williams, R. J. R., \& Falle, S. A. E. G., 2002, MNRAS, 337, 117

\bibitem[\protect\citeauthoryear{van Dishoeck}{2004}]{vanDishoeck04}
van Dishoeck, E.F. 2004, ARA\&A, 42, 119

\bibitem[\protect\citeauthoryear{Webber}{1998}]{Webber98}
Webber, W.R. 1998, ApJ, 506, 329

\bibitem[\protect\citeauthoryear{Williams et al}{1998}]{WilliamsElAl98}
Williams, J.P., Bergin, E.A., Caselli, P., Myers, P.C., \& Plume, R. 1998, ApJ, 503, 689

\bibitem[\protect\citeauthoryear{Wrathmall et al}{2007}]{WrathmallEtAl07}
Wrathmall, S.A., Gusdorf, A., \& Flower. D.R. 2007, MNRAS 382, 133

\bibitem[\protect\citeauthoryear{Xu \& McCray}{1991}]{XuMcCray91}
Xu, Y., \& McCray, R. 1991, ApJ, 375, 190

\end{thebibliography}
\end{document}